\documentclass[12pt]{article}

\usepackage{amsmath}
\usepackage{amssymb}
\usepackage{amsfonts}
\usepackage{fancyhdr}
\usepackage[dvips]{graphicx}
\usepackage{rotating}
\usepackage{psfrag}
\usepackage{epsfig}
\usepackage{url}

\numberwithin{equation}{section}

\def\lfig#1#2#3#4{
\begin{figure}
\centerline{\hfill \includegraphics[height=#3]{#2}\hfill}
\caption{#1 \label{#4}}
\end{figure}
}

\def\TEXfig#1#2#3{
\begin{figure}
\centerline{\hfill #1 \hfill}
\caption{#2 \label{#3}}
\end{figure}
}

\newcounter{tabl}
\newcommand{\be}{\begin{equation}}
\newcommand{\ee}{\end{equation}}
\newcommand{\beq}{\begin{eqnarray}}
\newcommand{\eeq}{\end{eqnarray}}
\newcommand{\bea}[2]{\be\label{#2}\begin{array}{#1}}
\newcommand{\eea}{\end{array}\ee}

\def\Cb{{\rm \bf C}}

\newcommand{\bR}{\mathbb{R}}

\newcommand{\Nint}{\mathbb{N}}
\def\Tr{\,{\rm Tr}\, }
\def\det{\,{\rm det}\, }

\def\tr{\,{\rm tr}\, }

\def\rangl{\right\rangle   }
\def\langl{\left\langle  }
\def\({\left(}
\def\){\right)}
\def\[{\left[}
\def\]{\right]}
\def\p{\partial}
\newcommand{\de}{\mathrm{d}}
\newcommand{\I}{\mathrm{i}}

\def\11{1\!\! 1}

\def\hf{\frac{1}{2}}
\def\hft{{\textstyle\frac{1}{2}}}

\def\eps{\varepsilon}

   \def\CC {{\cal C}}
   \def\CD {{\cal D}}

   \def\CG {{\cal G}}
   \def\CH {{\cal H}}
   \def\CI {{\cal I}}
   
   \def\CK {{\cal K}}
   
   \def\CM {{\cal M}}
   
   \def\CO {{\cal O}}

   \def\CS {{\cal S}}
   
   \def\CU {{\cal U}}
   \def\CV {{\cal V}}

\newcommand{\Cmat}{{\mathbb C}}

\newcommand{\Umat}{{\mathbb U}}
\newcommand{\Rmat}{{\mathbb R}}

\newcommand{\gmat}{\mathfrak{g}}

\newcommand{\hT}{\hat T}

\newcommand{\im}{\gamma}

\newcommand{\tE}{\lefteqn{\smash{\mathop{\vphantom{<}}\limits^{\;\sim}}}E}
\newcommand{\tP}{\lefteqn{\smash{\mathop{\vphantom{<}}\limits^{\;\sim}}}P}
\newcommand{\tQ}{\lefteqn{\smash{\mathop{\vphantom{<}}\limits^{\;\sim}}}Q}

\newcommand{\Pt}{\lefteqn{\smash{\mathop{\vphantom{\Bigl(}}\limits_{\sim}
\atop \ }}P}
\newcommand{\Qt}{\lefteqn{\smash{\mathop{\vphantom{\Bigl(}}\limits_{\sim}
\atop \ }}Q}

\newcommand{\SA}{{\cal A}}
\newcommand{\SAab}{^{(a,b)}\!\!{\cal A}}
\newcommand{\SApar}[2]{^{(#1,#2)}\!\!{\cal A}}
\newcommand{\SSA}{{\bf A}}

\newcommand{\Ab}[2]{{A^{(\im)#2}_{#1}}}

\newcommand{\tPb}{{\tP_{\smash{(\im)}}}}
\newcommand{\tEb}[2]{{\tE_{(\im)#2}^{#1}}}
\newcommand{\Abi}[2]{{A^{(\I )#2}_{#1}}}
\newcommand{\tEbi}[2]{{\tE_{(\I)#2}^{#1}}}

\newcommand{\Ppr}[2]{\pi^{(#1)}{(#2)}}
\newcommand{\pr}[1]{\pi^{(#1)}}

\newcommand{\oim}{\frac{1}{\im}}

\newcommand{\cohr}[2]{|{ #1},{\bf #2}\rangle}
\newcommand{\cohl}[2]{\langle{ #1},{\bf #2}|}
\newcommand{\Xbn}[2]{{\bf X}_{( #1,{\bf #2})}}

\def\jp{j^+}
\def\jm{j^-}
\def\jpm{j^\pm}

\def\np{n^+}
\def\nm{n^-}
\def\npm{n^\pm}

\newcommand{\gb}{{\rm g}}

\newcommand{\gx}{{\rm g}}
\newcommand{\gl}{g}
\newcommand{\Db}{{\bf D}}

\def\Int{\CI}

\newcommand{\subsubsubsection}[1]{
\vspace{0.3cm}
\noindent {\bf #1}
\vspace{0.3cm}
}

\begin{document}
%
%

\title{Critical Overview of Loops and Foams}

\vspace{0.7cm}

\author{Sergei Alexandrov and Philippe Roche}
\date{}

\maketitle

\vspace{-1cm}
\begin{center}
\emph{Laboratoire de Physique Th\'eorique \&
Astroparticules, CNRS UMR 5207,}\\
\emph{Universit\'e Montpellier II, 34095 Montpellier Cedex 05, France}
\end{center}
\vspace{0.1cm}
\begin{abstract}
This is a review of the present status of loop and spin foam approaches
to quantization of four-dimensional general relativity.
It aims at raising various issues which seem to challenge some of the methods and the results
often taken as granted in these domains.
A particular emphasis is given to the issue of diffeomorphism and local Lorentz symmetries
at the quantum level and to the discussion of new spin foam models.
We also describe modifications of these two approaches which may overcome their problems and
speculate on other promising research directions.

\end{abstract}

\newpage

\tableofcontents

\section{Introduction}

Whereas string theory remains the most developed and active approach to quantum
gravity, during last years Loop Quantum Gravity (LQG) and Spin Foam (SF) models
become more and more popular.
Following this growth of interest, there have appeared many reviews on this subject
\cite{Gaul:1999ys,Thiemann:2007zz,Thiemann:2001yy,Perez:2003vx,Oriti:2003wf,Rovelli:2004tv,Ashtekar:2004eh,Smolin:2004sx,Ashtekar:2007tv,Rovelli:2010wq}.
Most of these reviews give a quite optimistic picture of the developments
in this domain, so that one could think that one has already at our disposal
a theory (or at least a model) of quantum gravity, which is derived in a rigorous way following
the standard well established quantization rules, internally self-consistent
and able to produce physical predictions.
Besides, there have been several reviews comparing string theory and LQG/SF approaches
written mostly from the point of view of the latter
\cite{Horowitz:2000sh,Rovelli:2003wd,Smolin:2003rk}.
On the other hand, a critical overview of the recent advances almost does not exist in the literature.
A notable exception is the review \cite{Nicolai:2005mc}
(see also its answer and criticism \cite{Thiemann:2006cf}) which concentrates mostly on the Loop Quantum Gravity side.

The present review aims to fulfill this gap. It gives a picture of the present situation
in the domain of LQG and SF seen from the perspective which is based on the results obtained by the authors
during last years. Thus, it represents a personal viewpoint
which might not coincide with the viewpoint spread in the community.
Although we have raised the points already known by the experts in the field, they are rarely spelled out explicitly.
At the same time, their understanding is crucial for the viability of these theories.
Unfortunately, it is extremely difficult to obtain exact results which may
resolve the raised questions in one or another way.
Besides, even the quantization procedure which must be followed in the case of gravity
is not completely well determined and still subject to questioning.
Therefore, our analysis cannot be claimed to be a proof of any kind,
but it is supported by various arguments that we find compelling.

The central question to be answered below is whether one has
a model at Planck scale which is mathematically self-consistent, has a potential to reproduce
general relativity in the low-energy limit, and incorporates consistently
its fundamental gauge symmetries. The last condition is the main point of concern
in this review. We accept the viewpoint that the fundamental symmetries of general relativity,
such as space-time diffeomorphism invariance and local Lorentz invariance in the tangent space,
appearing as we are working in the first order formalism,
must {\it not} be anomalous in the corresponding quantum theory.

Of course, as soon as the notion of manifold may disappear at the quantum level
as it seems to happen in LQG, or be replaced by a simplicial complex as in SF models,
one should precise what is meant by diffeomorphism invariance.
For example, in the discrete setting this issue is very non-trivial and has been discussed in \cite{Dittrich:2008pw,Bahr:2009qc}.
In this review we do not enter the discussion of this extremely important problem,
but subscribe to the commonly accepted idea
that the diffeomorphism symmetry in SF models is recovered as a result of summing over all admissible triangulations.
In general, we take the viewpoint that a symmetry remains unbroken if
the quantum theory properly implements the corresponding constraints of the classical phase space.
All our conclusions are based on this assumption and therefore should be taken with care.

Thus, we will pay a particular attention to the imposition of constraints in both loop and SF quantizations.
Since LQG is supposed to be a canonical quantization of general relativity, in principle, it should be
straightforward to verify the constraint algebra at the quantum level. However, in practice, due to peculiarities
of the loop quantization this cannot be achieved at the present state of knowledge. Therefore, we use
indirect results to make conclusions about this issue. In particular, we use a Lorentz covariant approach
\cite{Alexandrov:2000jw}, which allows to put LQG in a broader context and thereby to get insights on the fate of
classical symmetries after quantization.

In contrast, the SF approach is based on a covariant path integral. Nevertheless, constraints
play an important role in this case too since the main idea of this approach is to get
quantum gravity by imposing certain constraints on the topological BF theory.
During  the three previous years there was a great activity in this area. It was initiated
by reconsidering the  old methods of imposing constraints and resulted in new techniques \cite{Livine:2007vk}
and new SF models \cite{Freidel:2007py,Engle:2007wy}.
These models have been shown to possess some attractive features and in particular
it was claimed that they are consistent with LQG at least at the kinematical level.
Thus, it might look like one has a beautiful coherent picture where two different
quantization approaches lead to equivalent consistent quantum gravity models.

However, we suggest to revisit the constraint imposition once more, now for the new SF models.
For this purpose it is extremely useful to take into account the canonical structure of the theory
so that the preceding covariant analysis of LQG will be very helpful.
Such approach immediately reveals various fallacies of the new models and weak points in the
interpretation of their results.

Our main conclusions regarding the status of the two quantization approaches are the following:
\begin{itemize}
\item
Although LQG can perfectly incorporate the full local Lorentz symmetry,
we find some evidences that LQG might have
problems to maintaining space-time diffeomorphism symmetry at the quantum level.
Thus, we argue that it is an {\it anomalous} quantization of general relativity
which is not physically acceptable.
\item
There is an alternative quantization following the same loop ideas, the so called Covariant LQG (CLQG),
which has a potential to resolve the drawbacks of LQG.
However, it is supplied with some serious technical obstacles
(consisting mainly  in finding a representation of the algebra of connections)
preventing yet  the realization of this quantization program.
\item
The claim \cite{Engle:2007wy} that the recently introduced spin foam models \cite{Freidel:2007py,Engle:2007wy}
have the same boundary states as the kinematical states of LQG cannot be formulated as such
because they have completely different representations as functionals of connection.
\item
The new spin foam models in the presence of
a {\it finite} Immirzi parameter represent quantizations which do not respect the
standard Dirac rules and we argue that they are  incompatible with a self-consistent  canonical quantization.
Moreover, any SF model derived by the usual strategy ``first quantize, then constrain" (see section \ref{subsubsec_strategy}),
including the models {\it without} the Immirzi parameter,
does not implement consistently all constraints of general relativity and therefore cannot properly
describe its quantum dynamics.
\item A spin foam quantization consistent with the canonical one can be achieved by
modifying the association of geometric bi-vectors to generators of the gauge algebra and
by relaxing the closure constraint. The vertex amplitude should also be modified
and in general is given by the integral formula \eqref{amplit_simplex} with a non-trivial measure
which however remains still unknown.
\end{itemize}

Given these statements, we have to conclude that neither the canonical loop approach nor its
spin foam cousin were able to provide so far a model which can be claimed
to be free from inconsistencies and anomalies.
This  does not mean however that the ideas behind these approaches are not reasonable.
They might well be relevant and even indispensable for a theory of quantum gravity.
For example, as  argued in \cite{Carlip:2009kf}, there is a remarkable convergence of various
modern approaches to quantum gravity, which all lead to an effective spacetime dimension 2 at the Planck scale.
This feature is exhibited very explicitly in the loop and spin foam quantizations.
However, in our opinion, their present realization is not satisfactory
and requires a serious reconsideration.
In fact, in this review we consider some of the modifications,
which have been already suggested, and point out their advantages and loopholes.

The review is not technical in the sense that the details of proofs and derivations, if they are not subject of
a critical discussion, are omitted. Instead, we concentrate mostly on the ideas behind
the loop and SF quantizations, their results, properties and interpretation.
On the other hand, the review does not address such important issues as the problem of time,
prediction power on the low energy limit, inclusion of matter, {\it etc.}
Besides, we do not consider some branches of LQG and SF such as, for example, Loop Quantum Cosmology (LQC)
\cite{Bojowald:2006da}
and evaluation of the graviton propagator \cite{Rovelli:2005yj}.
Since these branches are based on results and ideas of the two main approaches,
they seem to have even less firm ground than those approaches themselves.
Therefore, for example, if LQG in its present form
fails to provide a consistent quantization of general relativity,
it is highly unlikely that LQC can do better.

Let us briefly describe the content of the review.
Chapter \ref{sec_canan} is devoted to the loop approach to quantum gravity which is a type of canonical quantization.
In the first section of this chapter we recall the basics of LQG following the standard
presentation of this domain.
Then in section \ref{subsec_lorcf} we briefly review the Lorentz Covariant approach,
which allows to look at LQG from a different angle and suggests an alternative loop quantization,
as discussed in section \ref{subsec_twoq}.
The implications of these results are further analyzed in section \ref{subsec_LQGdiscus},
where we also discuss various controversial issues and constructions of LQG.
The closing section of this chapter presents a summary of our main conclusions regarding the status
of the loop quantization.

Chapter \ref{sec_path} deals with the SF quantization which can be seen as
a discretized path integral for general relativity.
First, we give a brief introduction to general ideas of the SF approach and
present the strategy followed in most of the SF models in 4 dimensions.
Then we introduce the main SF models existing in the literature and realizing
the strategy mentioned above: section \ref{subsubsec_BC} reviews the Barrett--Crane model
and section \ref{subsec_newvert} presents the new models.
Since we want to revise their derivation, we discuss its main steps in detail and
critically analyze their relation to LQG. Then in section \ref{subsec_imposeconstr}
we reconsider the imposition of constraints in the new models and observe a few sources of potential mistakes.
The main issue, in our opinion, is that the SF models quantize the symplectic structure not of general relativity,
but of BF theory. This results in various inconsistencies demonstrated on a simple example
in subsection \ref{subsubsec_example}. This does not however exhaust all problems which
become apparent under the thorough analysis of the constraint imposition in the following subsections.
At the same time this analysis reveals some connections to the structures appearing
in the covariant approach to the canonical loop quantization and suggests a way to cure the problems.
So we finish this chapter by section \ref{subsec_compar} comparing the situations
we arrived at in the spin foam and loop approaches to quantum gravity.

Finally, in chapter \ref{sec_where} we speculate on possibilities
to overcome the problems exposed in this review.
We consequently discuss the canonical (loop) approach, path integral (spin foam) quantization
and group field theory reformulation of SF models. The last section \ref{subsec_effect}
is devoted to a brief discussion of holography and the perspective on gravity as an emergent theory put
in the context of LQG and spin foams.

\newpage

\section{Canonical approach}
\label{sec_canan}

\subsection{SU(2) Loop Quantum Gravity}
\label{subsec_LQG}

We start by briefly reviewing the main elements of LQG.
Although it is possible to provide a mathematically precise construction
involving cylindrical functions, GNS construction, projective limits, {\it etc.},
which can be found for instance in \cite{Thiemann:2007zz,Thiemann:2001yy},
our exposition will be very elementary. It is nevertheless sufficient for introducing the main ideas
and for understanding  the strong and the weak points of this approach.

\subsubsection{Ashtekar--Barbero canonical formulation}
\label{subsubsec_classLQG}

LQG is based on a canonical formulation of general relativity obtained by
a combination of ideas of Ashtekar \cite{Ashtekar:1986yd,Ashtekar:1987gu}
and Barbero \cite{Barbero:1994an,Barbero:1994ap}.
In the following it will be called Ashtekar--Barbero (AB) formulation.
There are many ways to arrive at this formulation. We follow the way
which is the most direct and most suitable for generalizations presented below.

The starting point is the so called Holst action \cite{Holst:1995pc}
\be
S_{(\im)}[e, \omega]=\frac{1}{16\pi G} \int_{\cal M} \eps_{IJKL} e^I \wedge e^J \wedge
\(F^{KL}(\omega)+\frac{1}{\im}\, \star F^{KL}(\omega)\),
\label{Sd1}
\ee
where ${\cal M}$ is a four dimensional oriented manifold, the index $I$ belongs to $\{0, 1,2,3\}$,
$e^I$ is a set of one-forms giving the cotetrad, $\omega^{IJ}$ denotes the one-form of the spin-connection,
the Hodge star operator is defined on antisymmetric tensors as
$ \star B^{IJ}=\frac 12 {\eps^{IJ}}_{KL}B^{KL} $, and $F^{IJ}(\omega)$ is the curvature of
the spin-connection $\omega$. The metric is recovered by
$g_{\mu\nu}=\eta_{IJ}e^I_\mu e^{J}_{\nu}$  where $\eta=(\sigma,+,+,+)$ with $\sigma=\pm$ for Riemannian and Lorentzian
cases, respectively. Although in this chapter most of equations are written for the Lorentzian case, they are easily
generalized to the other signature.

The Holst action is a generalization of the usual Hilbert--Palatini action representing Einstein
gravity in the first order formalism.
The second term in \eqref{Sd1} does not change
the dynamics since the equations of motion following from it coincide with the usual
Cartan equations ensuring the vanishing of torsion. Moreover, it vanishes on the surface
of these equations. As a result, the coupling constant $\im$ in front of this term,
called Immirzi parameter, is a  real parameter  which is completely free
in the classical theory and nothing depends on it.
Its role in quantum theory is a controversial issue which will be discussed a lot in the following.

To construct the Hamiltonian formalism for the action \eqref{Sd1}, one assumes that
${\cal M}={\mathbb R}\times M$ where $M$ is a three dimensional manifold and introduces
a 3+1 decomposition of  ADM type, this time for the tetrad field.
Since the action \eqref{Sd1} possesses several gauge symmetries, 4 diffeomorphism
symmetries and 6 local Lorentz invariances in the tangent space, there will be 10
corresponding first class constraints in the canonical formulation.
Unfortunately, as will become clear in section \ref{subsec_lorcf},
there are additional constraints of second class which cannot be solved explicitly in
a Lorentz covariant way. To avoid these complications, one usually follows
an alternative strategy. It involves the following three steps:\label{page_LQGreduct}
\begin{enumerate}
\item the boost part of the local Lorentz gauge symmetry is fixed from the very beginning by
choosing the so called {\it time gauge}, which imposes a certain condition on the tetrad field
($e^0=Ndt$ so that the normal to the equal-time hypersurfaces is time directed);
\item the three first class constraints generating the boosts are solved explicitly
w.r.t. space components of the spin-connection;
\item the same is done for six second class constraints which leaves only nine independent
components for the space part of the spin-connection.
\end{enumerate}

The result of these steps is the following phase space.\footnote{We deviate
from the common notations in the literature and denote
$SU(2)$ indices in the tangent space by small letters $a,b,\dots$ from the beginning of the alphabet,
whereas the space coordinates are labeled by indices $i,j,\dots$ from the middle.}
The canonical coordinates and momenta are given by
\be
\Ab{i}{a}=\Gamma_i^a(\tE)-\im K_i^a, \qquad
\tEb{i}{a}=\oim \tE^i_a,
\label{change}
\ee
where $\Gamma_i^a(\tE)$ is the Christoffel connection compatible with the triad $E_i^a=e_i^a$
and $K_i^a=\omega_i^{0a}$ is the extrinsic curvature.
The spatial volume element is $\sqrt{h}=\det E_i^a$ from which one defines the
densitized triad $\tE^i_a=\sqrt{h} E^i_a$ where $E^i_a$ is the inverse of $E_i^a.$
By definition of the canonical coordinates, the only non-vanishing Poisson brackets are\footnote{We will omit
the dependence on the space coordinates and the corresponding $\delta^3(x,y)$ on the r.h.s. of the commutation relations.
Besides, we will not write explicitly the factors $8\pi G$, but will indicate
the dependence on the Planck constant $\ell_p^2=8\pi \hbar G$ in the spectra of quantum geometric operators.}
\be
 \{\tEb{j}{b}, \Ab{i}{a} \}=\delta_i^j\delta^a_b.
\label{brAB}
\ee
On this phase space the canonical analysis forces  to impose the three sets of first class constraints:
\beq
{\cal G}_a&=&\nabla_i \tEb{i}{a}=\partial_i \tEb{i}{a}
-{\eps_{ab}}^c  \Ab{i}{b}  \tEb{i}{c}\approx 0 , \nonumber \\
H_i&=&\tEb{k}{a} F^{(\im)a}_{ik}\approx 0,
\label{constrAB} \\
H&=&\frac12 \,\tEb{i}{a}\tEb{j}{b}\left( {\eps^{ab}}_c F^{(\im)c}_{ij}
-(1+\im^2) K^a_{[i}K^b_{j]}\right)\approx 0 ,
\nonumber
\eeq
with $F^{(\im)}_{ij}$ being the curvature of the $su(2)$ space connection $\Ab{i}{}.$

The first one is the Gauss constraint generating the $SU(2)$ gauge transformations in the tangent space.
It is easy to see that the canonical variables transform covariantly under its action:
$\tEb{i}{a}$ transforms as a vector, whereas $\Ab{i}{a}$ is a $SU(2)$ connection.
The constraints $H_i$ and $H$ are related to diffeomorphism symmetry. The former is the generator
of diffeomorphisms of the three-dimensional slice of constant time, whereas the latter
is responsible for time translations.

The constraints \eqref{constrAB} would be polynomial in the canonical variables if not the last
term in the Hamiltonian constraint. This term  disappears only for the choice $\im=\pm \I $, which corresponds
to the original self-dual Ashtekar formulation. However, this formulation drastically differs
from the above presented formulation put forward by Barbero. First, the fact that
the canonical coordinates are complex variables
requires to impose the so-called reality conditions which ensure that the metric is real
and the reality is preserved by the evolution.
They are formulated as
\be
\overline{\vphantom{\tilde E}\tEbi{i}{a}}=\tEbi{i}{a},
\qquad
\overline{\Abi{i}{a}}+\Abi{i}{a}=2\Gamma_i^a(\tE).
\label{realcon}
\ee
These conditions are highly non-linear and non-holomorphic in the canonical self-dual variables and therefore
cannot be formulated merely on the self-dual phase space.
Although the Dirac analysis can be extended in order to take them into account,
up to now no quantization have been found which would be capable of incorporating them at the quantum level.
The second difference is that the actual gauge group for $\im=\pm \I $ is extended
to $SO(3,\Cb)$ which is the same as the original Lorentz gauge symmetry.
As we will see, the self-dual Ashtekar gravity is much closer to the covariant canonical formulation of
\cite{Alexandrov:2000jw} than to the Ashtekar--Barbero formulation presented in this section.

All other values of the Immirzi parameter seem to be on equal footing.
Moreover, the classical formulations with different $\im$ can be related to each other
by a canonical transformation \cite{Thiemann:1995ug} generated by $e^{\beta\{K,\, \cdot\, \}}$ with
\be
K:=\int \de^3 x\, K_i^a \tE^i_a.
\label{cantrim}
\ee
This fact explicitly demonstrates that the introduction of the Immirzi parameter does not change
the classical dynamics of general relativity.

\subsubsection{Loop quantization: kinematical Hilbert space}
\label{subsubsec_quantLQG}

The loop quantization of the above phase space proceeds as follows.
One assumes that the Wilson loops of the SU(2) connection $\Ab{}{}$
\be
\CU_\alpha^{(j)}=\Tr_j \[U_{\alpha}\],
\qquad
U_{\alpha}[\Ab{}{}]={\cal P}
\exp\left(\int_{\alpha} \de x^i \Ab{i}{a} T_a\right),
\label{holAB}
\ee
where trace is taken in representation $j$ of $SU(2)$,
$\alpha$ is a loop, {\it i.e.}, a smooth closed curved immersed in $M$ (it may have self-intersection points)
and $T_a$ denote a basis of the Lie algebra of the gauge group,
are well defined operators in the Hilbert space of quantum gravity. Physically this means that the
excitations of quantum geometry are concentrated on one-dimensional structures, such as
loops in three-dimensional space.

\lfig{SU(2) spin network}{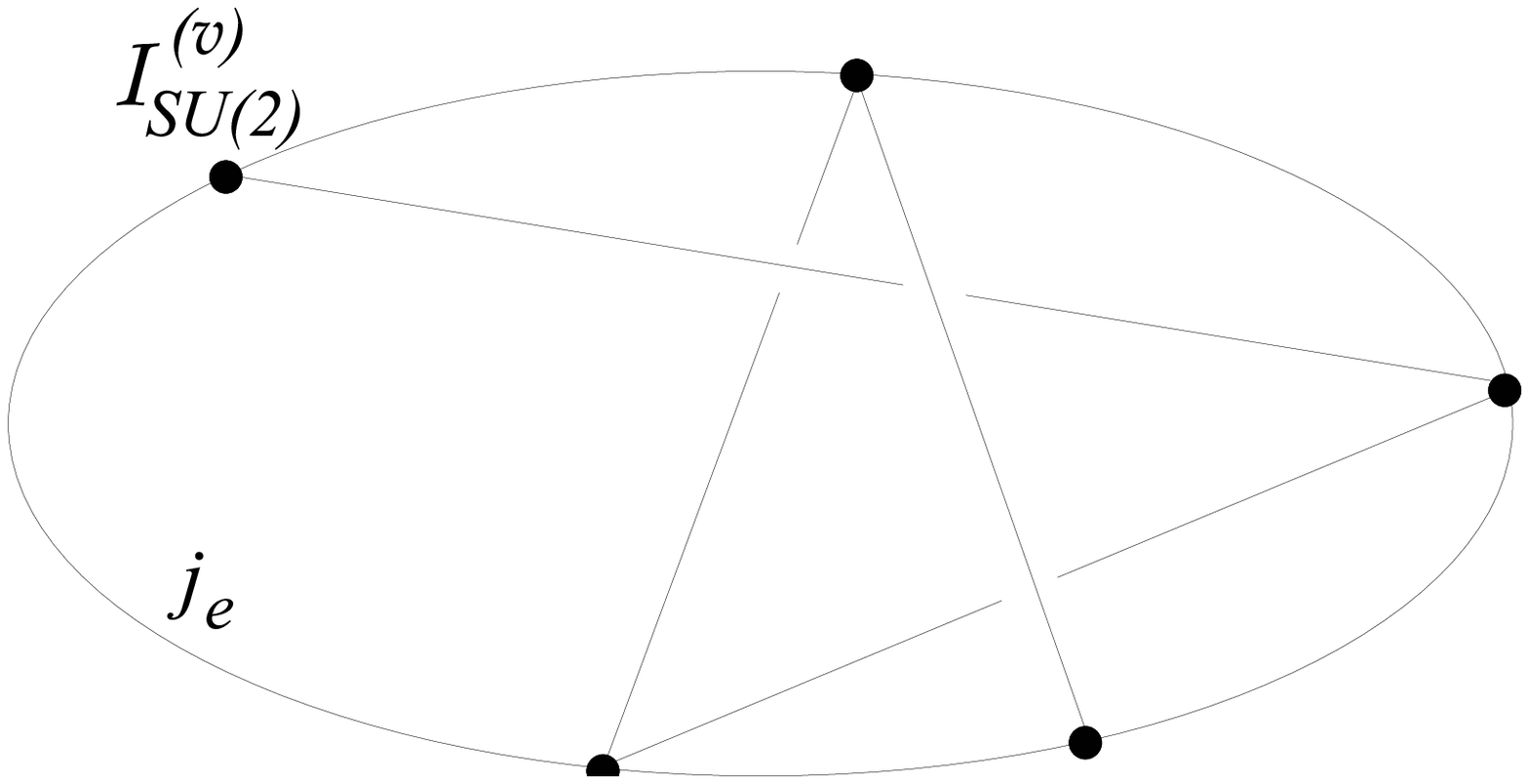}{4.5cm}{spinnetwork}

Considering the SU(2) invariant functionals,
one finds oneself immediately in a (kinematical) Hilbert space ${\cal H}_G$ where the Gauss
constraint has been already imposed.
The loops however are not very convenient to label the states of this Hilbert space
because they give rise to an overcomplete basis.
An orthonormal basis in ${\cal H}_G$ is found by performing harmonic analysis on the space of $SU(2)$
invariant functions of connections and is given by the so-called {\it spin network} states.
Such a state is labeled by a colored graph $\Gamma$ embedded in $M$.
The embedded graph is just a finite number of points $\{ v\}$ connected
by a finite number of smooth embedded curves $\{e\}$ in $M$,
whereas the coloring associates irreducible representations
of $SU(2)$  (half-integer spins $j_e$) to the edges $e$ and $SU(2)$
invariant intertwiners $\Int_{SU\!(2)}^{(v)}$ to the vertices $v$
(see Fig. \ref{spinnetwork}).
The corresponding state $\Psi_{\Gamma}$ is  constructed by contracting holonomies of $\Ab{}{}$ along edges in
representations $j_e$ with invariant intertwiners at vertices
\be
\Psi_{\Gamma}(\Ab{}{})= \langl\bigotimes\limits_e R_{SU(2)}^{(j_e)}\(U_{e}[\Ab{}{}]\),
\bigotimes\limits_v  \Int_{SU\!(2)}^{(v)}\rangl.
\ee

These states are orthonormal with respect to the scalar product defined as follows.
First we introduce it on {\it cylindrical functions} $Cyl_{\Gamma}$  defined by a graph $\Gamma$ and
a function $f$ on $E$ copies of the gauge group $(SU(2))^E$ where $E$ is
the number of edges of $\Gamma$
\be
\Psi_{\Gamma,f}(\Ab{}{})=
f(U_{e_1}[\Ab{}{}],\dots,U_{e_E}[\Ab{}{}]).
\label{defcyl}
\ee
For two such cylindrical functions, the scalar product is given simply
by an integral over the normalized  Haar measure of functions trivially extended to the union of $\Gamma$ and $\Gamma'$
\be
\langle \Psi_{\Gamma,f}|\Psi_{\Gamma',f'}\rangle= \int_{[SU(2)]^{E_{\Gamma\cup\Gamma'}}}
\de\mu(h) \overline{f(h_1,\dots,h_E)}f'(h_1,\dots,h_E).
\label{scSU2}
\ee
Since the cylindrical functions are dense in the space of all functions of connection,
the kinematical Hilbert space ${\cal H}_G$ is defined by completion of the former space with respect to the measure
induced by the scalar product \eqref{scSU2}. Note that this Hilbert space is not separable.

On the Hilbert space ${\cal H}_G$ one still has to impose the first class constraints $H_i$ and $H$.
We postpone this to subsection \ref{subsubsec_diffcon} and before we discuss various
geometric operators defined on ${\cal H}_G$.

\subsubsection{Geometric operators}
\label{subsubsec_geomoper}

One of the most elaborated aspects of LQG is the study of geometric operators
associated to the process of measuring area, volume and length.
Although these operators are defined only on the kinematical Hilbert space ${\cal H}_G$,
they are extremely important for the interpretation of LQG results
as well as for the implementation of the remaining constraints.
Below we mainly concentrate on the area and volume operators.

It is also possible to define a length operator which has been done in \cite{Thiemann:1996at}
(see \cite{Bianchi:2008es} for a new, spin foam motivated version of this operator).
However, it has not found so far an important application in the LQG analysis and
therefore we do not consider it here.

\subsubsubsection{Area operator}

The most studied geometric operator is the area \cite{Rovelli:1994ge}, mainly due to its relative simplicity,
unambiguiousness, its relevance for the black hole entropy calculation and various other applications.
This operator, acting on the kinematical Hilbert space ${\cal H}_{G},$  is a quantization of the classical
expression for the area of a two-dimensional surface $\Sigma$
embedded into $M$
\be
{\cal S}_\Sigma=\int_{\Sigma} \de^2 \sigma  \sqrt{n_i n_j g^{ij}},
\qquad g^{ij}=\delta^{ab}\tE^i_a\tE^j_b,
\label{areaop}
\ee
where $n_i$ is the normal to the surface.
The quantization of this operator amounts to still consider $\Sigma$ as a classical
embedded surface in $M$ and to define $\hat{{\cal S}}_\Sigma$  in terms of the smeared triad operators
\be
\widehat{\vphantom{\tilde E}\tE}_{a}(\Sigma')=\int_{\Sigma'} \de^2\sigma  \,
n_i(\sigma) \widehat{\vphantom{\tilde E}\tE}{}^i_{a}(\sigma)
\label{smtriad}
\ee
associated to a surface $\Sigma'\subset \Sigma$. The smearing ensures that
their action on spin network states is well defined.
To define the area operator, one then uses a decomposition of the measured surface into small pieces and
takes the limit of infinitely small partition, $\rho:\ \Sigma=\bigcup_n \Sigma_n$, of
a regularized expression for the area
\be
\hat{{\cal S}}_\Sigma=\lim\limits_{\rho \to \infty}\sum\limits_{n}
\sqrt{\hat{g}(\Sigma_n)},
\qquad
\hat{g}(\Sigma_n)=\delta^{ab} \widehat{\vphantom{\tilde E}\tE}_{a}(\Sigma_n) \widehat{\vphantom{\tilde E}\tE}_{b}(\Sigma_n).
\label{3-met}
\ee
Applying the resulting operator to a spin network state $\Psi_\Gamma$, one finds that it is an eigenstate
with the eigenvalue given by the following expression\footnote{We ignore situations where
the intersections of the surface with the spin network happen at vertices of the latter.
The full spectrum taking into account all possible cases can be found in \cite{Ashtekar:1996eg}.}
\be
{\CS}_{\Sigma,\Gamma}= \im\ell_p^2\sum\limits_{e\cap\Sigma\not=\emptyset}\sqrt{j_e(j_e+1)},
\label{spsu2}
\ee
where we restored the dependence on the Planck constant and
the sum goes over intersections of the surface with the graph determining the spin network (Fig. \ref{figarea}).
The expression in the square root is nothing else but the Casimir operator of $SU(2)$.
Thus the LQG spectrum of the area operator is discrete and has a minimal non zero eigenvalue.

\lfig{Intersection of a surface with a spin network.}{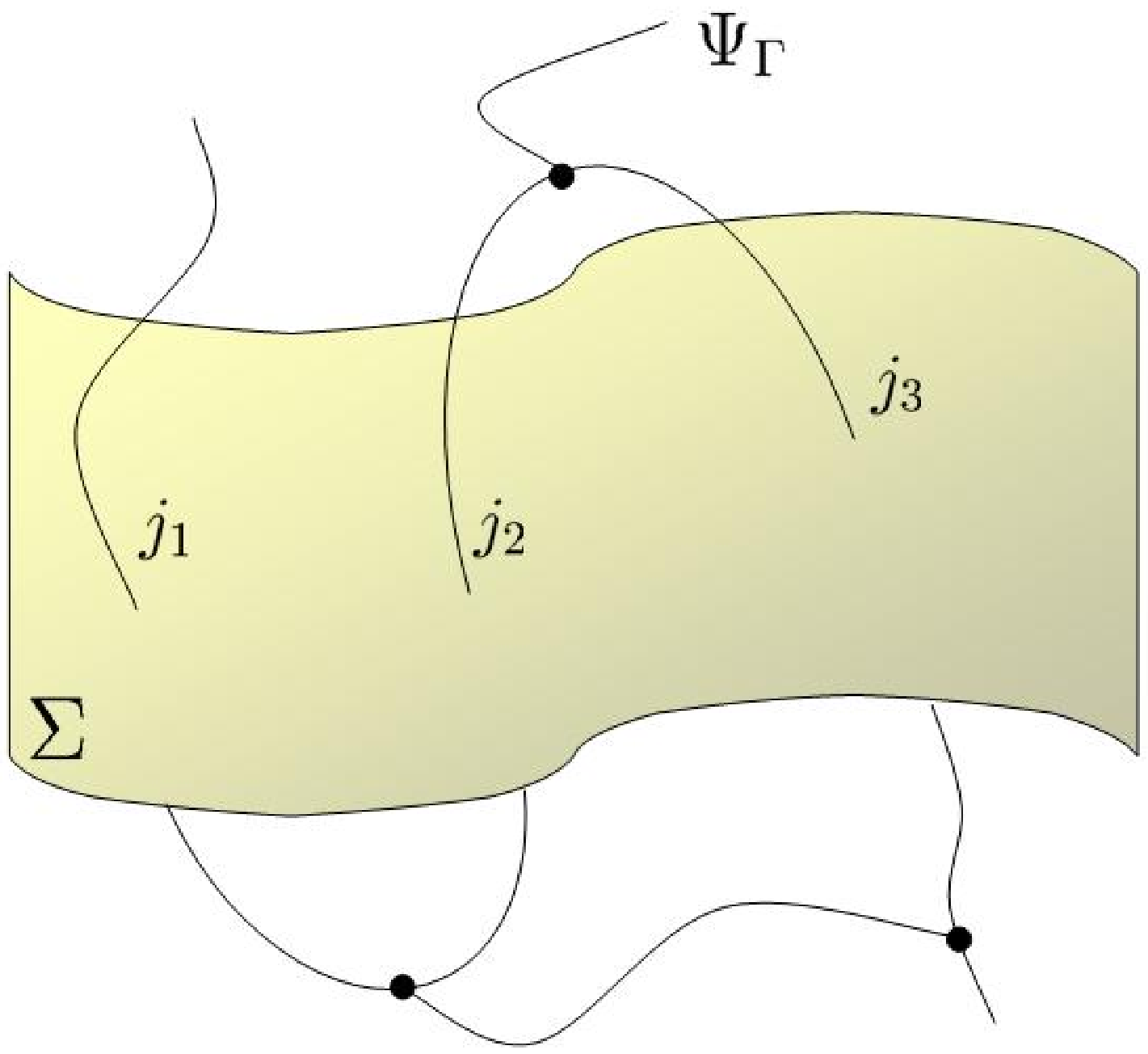}{5.7cm}{figarea}

An important observation is that the spectrum \eqref{spsu2} is proportional to the Immirzi parameter $\im$.
This proportionality arises due to the difference between $\tE$ and the variable $\tEb{}{}$ having
canonical commutation relations with the connection. It signifies that this parameter, which
did not play any role in classical physics, becomes a new fundamental physical constant in quantum theory.
This fact obviously requires an explanation how this could happen.
A usual explanation is that it is similar to the $\theta$-angle in QCD \cite{Rovelli:1997na}.
However, in contrast to the situation in QCD, the formalism of LQG does not even exist for the most natural
value $\im=\infty$ corresponding to the usual Hilbert--Palatini action.
Moreover, the Immirzi parameter enters the spectra of geometric operators
in LQG as an overall scale, which is a quite strange effect.
Even stranger is that the canonical transformation \eqref{cantrim},
mapping classical formulations with different $\im$
to each other, turns out to be implemented non-unitarily,
so that the area operator is sensitive to the choice of canonical variables.
To our knowledge, there is no example of such a phenomenon in quantum mechanics.
Although \cite{Rovelli:1997na} suggests a special quantization of a three dimensional
non-relativistic particle of zero orbital momentum which is supposed
to give rise to such an effect with the role of the area
played by the kinetic energy, a correct analysis of this example
given in \cite{Samuel:2001pq} shows the physical inadequacy of this quantization.
Anyway, the issue is not settled until an explanation is not given.
Below we will argue that the dependence on the Immirzi parameter is due to a quantum anomaly
in the diffeomorphism symmetry, which in turn is related to a particular choice of the connection
used to define quantum holonomy operators \eqref{holAB}.
We return to the discussion of the area operator in section \ref{subsec_LQGdiscus}.

\subsubsubsection{Volume operator}

The next operator to be considered is the volume operator which has been defined in \cite{Rovelli:1994ge,Ashtekar:1997fb}.
This operator is very important because it is at the heart of the construction by Thiemann of the quantization of
the Hamiltonian constraint \cite{Thiemann:1996aw} to be considered below.
Like the area operator, it is defined on the kinematical Hilbert space as a quantization of the classical
volume of a region $R\subset M$ which is given by the following integral
\begin{equation}
V_R=\int_R \de^3 x \sqrt{h}=
\int_R \de^3 x \(\left|\frac{1}{3 !}\,\eps_{ijk}\eps^{abc} \tilde{E}^i_a\tilde{E}^j_b\tilde{E}^k_c \right| \)^{1/2}.
\end{equation}
There exist two different regularizations of this classical expression, leading to two different
versions of the quantum volume operator. The first one is due to Rovelli and Smolin \cite{Rovelli:1994ge}
and the second one is due to Ashtekar and Lewandowski \cite{Ashtekar:1997fb},
so that we denote them as ${\hat{V}}_R^{\rm RS}$ and ${\hat{V}}_{R}^{\rm AL}$, respectively.

To give the action of these operators on a spin network state, let us first introduce
left (right) derivatives on the space of cylindrical functions \eqref{defcyl}.
If $\Gamma$ is a colored graph immersed in $M$ and $e_I$ is an oriented edge starting at a vertex $v$,
we define the operator $X_{v,e_I}^a$ as
\begin{equation}
X_{v,e_I}^a \Psi_{\Gamma,f}=
\left.\frac{\de}{\de t}f(U_{e_1}[A^{(\gamma)}],...,e^{t\tau_a}U_{e_I}[A^{(\gamma)}],..., U_{e_E}[A^{(\gamma)}])\right|_{t=0}.
\end{equation}
If $v$ is the arrival point of $e_I$, one writes a similar formula with the exponential on the right.
In terms of these derivative operators,
the action of the volume operators on spin network states are given by
\begin{equation}
{\hat{V}}_R^{\rm RS}\Psi_{\Gamma}=
\im^{3/2}\ell_p^3\sum_{v\in R\cap \Gamma}
\sum_{I,J,K}\left|\frac{\I C_{\rm reg}}{8}\,\eps_{abc}X_{v,e_I}^aX_{v,e_J}^bX_{v,e_K}^c\right|^{1/2}
\Psi_{\Gamma}
\end{equation}
and
\begin{equation}
{\hat{V}}_R^{\rm AL} \Psi_{\Gamma}=
\im^{3/2}\ell_p^3 \sum_{v\in R\cap \Gamma}
\left|\frac{\I C_{\rm reg}}{8}\,\sum_{I,J,K}\epsilon_v(e_I,e_J,e_K)\eps_{abc}X_{v,e_I}^aX_{v,e_J}^bX_{v,e_K}^c\right|^{1/2}
\Psi_{\Gamma},
\end{equation}
where $C_{\rm reg}$ is a constant dependent on the regularization scheme,
the second sum goes over all triples of edges meeting at vertex $v$, and
in the Ashtekar--Lewandowski version
$\epsilon_v(e_I,e_J,e_K)\in \{-1,1,0\}$ is the sign of the orientation of the
three tangent vectors at $v$ of the curves $e_I,e_J,e_K$.

\begin{figure}
\psfrag{e1}{$j_{e_1}$}
\psfrag{e2}{$j_{e_2}$}
\psfrag{e3}{$j_{e_3}$}
\psfrag{e4}{$j_{e_4}$}
\psfrag{v}{$v$}
\psfrag{Rn}{$R_n$}
\centering
\includegraphics[scale=0.51]{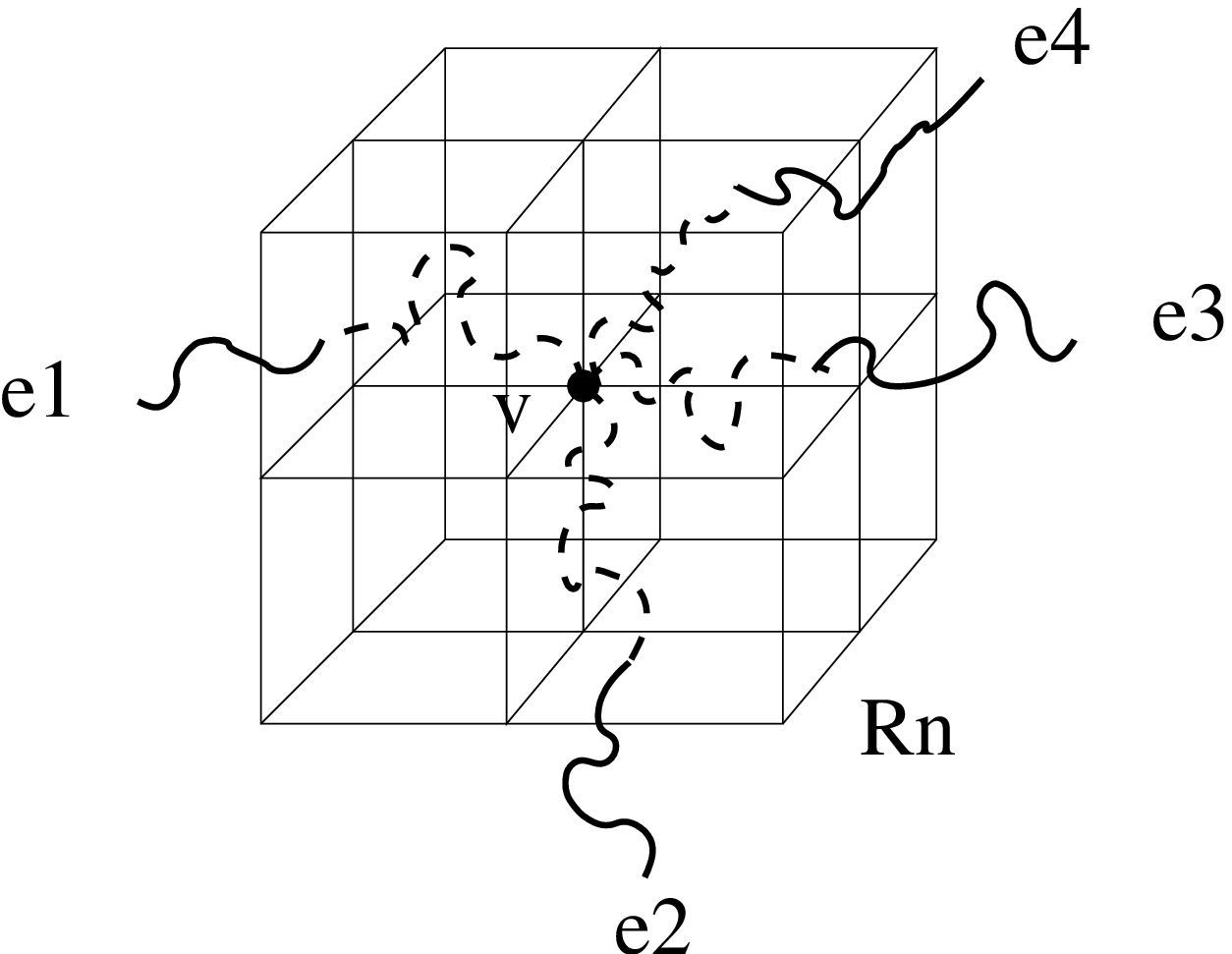}
\caption{Infinitesimal volume element centered on a vertex $v$}
\label{volumefig}
\end{figure}

This result is derived using a procedure similar to the one used for the area operator, {\it i.e.},
partitioning the region $R$ in sufficiently small cubes $R_n$
and expressing their volume elements through the densitized triad.
The two versions are called external and internal regularizations, respectively, because the second one
is such that it depends on the orientation of the tangent vectors inside these small cubes,
whereas the first is independent on these data.
Besides, note that in the first version, the operator is given as a sum of square root factors, whereas in the second
it appears as the square root of the sum over triples of edges.
Thus, the two versions realize inequivalent quantizations of the classical volume function.

In fact, in \cite{Giesel:2005bk} it has been demonstrated that only ${\hat{V}}_R^{\rm AL}$ passes the so called
{\it triad test}. This is the requirement that the action of the smeared triad operator \eqref{smtriad}
coincides with the action of its classical equivalent constructed from commutators of the volume with holonomies
of the gauge connection. This ensures the self-consistency at the kinematical level
of the loop quantization and the regularization procedures involved in the definitions.
Remarkably, only the Ashtekar--Lewandowski version of the volume operator satisfies
this condition, whereas the Rovelli--Smolin version turns out to be ruled out.\footnote{In fact,
the Ashtekar--Lewandowski volume operator satisfies the triad test {\it only} on 6-valent spin network graphs.
Therefore, the actual meaning of this test is still unclear and is a subject of debate.}
Besides, this test allows to fix the scheme dependent constant left undetermined above.
It has to be chosen as $C_{\rm reg}=\frac{1}{3!8}$.

The important fact is that the contribution to the volume operator of a vertex of valence less or equal to 3 vanishes
so that only vertices of valence greater than 3 contribute to the volume.
This is important because in the old formulation of LQG expressed in terms of Ashtekar variables, states were
labeled by loops (eventually with intersection points) and it was shown that
loop states without intersections (called regular loops) satisfy all
the constraints of LQG including the Hamiltonian constraint \cite{Jacobson:1987qk}.
However, it was soon realized that with these regular loop states one could not reproduce
in any limit a classical geometry because the action of the volume operator on these states is zero.
Therefore the weave state of \cite{Ashtekar:1992tm} constructed using regular loops cannot be considered
as states satisfying all constraints and approximating a classical metric.
These weave states have been corrected in \cite{Grot:1997cx} by including 4-valent
intersections, but the resulting state is only a kinematical state approximating a classical
metric and it does not satisfy the Hamiltonian constraint.

Another important property of the volume operator is that it is not diagonal on general spin network states
contrary to the area operator.
Therefore, although it is easy to see that its spectrum is discrete
\cite{Rovelli:1994ge}, computing its precise form
is a much more complicated problem than it is for the area operator.
However, this is just a technical difficulty which comes from the various sign possibilities for
$\epsilon(e_I,e_J,e_K)$ in ${\hat{V}}_R^{\rm AL}$.
Numerically the spectrum of ${\hat{V}}_R^{\rm AL}$ has been extensively studied and much about it can be found,
for example, in \cite{Thiemann:1996au,Brunnemann:2004xi,Brunnemann:2007ca}.
In particular, these computations indicate that
the lower bound of this spectrum tends to zero when evaluated on spin networks having vertices of valence $5$ and $6$.

\subsubsection{Loop quantization: diffeomorphism and Hamiltonian constraints}
\label{subsubsec_diffcon}

The imposition of $H_i$ and $H$ on the kinematical Hilbert space is implemented
in the LQG approach in two very different schemes.
One first implements the spatial diffeomorphisms by identifying spin network states which are in the same orbit of $Diff(M)$.
This is usually done as follows: let $\CD$ be a set of spin networks embedded in $M$.
For each spin network $\Gamma\in \CD$ we define the element $\langle \Psi_{\Gamma},\cdot \rangle$ of the algebraic dual $\CD^{\star},$
and denote \begin{equation}
\Psi_{\[\Gamma\]}= \sum_{\varphi(\Gamma), \varphi\in Diff(M)} \langle \Psi_{\varphi(\Gamma)},\cdot\rangle.\end{equation}
Although the sum is infinite, this gives a well defined element of $\CD^*$
because the action on an element of $\CD$ has only a finite number
of non-vanishing contributions.
This element only depends on the orbit  $\[\Gamma\]$ of  $\Gamma$ under $Diff(M).$
One can define a preHilbert space on the subvector space of $\CD^*$ generated by $\Psi_{\[\Gamma\]}$ with
the inner product:\footnote{We have simplified a little bit the definition here by assuming that $\Gamma,\Gamma'$ have no symmetries.
It is easy to correct the picture in the general case.}
\be
\langle \Psi_{\[\Gamma\]}, \Psi_{\[\Gamma'\]}\rangle_{Diff}:=\langle \Psi_{\[\Gamma\]}, \Psi_{\Gamma'}\rangle.
\ee
After completion of this pre-Hilbert space, one obtains the Hilbert space
${\cal H}_{GDiff}$ on which $Diff(M)$ acts trivially.
A Hilbertian basis of ${\cal H}_{GDiff}$ is therefore labeled by the equivalence
classes of embedded spin networks in $M$ under diffeomorphisms.

The imposition of the Hamiltonian constraint causes much more trouble mainly because its action on spin networks
does not have an easy geometric interpretation. But the first problem to be solved is actually
to construct an operator $\hat{H}$ which could be considered as a quantization of the classical
function $H$ given in \eqref{constrAB}.
On general ground, this operator has to satisfy the following requirements:
\begin{itemize}
\item
$\hat{H}$ should give back $H$ in the classical limit and its definition should not be too much dependent
on a regularization;
\item
$\hat{H}$ should properly implement the quantization of the Dirac algebra of constraints;
\item
there should exist normalizable states annihilated by $\hat{H}$ whose classical limit
is described by solutions of general relativity.
\end{itemize}

A proposal for such quantum Hamiltonian constraint operator has been given by Thiemann
in the series of works \cite{Thiemann:1996aw,Thiemann:1996av,Thiemann:1997rv} and it
essentially relies on the definition of the volume operator
described in the previous subsection.
Note that the classical Hamiltonian constraint \eqref{constrAB} is naturally split into two pieces.
The first one is polynomial in canonical variables and can be written as
\begin{equation}
H_E=\hf\, \tEb{i}{a}\tEb{j}{b} {\eps^{ab}}_c F^{(\im)c}_{ij}=
-\im^{-3}\eps^{ijk}\delta_{ab} F^{(\im)a}_{ij}\{\Ab{k}{b}, V_{M} \},
\label{defEH}
\end{equation}
where ${V}_M$ is the volume function of the whole spacelike hypersurface.
It is called sometimes ``Euclidean'' Hamiltonian constraint because in Euclidean theory
the second non-polynomial term of the Hamiltonian constraint in \eqref{constrAB}
is weighted by the factor $1-\im^2$ and thus vanishes for the real value $\im=\pm 1$,
corresponding to self-dual theory, leaving us with $H_E$.
The classical expression \eqref{defEH} is easy to quantize, whereas the second term dependent on the extrinsic curvature
is much more difficult to deal with. A beautiful insight of Thiemann was that
classically it can be recovered from the Poisson algebra generated by the canonical variables and the volume operator.
More precisely, the extrinsic curvature can be obtained from the commutation relation with
its integrated trace $K$ defined in \eqref{cantrim} as $K^a_i=\im^{-1}\{K,\Ab{i}{a}\}$, whereas the latter
can be written using the above defined Euclidean Hamiltonian
\begin{equation}
K=\left\{V_{M}, \int_{M} H_E(x) \de^3 x\right\}.
\label{extr_vol}
\end{equation}
Thus, the quantum Hamiltonian constraint operator $\hat{H}$ can be constructed by
''quantizing'' these classical relations.

Thus, the quantization of the Hamiltonian constraint and its representation on the kinematical Hilbert space
follows the following steps:
\begin{enumerate}
\item
choose a triangulation $\Delta$ of the spacelike hypersurface $M$
and for each tetrahedron $t\in T$ select one of its vertices $v_t$;
\item
define $\hat{V}=\hat{V}^{AL}_{M}$;
\item
define $\hat{H}_E[N]=\sum_{t\in \Delta} \hat{H}_{E,\,t}$ where $N$ is a lapse function and
\begin{equation}
\hat{H}_{E,\,t}= -\frac{N(v_t)}{\I\hbar \im^3}\eps^{ijk}\,\tr\(U_{\alpha_{ij}^t}U_{e_k^t}\[U_{e_k^t}^{-1}, \hat{V}\]\),
\label{qHE}
\end{equation}
where $e_i^t$ are the three edges of the tetrahedron $t$ meeting at $v_t$ and
$\alpha_{ij}^t$ is the closed loop originating from $v_t$ and bounding the face of $t$ defined by $e_i^t, e_j^t$;
\item
define $\hat{K}=\frac{1}{\I\hbar}\[\hat{V}, \hat{H}_E[1]\]$;
\item
finally, define the quantum Hamiltonian constraint to be
$\hat{H}[N]=\sum_{t\in \Delta} \hat{H}_t$ where
\begin{equation}
\hat H_t=\hat H_{E,\, t}+\frac{1+\im^2}{(\I\hbar)^3\im}\,N(v_t)
\eps^{ijk}\tr\(h_{e_i^t}\[h_{e_i^t}^{-1},\hat{K}\]
h_{e_j^t}\[h_{e_j^t}^{-1},\hat{K}\]
h_{e_k^t}\[h_{e_k^t}^{-1},\hat{V}\]\).
\end{equation}
\end{enumerate}
It is easy to see that the action of $\hat{H}[N]$ on a spin network state is a finite linear combination
of spin networks. By taking $\Delta$ sufficiently small and adapted to $\Gamma$, it can be shown that
$\hat{H}[N]\Psi_{\Gamma}$ does not depend on $\Delta$ under refinement and thus the operator is well defined.
Its action can be computed explicitly in terms of matrix elements of the volume operator \cite{Thiemann:1996av}.

We refer to \cite{Nicolai:2005mc,Thiemann:2006cf} for discussion of weak points of this construction.
Even our second requirement that $\hat{H}$ properly implements the Dirac algebra is not evidently satisfied.
But the most difficult problem is to say something about the third requirement, essentially equivalent to that
one should recover Einstein gravity in the classical limit. However, even before taking the limit,
not much progress has been done in finding the physical Hilbert space of LQG by the canonical method
(see however the master constraint program \cite{Thiemann:2003zv}). Therefore, at present
the hope relies on covariant methods of spin foams\footnote{Note in this relation that recently
a new quantization of the Hamiltonian constraint has been proposed in \cite{Alesci:2010gb}.
It goes essentially along the same lines as Thiemann's quantization with the only modification in
the Euclidean Hamiltonian \eqref{qHE} where the curvature is approximated using a certain tetrahedron instead
of the triangle $\alpha_{ab}^t$. Such regularization was argued to be more compatible with the spin foam
approach than the original one.}
appearing in this context
from expansion of the projection operator, which is determined by the Hamiltonian constraint
and maps states from ${\cal H}_{GDiff}$ to the physical states \cite{Reisenberger:1996pu}.
This approach will be the main subject of chapter \ref{sec_path}.

\subsection{Lorentz covariant approach}
\label{subsec_lorcf}

\subsubsection{Covariant canonical formulation}
\label{subsubsec_covcan}

Although the construction leading to LQG is straightforward and somewhat elegant, it possesses one feature
which is not completely satisfactory. Whereas general relativity in the first order formulation
has the Lorentz group as a local symmetry in the tangent space, the gauge group of LQG is only
its $SU(2)$ subgroup. If this is a consistent quantization, it must be possible to formulate it
in a Lorentz covariant way.
Below we shall see that this is indeed possible, but the corresponding construction
reveals other issues hidden in the original framework.

The reduction of the gauge group originates from the first two steps in the procedure leading to
the AB canonical formulation on page \pageref{page_LQGreduct}.
Therefore, it is natural to construct a canonical formulation and to quantize it avoiding any
partial gauge fixing and keeping all constraints generating Lorentz transformations in the game.
The third step in that list (solution of the second class constraints)
can still be done and the corresponding canonical formulation can be found in \cite{BarroseSa:2000vx}.
However, this necessarily breaks the Lorentz covariance. On the other hand, it is natural to keep
it since the covariance usually facilitates analysis both at classical and quantum level.
Thus, we are interested in a canonical formulation of general relativity with the Immirzi parameter,
which preserves the full Lorentz gauge symmetry and treats it in a covariant way.

Such a formulation was constructed in \cite{Alexandrov:2000jw}.
It originates from the following $3+1$ decomposition of the cotetrad
\be
\begin{split}
e^0 = N dt+\chi_aE^a_idx^i,
\qquad
e^a  =E^a_i N^i
dt+E^a_idx^i.
\end{split}
\label{decomp}
\ee
As usual, $E_i^a$ is the triad, whereas $N$ and $N^i$ are the usual laps and shift, respectively, which
however should be properly redefined to appear precisely
as Lagrange multipliers for the constraints generating diffeomorphisms.
The new field which appears here is $\chi^a$. It has a very clear geometric meaning:
$\chi^a$ determines the direction of the normal to the three-dimensional slices of constant time.
In particular, the slice is spacelike, timelike or lightlike depending on whether $\chi^2=\chi^a\chi_a$ is less,
larger or equal to 1. In the following we restrict ourselves to the case of
spacelike foliation, although the timelike case can be treated in the same way \cite{Alexandrov:2005ar}.

It is useful to define a four-dimensional unit vector ($x^I x_I=\sigma$) determined by $\chi^a$
\be
x^I(\chi)=\(\frac{1}{\sqrt{1+\sigma \chi^2}},\frac{\chi^a}{\sqrt{1+\sigma\chi^2}}\)
\label{xxx}
\ee
and two orthogonal projectors acting in the adjoint representation of the gauge algebra $\gmat$
\be
I_{(Q)}^{IJ,KL}(x)=
\eta^{I[K}\eta^{L]J}-2\sigma\, x^{[J}\eta^{I][K} x^{L]},
\qquad
I_{(P)}^{IJ,KL}(x)=2\sigma\,x^{[J}\eta^{I][K} x^{L]}.
\label{projxxx}
\ee
The vector $x^I$ is precisely the direction of the normal mentioned above.
It defines a subgroup\footnote{In this review we do not distinguish between SU(2) and SO(3) groups.
The choice of the correct group is a controversial issue in LQG (see, for example, \cite{Dreyer:2002vy}), but
most of the analysis is done at the algebraic level anyway and does not depend on this choice.}
$H_x=SU_x(2)$ of the gauge group $G=SO(\eta)$
which is the isotropy  subgroup of $x^I$ with respect to the standard
action of $G$ on $\Rmat^4$.
Then the geometric meaning of $I_{(Q)}$ and $I_{(P)}$ is
that they project on the Lie subalgebra of $H_x$ (rotations) and on  its orthogonal
complement (boosts) with respect to the Killing form $g_{IJ,KL}$, respectively.
These projectors play an important role and appear both in canonical and spin foam approaches to quantum gravity.

The time gauge heavily used to obtain the AB canonical formulation corresponds to the particular choice $\chi^a=0$.
The subgroup $H_x$ coincides in this case with the canonically embedded $SO(3)$, which
is the central object of LQG defining its structure.

As usual, various components of the tetrad play different role in the canonical formulation. The laps and shift
appear as Lagrange multipliers for the diffeomorphisms and Hamiltonian constraints, respectively, whereas
the triad and the field $\chi^a$ enter canonical variables.
The latter are given by the space components of the spin-connection $\omega_i^{IJ}$
and by the field
\be
\tPb_{IJ}^i=\(1+\frac{1}{\im}\, \star\)\tP_{IJ}^i
\ee
with
\be
\tP_{IJ}^i=
\hf\, \eps^{ij k}\eps_{IJKL}e_{j}^{K}e_{k}^{L}
=\left\{
\begin{array}{ll}
\tE^i_a &\ \ I=0,\ J=a\\
\tE^i_a\chi_b-\tE^i_b\chi_a &\ \ I=a,\ J=b
\end{array}
\right.
\label{multPIJ}
\ee
It is clear that not all components of the  field $\tPb_{IJ}^i$ are independent.
By a counting argument, there should be six constraints which can be written as
\be
\phi^{ij}=\eps^{IJKL}\tP^i_{IJ}\tP^j_{KL}\approx 0.
\label{phi}
\ee
They are of second class because commuting them with the Hamiltonian, one generates secondary constraints
\be
\psi^{ij}=f^{IJ,KL,MN}\tQ_{IJ}^{l}\tQ_{KL}^{\{ j}\nabla_l \tQ_{MN}^{i\} }
\approx 0,
\label{psi}
\ee
where $\tQ^i_{IJ}=-\star\tP^i_{IJ}$
and  $f_{IJ,KL}^{MN}$ are $so(3,1)$ structure constants,
and these constraints do not commute with the primary ones, $\{\phi^{ij},\psi^{kl}\}\ne 0$.
The conditions \eqref{phi} are the famous simplicity constraints ensuring that the bi-vectors $\tP^i_{IJ}$
are constructed from tetrads as in \eqref{multPIJ}. At the same time, the conjugate constraints
\eqref{psi} impose certain conditions on the spin-connection sitting in the covariant derivative.
In the Lagrangian picture they arise as a part of Cartan equations.

The presence of the second class constraints is the main complication of the covariant canonical formulation.
They change the symplectic structure on the phase space which must be determined by the corresponding
Dirac bracket. It can be computed straightforwardly. In particular, one has the following commutation relation
\be
\{ \omega_i^{IJ},\tPb_{KL}^j\}_D=\delta_i^j\delta^{IJ}_{KL}
-\frac{1}{2} \(g^{IJ,MN}-\frac{1}{2\im}\,\eps^{IJMN}\)\left(\tQ^j_{MN}\Qt_i^{PQ}+\delta^j_i I_{(Q)MN}^{PQ}
\right) g_{PQ,KL},
\label{comomP}
\ee
where the definition of $\Qt_i^{IJ}$ can be found in \cite{Alexandrov:2000jw}.
Another new feature is that the spin-connection becomes "non-commutative" in the sense that the
Dirac bracket of two spin-connections is non-vanishing.

This phase space carries the action of ten first class constraints. Six of them, $\CG_{IJ}$, generate local
Lorentz transformations and four, $H_i$ and $H$, generate space-time diffeomorphisms. The constraints have essentially
the same form as the ones of AB formulation \eqref{constrAB} with $\tEb{i}{a},\Ab{i}{a}$ being
replaced by $\tPb_{IJ}^i, \omega_i^{IJ}$, the structure constants of $SU(2)$ being replaced by the structure constants
of $SO(3,1)$ and the last term  in the Hamiltonian constraint \eqref{constrAB} involving the intrinsic curvature
being dropped.\footnote{Besides, the first term in the Hamiltonian constraint contains
additional factor $\frac{1-\frac{1}{\im}\, \star}{1+1/\im^{2}}$.}
Thus, all constraints are polynomial in the canonical variables as in the self-dual Ashtekar gravity.

This completes the description of the resulting canonical formulation.
But this is not the end of the story yet. Unfortunately, it turns out
that the spin-connection is not appropriate for the loop quantization \cite{Alexandrov:2001pa}.
Indeed, if one considers spin network states constructed from the spin-connection (forgetting
about its non-commutativity), they are not eigenstates of the area operator \eqref{3-met},
where the metric in our variables reads
\be
g^{ij}=-\hf\, g^{IJ,KL} \tP^i_{IJ}\tP^j_{KL}.
\ee
This happens due to the complicated structure of the commutator \eqref{comomP} which
replaces the simple canonical commutation relation of LQG \eqref{brAB}.
Thus, one looses one of the fundamental results of LQG together with the geometric interpretation of the
spin network states.

Quite remarkably, there is a way to overcome this problem. The solution is very simple: one should construct
loop states using a different connection appropriately chosen so that the area operator is diagonal
on the new states. To achieve such a diagonalization, one should assure that the action of
$\tP$ smeared over a surface is purely algebraic on these states, {\it i.e.}, it does not depend
on embedding of the underlying graph and the surface. The necessary and sufficient condition
for that is the proportionality of the Dirac bracket between $\tP$ and the new connection to $\delta_i^j$,
which is not the case for the spin-connection due to the second term in \eqref{comomP}.

But what are these connections which can be taken to define the loop states?
It is clear that they should be functions on the phase space, let us call them $\SA^{IJ}_i$,
transforming under gauge symmetries of the theory in the same way as the spin-connection.
Adding to this condition the above requirement that the area operator is diagonal,
one arrives at the following list of conditions on these quantities:
\beq
i) &\ & \{ \CG(n),\SA^{IJ}_i\}_D=
\partial_i n^X +f_{KL,MN}^{IJ} \SA_i^{KL} n^{MN},
\label{gtr}
\nonumber\\
ii) &\ & \{ {\cal D}(\vec N), \SA_i^{IJ} \}_D=
\SA_j^{IJ}\partial_i N^j+N^j\partial_j \SA_i^{IJ}, \label{difftr}
\\
iii) &\ & \{ \SA^{IJ}_i,\tP_{KL}^j\}_D \sim \delta_i^j,
\label{newAP}
\nonumber
\eeq
where ${ \cal D}_i=-H_i+\SA_i^{IJ}\CG_{IJ}$ are the generators of spatial diffeomorphisms
and  $\sim$ means up to a
factor which can be arbitrary tensor in the tangent space indices.
It is not a very difficult exercise to find all $\SA_i^{IJ}(\omega,\tP)$ satisfying \eqref{difftr}.
Once the second class constraints \eqref{psi} are taken into account, there is a two-parameter family of
such objects $\SAab_i^{IJ}$ labeled by $(a,b)\in\bR^2$ \cite{Alexandrov:2001wt}.
What we need to know about them is their Dirac brackets with $\tP$ which are given by
\be
\{ \SAab^{IJ}_i,\tP_{KL}^j\}_D=\delta_i^j \left(
(1-b)\delta^{IJ}_{MN} -\frac{a}{2}\,{\eps^{IJ}}_{MN} \right) I_{(P)KL}^{MN} .
\label{AP2}
\ee

An important consequence of this Dirac bracket, which will play an essential role in the following,
is that all new connections commute with the projectors \eqref{projxxx} and therefore with the field $\chi^c$
\be
\{ \SAab^{IJ}_i,\chi^c\}_D=0.
\label{commAabchi}
\ee
This is possible only if $\SAab_i^{IJ}$ have three independent components less than the spin-connection. And indeed,
only nine of their components are independent. Six components are fixed by the constraints $\psi^{ij}$
and the three missing components can be recovered from the Gauss constraint $\CG_{IJ}$.
Thus, the parametrization of the phase space, which we would like to use
as the starting point for quantizing the theory,
is provided by $\SAab_i^{IJ},\tP^i_{IJ}$ and $\CG_{IJ}$ subject to various first and second class constraints.

\subsubsection{Loop quantization}
\label{subsubsec_loopq}

Let us now try to quantize the covariant canonical formulation presented above following the ideas
of the loop approach. This means that we assume the loop functionals of the connection $\SAab$
to give rise to well defined states of quantum gravity. There are however several differences of the
present situation comparing to the one in LQG, which must be taken into account.
\begin{itemize}
\item First of all, the connection lives now in the Lorentz Lie algebra so that its holonomy
operators belong to a non-compact group. This is a striking distinction from LQG where
the compactness of the structure group $SU(2)$ is crucial for the discreteness of geometric operators
and the validity of the whole construction.
\item The symplectic structure is not anymore provided by the canonical commutation relations
of the type \eqref{brAB} but is given by the Dirac brackets. In particular, the commutator relevant
for the evaluation of the area spectrum is \eqref{AP2}. This means that one has to quantize a much more
complicated system than one had previously.
\item As a consequence of the new symplectic structure, all (except one) connections $\SAab$
are non-commutative. This questions the use of the loop or spin network functionals
to span the Hilbert space of quantum theory.
\item The fact that passing to the new connection one lost three independent components
and as a result $\SAab$ commutes with $\chi$ \eqref{commAabchi}, indicates that it is insufficient
to consider the functionals of only $\SAab$. The full configuration space is spanned by
functionals dependent on both, $\SAab$ and $\chi$.
\item In addition to the first class constraints generating gauge symmetries,
the phase space to be quantized carries second class constraints. Although they are already taken into account
in the symplectic structure by means of the Dirac brackets, they lead to a degeneracy in the Hilbert space
constructed ignoring their presence \cite{Alexandrov:2005ar}. It is
 a non-trivial problem to remove
such a degeneracy.
\end{itemize}

How do these differences affect the construction? As we will see, the non-compactness of the Lorentz group
is actually not a serious obstacle and moreover in some cases it does not imply that the discreteness of LQG is lost.
Also the commutation relation \eqref{AP2} can be easily realized since the action of the smeared $\tP$ operator
on holonomies of $\SAab$ is purely algebraic. A really difficult issue is the non-commutativity
of the connection. But even this property does not prevent from considering loop states because of
the path ordering in the definition of holonomies, which makes them well defined.
The loop states
defined by a non-commutative connection  associated to the Lorentz group  are well known in the context
of the Chern-Simons approach to 2+1 gravity with positive  cosmological constant  \cite{Buffenoir:2002tx}.
However, here the situation is more complicated due to a more complicated from of the commutator of
two connections \cite{Alexandrov:2002xc,Alexandrov:2005ar}.
We will discuss this issue in more detail later in section \ref{subsec_canway}.

The fourth point in the above list does lead to something new. Since the wave functionals have to depend
now on two variables of different origin, the basis elements of the corresponding Hilbert space
are going to have a richer structure than the usual spin networks.
To define the Hilbert space structure, one considers {\it generalized} cylindrical functions
which, as the usual ones, are associated with graphs and whose
dependence on the connection is supposed to be through the Lorentz group elements represented by holonomies.
In addition, they also depend on the values of the field $\chi$ at vertices.
Note that $\chi$ is naturally encoded in the unit vector $x^I$ \eqref{xxx} which can be considered as
an element of the quotient space $X=G/H$. This implies that the generalized cylindrical functions
live on the homogeneous space $G^E\times X^V$.
The scalar product is then given by a natural generalization of the scalar product \eqref{scSU2}
\be
\langle \Psi_{\Gamma,f}|\Psi_{\Gamma',f'}\rangle= \int_{G^{E_{\Gamma\cup\Gamma'}}}
\de\mu(g) \overline{f(g_1,\dots,g_E;x_1,\dots,x_V)}f'(g_1,\dots,g_E;x_1,\dots,x_V),
\label{sc_proj}
\ee
where the two functions are evaluated on arbitrary, but same $x_v$ and the scalar product does not depend
on their choice due to the gauge invariance of the wave functionals.

\lfig{Projected spin network and the structure of its intertwiners.}{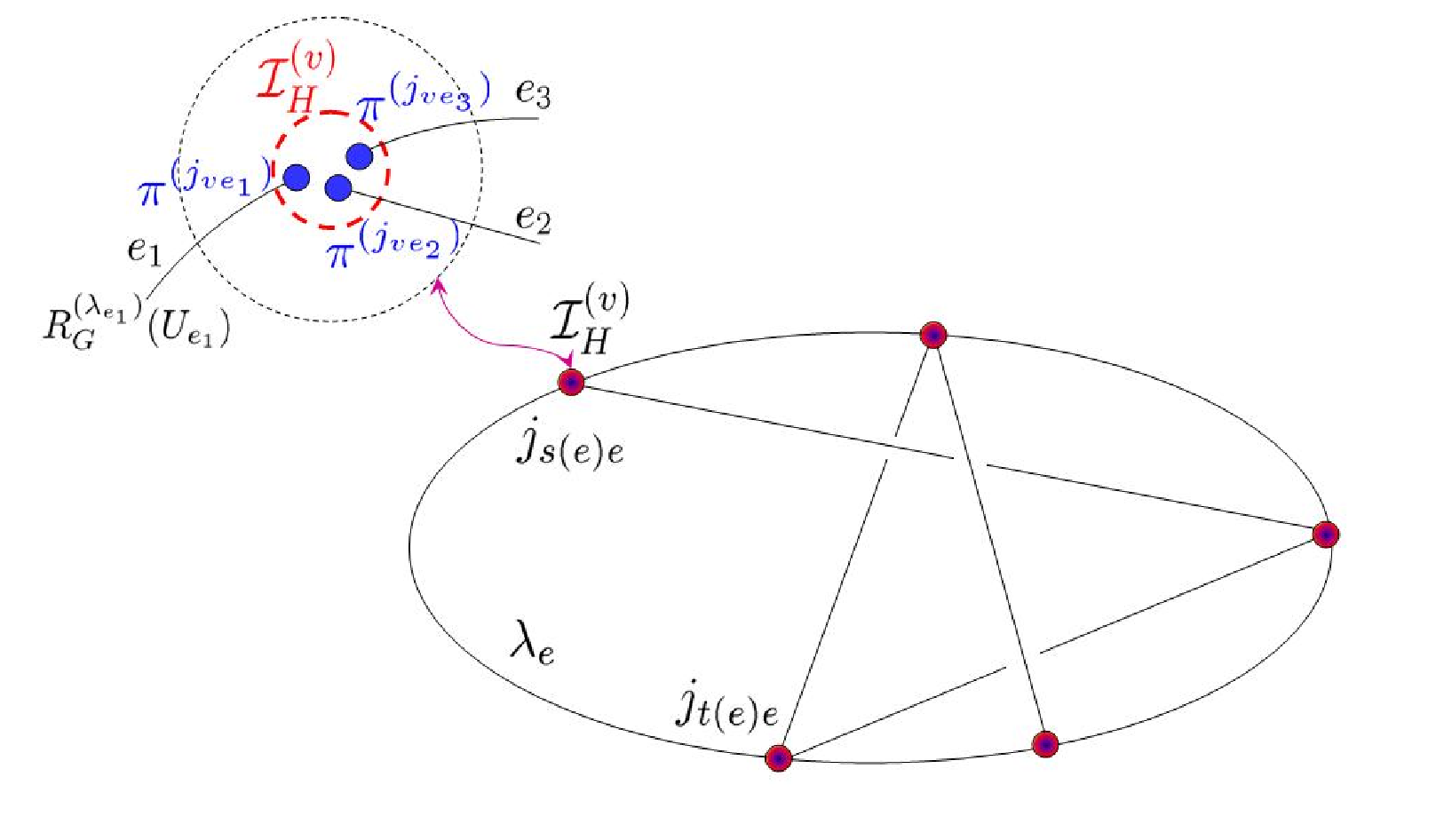}{7cm}{figprojsp}

A basis in this Hilbert space can be found by the method of harmonic analysis
and is given by the so called {\it projected spin networks} \cite{Livine:2002ak}.
As the usual spin networks, they are represented by graphs colored by representations and intertwiners.
But now the coloring is richer. One assigns an irreducible representation $\lambda_e$ of the group $G$ to each edge,
a representation $j_{ve}$ of the subgroup $H$ to each couple of edge and vertex belonging to this edge,
and an intertwiner $\Int_{H}^{(v)}$ of $H$ which couples the representations $j_{ve}$ (see Fig. \ref{figprojsp}).
Remarkably, despite the appearance of the subgroup, the full state is invariant with respect to the full
gauge group $G$. This is possible because the subgroup to be considered is
actually defined by the element $x$ of the factor space
and is transformed together with the group elements defined by holonomies of the connection.

More specifically, let $H_x$ be the stationary subgroup of $x\in X$ (as below \eqref{projxxx}) and
\be
\CH_{G}^{\lambda_{e}}=
\mathop{\bigoplus}\limits_{j_{ve}} \CH_{H_{x_v}}^{j_{ve}}
\label{repsp}
\ee
is the decomposition of the representation $\lambda_e$ to the subgroup defined by the value of $x$ at the vertex $v$.
Let $\Ppr{j_{ve}}{x_v}$ be the projector to the subspace of representation $j_{ve}$.
Then the state associated with the projected spin network can be written as
\be
\Psi_{\Gamma}(\SAab,\chi)=\langl\bigotimes\limits_e
\(\Ppr{j_{t(e)e}}{x_{t(e)}}\, R_G^{(\lambda_e)}\(U_{e}[\SAab]\)\,
\Ppr{j_{s(e)e}}{x_{s(e)}}\), \bigotimes\limits_v  \Int_{H}^{(v)} (x_v)\rangl,
\label{projspnet}
\ee
where $t(e)$ and $s(e)$ denote target and source vertices, respectively, of the edge $e$.
The projectors and intertwiners carry dependence on $x_v$ because they are defined with respect to the subgroup
dependent on it. Due to this, they transform under a general gauge transformation precisely in such a way
to leave the whole state invariant.

Thus, we see that the naive generalization of the $SU(2)$ spin networks to their Lorentz analogues
is not the correct way to proceed. A more elaborated structure is required.
The origin of this novelty can be traced back to the presence of
the second class constraints which modified the symplectic structure and invoked a connection
different from the usual spin-connection.
The projection appearing in the definition \eqref{projspnet} solves also some problems arising for the usual
spin networks defined for a non-compact gauge group \cite{Freidel:2002xb}, because it effectively reduces
the holonomies to the compact subgroup in the sense
that they now live in finite dimensional representation spaces of $H$.
In particular, the projected spin network states can be well evaluated on a vanishing connection.

These states turn out to be extremely important also in the context of spin foams since
they describe boundary states of any SF model of four-dimensional general relativity based on
Plebanski formulation (see the next chapter) \cite{Alexandrov:2008da}.
In particular, the boundary states associated with the new vertices of \cite{Freidel:2007py,Engle:2007wy}
belong to particular subsets of projected spin networks.

Using the commutation relation \eqref{AP2}, one can evaluate the action of the area operator
on the states \eqref{projspnet}. The result is given in terms of Casimir operators
of both the full gauge group and the subgroup  \cite{Alexandrov:2001wt}
\begin{equation}
\CS =\ell_p^2 \[(a^2 + (1-b)^2) C_{SO(3)} -(1-b)^2 C_{SO(3,1)}^{(1)}
+a(1-b) C_{SO(3,1)}^{(2)}\]^{1/2},
\label{oper}
\end{equation}
where the two Casimir operators of $so(3,1)$ are defined in terms of the generators and
evaluated on a principal series representation $\lambda=(n,\rho)$ as follows
\be
\begin{split}
C_{SO(3,1)}^{(1)}&=-\hf\, g^{IJ,KL}\hT_{IJ}\hT_{KL}=n^2-\rho^2-1 ,
\\
C_{SO(3,1)}^{(2)}&=-\frac{1}{4}\,\eps^{IJKL}\hT_{IJ}\hT_{KL}=2n\rho,
\end{split}
\label{CasimirSL}
\ee
and the representations of the subgroup are restricted to satisfy $j\ge n$.
The spectrum \eqref{oper} depends explicitly on the parameters $a,b$
entering the definition of the connection.
This implies that the quantizations based on different connections of the two-parameter
family are all inequivalent.

Finally, we notice that the projected spin networks are obtained by quantizing the phase space of the covariant
canonical formulation ignoring the second class constraints. Therefore they form what can be called
enlarged Hilbert space and as we mentioned above this space contains many states which are physically indistinguishable.
To remove this degeneracy one has to somehow implement the second class constraints at the level of the Hilbert space.
The idea is that this can be done by appropriately restricting the labels of spin networks
\cite{Alexandrov:2005ar,Alexandrov:2007pq}. How this is done in detail depends of course on
the explicit form of the second class constraints in question, which in turn is determined by the choice of connection.
Below we will see how this works in some particular cases.

\subsection{Two quantizations}
\label{subsec_twoq}

As we saw, different connections of the two-parameter family give rise to inequivalent quantizations.
Of course, such situation is unsatisfactory until we resolve this quantization ambiguity.
For this purpose some additional physical arguments should be invoked.

It turns out that there are two natural additional requirements to the list \eqref{difftr},
each selecting one particular connection.
Thus, there are two distinguished quantizations which are in fact different and in some sense even
orthogonal to each other. Here we describe their main features and in the next subsection we discuss
which of the two is the physically relevant one.

\subsubsection{LQG in a covariant form}
\label{subsubsec_LQGcov}

Above we mentioned that all connections $\SAab$ except one are non-commutative with respect to
the symplectic structure induced by the Dirac bracket.
The non-commutativity is a serious obstacle to find a representation of the classical commutation relations.
Therefore, it is natural to require the commutativity as an additional condition on the connection
to be used in the loop operators.
This condition fixes parameters $a$ and $b$ as follows \cite{Alexandrov:2002br}
\begin{equation}
a=-\im, \qquad b=1,
\label{parSU2}
\end{equation}
and the corresponding connection can be written as
\be
\SSA_i^{IJ}\equiv \SApar{-\im}{1}_i^{IJ} =
I_{(Q)KL}^{IJ}(1-{\im}\star) \omega_i^{KL}
+2(1+\im\star)x^{[J}\p_i x^{I]}.
\label{conSU2}
\ee
It possesses the following properties:
\begin{itemize}
\item
It is commutative
\be
\{ \SSA_i^{IJ}, \SSA_j^{KL}\}_D = 0.
\ee
\item
In the time gauge $\chi=0$, it coincides with
the Ashtekar-Barbero $SU(2)$ connection,
thus being its Lorentz generalization:
\be
\SSA_i^{IJ}\mathop{=}\limits_{\chi=0}
\left\{
\begin{array}{ll}
0 &\ \ I=0,\ J=a\\
\omega_i^{ab}-\im{\eps^{ab}}_{c}\omega_i^{0c} ={\eps^{ab}}_{c}\Ab{i}{c}
&\ \ I=a,\ J=b
\end{array}
\right.
\label{conAB}
\ee
\item
Its commutator with the bi-vector $\tQ$ is
\be
\{ \SSA_i^{IJ},\tQ_{KL}^j\}_D=\im \delta_i^j I_{(Q)KL}^{IJ},
\label{comSSA}
\ee
where $I_{(Q)}$ is the projector on the $SU_{x}(2)$
part of the Lorentz group. In the time gauge it reproduces the basic commutator
of AB formulation \eqref{brAB}.
\end{itemize}
Due to this last relation and as follows from \eqref{oper} with \eqref{parSU2},
the area spectrum corresponding to this $SL(2,\Cmat)$ connection
coincides exactly with the one coming from LQG
given by the Casimir operator of $SU(2)$ \eqref{spsu2}. Thus, despite the fact that
the connection lives in the Lorentz algebra and the full Lorentz symmetry is preserved,
the spectrum is discrete and one recovers the standard LQG results.

Moreover, one can show that once the second class constraints are taken into account at
the level of the Hilbert space, the kinematical states reduce to the usual
$SU(2)$ spin networks \cite{Alexandrov:2002br}.
Indeed, the second class constraints read
\be
I_{(P)KL}^{IJ} \SSA_i^{KL}=2\,x^{[J}\p_i x^{I]}.
\label{su2con}
\ee
Assuming that $x^I={\rm const}$, they imply that only the $SU_x(2)$ part of the connection
is non-trivial. Due to this its holonomy belongs to this subgroup and
therefore
\be
\Ppr{j_1}{x}\cdot R_G^{(\lambda)}\(U_{\alpha}[\SSA]\)\cdot \Ppr{j_2}{x}
=\delta_{j_1j_2}R_{H_x}^{(j_1)}\(U_{\alpha}[\SSA]\).
\label{wl}
\ee
The r.h.s. does not depend on the representation $\lambda$ and is non-vanishing only for $j_1=j_2$.
Substituting this into \eqref{projspnet}, one obtains an $SU(2)$ spin network labeled by representations
$j_e$ (the label $v$ becomes auxiliary) and intertwiners $\Int_{SU(2)}^{(v)}$.
It is clear that the case of arbitrary $x^I$ can be obtained by performing a Lorentz gauge  transformation.
Since the initial projected spin networks are gauge invariant, the result does not change.
Thus, choosing the parameters as in \eqref{parSU2}, we obtain LQG in the Lorentz covariant form.

\subsubsection{CLQG}
\label{subsubsec_CLQG}

Although the commutativity of the connection  is a nice property, there is another possibility of imposing  an
additional condition to resolve the quantization ambiguity, which has a clear physical origin.
Notice that the Lorentz transformations and spatial diffeomorphisms, which appear in the list
of conditions \eqref{difftr}, do not exhaust all gauge transformations. What is missing is the requirement
of correct transformations under time diffeomorphisms generated by the full Hamiltonian.
Only the quantity transforming as the spin-connection under {\it all} local symmetries of the theory
can be considered as a true spacetime connection.
Thus, another possible condition on $\SAab$ is that on mass shell it should satisfy
\be
\begin{split}
iv)\ \delta_{(\xi^0)}\SA_i^{IJ}=\,&\xi^0\p_0 \SA_i^{IJ}+\SA_0^{IJ}\p_i \xi^0,
\label{trtime}\\
\delta_{(\xi^0)}\SA_0^{IJ}=\,&\p_0(\xi^0\SA_0^{IJ}).
\end{split}
\ee
Remarkably, there is a unique member of the two-parameter family which solves \eqref{trtime}.
It corresponds to vanishing $a$ and $b$ and is given explicitly by \cite{Alexandrov:2001wt}
\be
\begin{split}
\SA_i^{IJ}\equiv \SApar{0}{0}_i^{IJ} =\, &  \omega_i^{IJ}+\hf \(1-\im^{-1}
\,\star\) I_{(Q)}^{IJ,\,KL}f^{ST}_{KL,\,PQ}\Pt_i^{PQ}\nabla_k\tP^k_{ST}
\\
=\, & I_{(P)KL}^{IJ} \(1+\im^{-1}
\,\star\) \omega_i^{KL}
+ \(1-\im^{-1}
\,\star\) \Gamma_i^{IJ}(\tP),
 \label{spconnew}
\end{split}
\ee
where
$\Gamma_i^{IJ}$ is the $SL(2,\Cmat)$ connection compatible with $\tP$.
{}From this expression one finds the following properties:
\begin{itemize}
\item
$\SA_i^{IJ}$ is non-commutative.
\item
On the surface of the Gauss constraint it coincides with the spin-connection.
\item
Its commutator with the bi-vector $\tP$ is
\be
\left\{ \SA_i^{IJ},\tP^j_{KL}\right\}_D
=\delta_i^j I_{(P)KL}^{IJ}.
\label{commPleb}
\ee
\end{itemize}
The last property, or equivalently \eqref{oper} with vanishing $a$ and $b$, implies that the area spectrum
in this case is completely different from the one of LQG and reads as
\begin{equation}
{\cal S}= \ell_p^2 \sqrt{C_{SO(3)} -C_{SO(3,1)}^{(1)}}.
\label{as}
\end{equation}
In particular, it involves a Casimir of the Lorentz group and hence this spectrum is {\it continuous}.
But the most striking and wonderful result is that the spectrum does {\it not} depend on the Immirzi parameter!
Moreover, one can show that this parameter drops out completely from
the symplectic structure written in terms of $\SA$ and $\tP$ \cite{Alexandrov:2005ar}.
Thus, it remains unphysical as it was in the classical theory, at least at this kinematical level.

The differences between the two cases are in fact very deep.
Comparing \eqref{commPleb} with \eqref{comSSA}, one observes that
the dynamical parts of $\SSA$ and $\SA$ are orthogonal to each other: for $\SSA$
these are the $SU(2)$ components that are dynamical, whereas in the case of $\SA$
these are the boost components. This is confirmed by the second class constraints
which for $\SA$ read as follows
\be
I_{(Q)KL}^{IJ}\SA_i^{KL}=\Gamma_i^{IJ}(\tP)
\label{contA}
\ee
and show that its $SU(2)$ part is fixed by the conjugate variables. Note the similarity of these
constraints with the reality conditions \eqref{realcon} for the complex Ashtekar connection. It is not accidental
since for $\im=\I$ the chiral component of the shifted connection coincides with the Ashtekar connection
and the reality conditions play the role of the second class constraints on an extended phase space \cite{Alexandrov:2005ng}.

The quantization relying on the use of the connection $\SA$ was called Covariant Loop Quantum Gravity
(CLQG) \cite{Livine:2006ix}. Unfortunately, there are two (related) problems which put this quantization
on unsteady ground. First, due to the complicated nature of the commutator of two connections,
it is not known how to represent the full Dirac algebra. In principle, the Dirac bracket of two connections is known explicitly
\cite{Alexandrov:2002xc,Alexandrov:2005ar} and has some nice properties, which can be easily derived from
the constraints \eqref{contA}. In particular, the commutator does not involve the connection
and can be seen as a first order differential operator with $\tP$-dependent coefficients acting on $\delta(x,y)$.
However, this information did not help so far to understand its underlying geometric meaning.
The second problem is that it is not known how the constraints
\eqref{contA} can be taken into account by a restriction of labels of projected spin networks
similarly to how the constraints \eqref{su2con} have been treated.
Nevertheless, we would like to take the above results seriously and to discuss what they imply for the status of LQG.

\subsection{Discussion}
\label{subsec_LQGdiscus}

The results presented in the previous subsection imply that LQG is a mathematically well
established quantization which can be formulated in a Lorentz covariant form.
But the basic holonomy operators are defined using a connection which transforms properly
only under space diffeomorphisms, whereas the action of time diffeomorphisms on it
is  extremely complicated. In other words, this connection is not a pull-back
of a spacetime connection, as was noticed in \cite{Samuel:2000ue}.

What does this mean? Is it just an inessential feature of the formalism or a serious problem?
From our point of view, this is a very important fact which indicates that LQG may have
troubles with the diffeomorphism invariance at the quantum level. The situation is in fact similar
to the one with space diffeomorphisms. Why do we need that the connection be a one form?
Because then the action of a diffeomorphism $\varphi$  on the holonomy along a loop is
another holonomy defined along  the  shifted loop
\be
U_{\alpha}[\varphi^*(A)]= U_{\varphi(\alpha)}[A].
\ee
If this property is not satisfied, it seems to be impossible to realize the symmetry
preserving all commutation relations. The constraint algebra
almost unavoidably is going to acquire an anomaly. And this is what we think does happen in LQG.
Some old observations supporting this conclusion can be found in \cite{Lewandowski:1997ba,Gambini:1997bc,Nicolai:2005mc}.

A breaking of some symmetry is not necessarily excluded. A notable example is the chiral symmetry.
However, here we deal with a {\it local} symmetry which is supposed to be fundamental and we do not consider
possibilities where it arises only in the low energy limit, as it happens, for example, in
the recent proposal \cite{Horava:2009uw}. The diffeomorphism invariance is one of the starting points of LQG
and it would be against its basic principles to give it up.

In fact, one could at least in principle consider such a possibility, if it was impossible to preserve
it at the quantum level. But as we saw in the previous subsection, there is an alternative choice of connection,
suitable for the loop quantization, which respects {\it all} gauge symmetries.
Besides, the latter approach, which we called CLQG, leads to results which seem to us much more natural.
For example, since it predicts the area spectrum independent on the Immirzi parameter, there is nothing
special to be explained and there is no need to introduce an additional fundamental constant.
Moreover, the spectrum appears to be continuous which is very natural given the non-compactness
of the Lorentz group and results from $2+1$ dimensions (see below).

Although these last results should be taken with great care as they are purely kinematical
and obtained ignoring the connection non-commutativity,
in our opinion, the comparison of the two possibilities to resolve the quantization ambiguity
points in favor of the second choice.
The only disadvantageous feature of this choice is that it is not so developed and experiences serious
technical difficulties. On the other hand, there is no reason to expect quantum gravity to be simple.
We believe that this feature cannot be considered as a physical argument in favor to neglect this possibility
and to take LQG as the only reasonable canonical quantization.
Moreover, the issue of diffeomorphism invariance harms the $SU(2)$ approach independently
on whether or not CLQG can be given a consistent realization, and does not allow
to view it as a physically acceptable theory.

Let us put now some of the LQG issues into a broader context.

\subsubsection{Immirzi parameter}

Since in LQG the Immirzi parameter becomes a new fundamental constant, it was asked whether
there are some effects where it appears already at the classical level, so that
it could be measured and compared, for example, with the value necessary to match
the black hole entropy \cite{Meissner:2004ju}.
At the same time, this would make its appearance in quantum theory not so surprising.

In \cite{Perez:2005pm} it was proposed that the role of such classical effect
may be played by the coupling with fermions. If one assumes that at the fundamental level
the fermions can be coupled to gravity only in a minimal way through the spin-connection,
as is well known, integrating it out leads to a four-fermion interaction with the coupling
determined by the Newton constant.
It turns out that the presence of the Immirzi parameter modifies the coupling and thus, in principle,
is measurable.

However, later it was noticed that this effect can be easily undone by a simple modification
of the fermion kinetic term \cite{Mercuri:2006um}. Although the modified action deviates
from the standard minimal coupling, it is still quadratic in fermions.
This observation suggested to consider the most general coupling of this type \cite{Alexandrov:2008iy}.
Then, the resulting effective action contains several current-current interaction terms.
The corresponding coupling constants are in general dependent on the Immirzi parameter,
but this dependence can be absorbed by a reparametrization of the initial coupling constants
appearing in the original action.
Thus, it is impossible to measure the Immirzi parameter once the fundamental couplings are not fixed.

Moreover, for a special choice of these couplings identical to the one from \cite{Mercuri:2006um},
the Immirzi parameter also drops out from the torsion removing CP violation effects. This choice is clearly distinguished,
which is confirmed also by an attempt to include the Immirzi parameter into supergravity \cite{Kaul:2007gz}.
It turns out to be possible, but local supersymmetry imposes so strong restrictions that
there is a unique way to achieve this goal. It is a straightforward generalization
of the coupling suggested in \cite{Mercuri:2006um} and leads to no effects of the parameter.

These results show that the Immirzi parameter strongly resists against any attempt to make it observable.
Although this is a classical story, one might expect it to continue in quantum theory as well.

\subsubsection{Area spectrum}

One of the striking differences between LQG and CLQG is the nature of the area spectrum: whether
it is discrete or continuous. Can one find some hints from simplified models
about which type of the spectrum one should expect in 4d gravity?

A very useful model for this purpose is general relativity in $2+1$ dimensions.
Since this is a topological theory\footnote{It is commonly said that 2+1 gravity is
a topological theory completely equivalent to Chern-Simons theory  but this has to
be taken with caution because of the problem of degenerate metrics which is
well analyzed in \cite{Matschull:1999he} }
it is much easier comparing to gravity in 4 dimensions.
In fact, there exist many successful approaches to its quantization. However, different approaches are
suitable for different types of questions. Here we are interested in the spectrum of the length operator,
which is the 3-dimensional analogue of the area spectrum in 4 dimensions.
Such spectrum was investigated in \cite{Freidel:2002hx} and, not surprisingly, it was found that
the spectrum is continuous for spacelike intervals and discrete for timelike intervals.
This is precisely the pattern which one has in CLQG \cite{Alexandrov:2005ar}!

In fact, there is no way to avoid this conclusion in 3 dimensions. In this case there are no second class
constraints and therefore there is no reason to introduce a time gauge or to change the connection.
As a result, one has to deal with the usual spin-connection and the structure group
coincides with the full gauge group $SO(2,1)$.
It is non-compact and naturally leads to a continuous spectrum.\footnote{There are many works where
people make contact between loop quantizations of 4 and 3-dimensional general relativity. However,
very often in 3d they consider Riemannian gravity without even mentioning this. Of course, then the
gauge group
is $SU(2)$ like in LQG and the spectra of kinematical length operators are discrete.
But one should remember that this is a different theory and results found there cannot
be used to support the results of LQG.}

In fact, there are two other more general issues which show that
the LQG area spectrum is far from being engraved into marble.
First, the area operator is a quantization of the classical area function and, as any quantization,
is supplied with ordering ambiguities. In fact, it is possible to define
other quantizations of this classical observable leading to other
spectra, even in the standard LQG framework.
For example, in \cite{Alekseev:2000hf} the equidistant form of the spectrum, $\CS_\Sigma\sim j+\hf$, was advocated.
Moreover, a similar ambiguity is used in the new SF models \cite{Freidel:2007py,Engle:2007wy}
to adjust constraint operators so that they would have non-trivial solutions.
Thus, the possibility of such renormalization effects should not be ignored.

Second, the computation of the area spectrum has been done only at the kinematical level.
The problem is that the area operator is not a Dirac observable. It is only gauge invariant,
whereas it is not invariant under spatial diffeomorphisms
and does not commute with the Hamiltonian constraint. This fact raises questions and suspicions
about the physical relevance of its spectrum and in particular about the meaning of its discreteness,
even among experts in the field \cite{Dittrich:2007th,Budd:2009kf}.

One can distinguish two viewpoints on this very important issue. The first one is that one must stick to the
Dirac formalism of constrained systems and to its standard quantization scheme. In this scheme
only Dirac observables have a physical meaning. Therefore, the area
operator should be promoted to some true Dirac observable. This can be done, for example,
by coupling matter fields and using the theory of partial and complete
observables in general relativity developed in \cite{Rovelli:2001bz,Dittrich:2005kc}.

The second strategy is to abandon the Dirac scheme and to use instead a relational interpretation
of relativistic Quantum Mechanics as advocated in \cite{Rovelli:2004tv}. In this case the quantum theory is defined by a
kinematical Hilbert space (in our case it is ${{\cal H}_G}$), whereas the dynamics is implemented by a projection operator
$P:{\cal H}_G\rightarrow {\cal H}_G $ whose image
lies in the physical Hilbert space. This implies that if $Q_n$ is a family of quantum partial observables,
then the probability of observing $q_n$ when $q_n'$ has been observed is given by
$\vert \langle q_n\vert P \vert q_n'\rangle\vert^2$.

The difference between the two interpretations and the importance of this issue
has been clarified in \cite{Dittrich:2007th,Rovelli:2007ep}.
Namely, the authors of \cite{Dittrich:2007th} proposed several examples of low
dimensional quantum mechanical constrained systems where the spectrum of the physical observable
associated to a partial observable is drastically changed. This is in contradiction with the expectation of LQG that the
spectrum should not change. Then in \cite{Rovelli:2007ep} it was argued that one should not stick to
the Dirac quantization scheme but to the relational scheme. Accepting this viewpoint allows to keep
the kinematical spectra unchanged.
Thus, the choice of interpretation for physical observables directly affects
predictions of quantum theory and clearly deserves a precise scrutiny.

Whereas the relational viewpoint seems to be viable, the work \cite{Dittrich:2007th} shows
that if we adhere only to the first interpretation, which is the most commonly accepted one,
then it is of upmost importance to study the spectrum of complete observables.
Unfortunately, up to now there are no results on the computation of the spectrum
of any complete Dirac observable in full LQG.
On the other hand, in the $2+1$ dimensional case,
some examples of complete observables associated to length variables of spacelike and timelike distances
were exhibited in \cite{Budd:2009kf}.
The spectra of these variables are both continuous and not bounded from below.

Given all complications with the search and quantization of complete Dirac observables,
one can restrict oneself to the study of partial observables which are only invariant under spatial diffeomorphisms.
For example, the partial observables corresponding to geometric operators can be defined
identifying the measured space region by some values of a matter field \cite{Rovelli:1994ge}.
As above, such an analysis can be performed exactly in $2+1$ gravity coupled to point-like particles
\cite{Matschull:1997du}. In this case the role of the length operator is replaced by an operator
measuring the position of the particle. It was shown that its spacelike component is continuous,
whereas the timelike spectrum is discrete. Again, this is precisely the same qualitative picture
which one finds in CLQG and for the length operator in three dimensions.

In our opinion, all these findings and the above mentioned issues clearly make the discreteness found in LQG
untrustable and suggest that the CLQG spectrum \eqref{as} is a reasonable alternative.

\subsubsection{Black hole entropy}

The area spectrum and its discreteness are closely related to another result of LQG
which is often presented as one of its main achievements --- the derivation
of the Bekenstein--Hawking formula for the black hole entropy
\cite{Rovelli:1996dv,Krasnov:1996tb,Ashtekar:1997yu}. Can one trust this derivation and
how rigid is it?

The LQG derivation can be summarized as follows.
The horizon of a black hole is considered as a spacetime boundary.
The condition that the boundary is an (isolated) horizon is formulated in terms of some boundary
conditions which induce a dynamics of the boundary degrees of freedom described by a Chern--Simons theory
with the structure group either $U(1)$ \cite{Ashtekar:1999wa} or $SU(2)$ \cite{Smolin:1995vq,Engle:2009vc}.
The system is quantized by following the loop quantization in the bulk and the standard
Chern--Simons technique on the boundary. After that a quantum version of the boundary conditions
couples the bulk and boundary quantum states.
As in the bulk the states are given by the usual spin networks, they puncture the horizon and the boundary conditions
impose some restrictions on the holonomies around these punctures.
Finally, a suitable counting of distinguishable states determined by the area spectrum produces
the entropy linear in the horizon area. Since the overall coefficient is proportional
to the Immirzi parameter, one can adjust the latter to reproduce the famous coefficient 1/4 in
the Bekenstein--Hawking formula.

{}From the first sight the discreteness of the area spectrum is crucial for this derivation.
Without it one would have a continuum of states to be counted. However in
\cite{Alexandrov:2004fh} (see also \cite{Dreyer:2004jy}) it was argued that
the counting should actually be restricted to the states producing the minimal area quantum
at each puncture because all other states describe a different bulk geometry which is not stationary.
Of course, this still requires the existence of a non-vanishing area quantum.
Remarkably, for the CLQG spectrum \eqref{as} it is non-vanishing, ${\cal S}_{\rm min}=8\pi\hbar G $,
once one restricts to the principal series representations of $SL(2,\Cmat)$.

However, since the spectrum \eqref{as} is independent of the Immirzi parameter, the challenge
now is to find such counting which gives the exact coefficient 1/4, and not just the proportionality to
the horizon area. In fact, the last point is the weakest place of the LQG derivation
comparing to all other derivations existing in the literature. All of them are able to get
the exact coefficient without invoking any additional parameter fitting.
Usually, it is not a big deal to get the proportionality to the area. It is the coefficient that is non-trivial.
See, for example, numerous entropy countings in string theory where the restriction to extremal
or near-extremal geometries is compensated by remarkable coincidences for plenty of charge combinations
\cite{Mohaupt:2000mj}.

Besides, there are two other points which make the LQG derivation suspicious.
First, it is not generalizable to any other dimension.
If one draws direct analogy with the 4-dimensional case, one finds a picture which is
meaningless in 3 dimensions and does not allow to formulate any suitable boundary condition
in higher dimensions.
Indeed, in the former case the punctures of the horizon would split it into a set of
disjoint segments. It is unclear what they can be used for. In the latter situation
the problem appears because any loop is contractible on the punctured $n$-dimensional sphere
with $n>2$. As a result, a boundary theory formulated in terms of connections
allows to write a quantum boundary condition only if a spacelike section of the horizon is
two-dimensional.

This situation should be contrasted with the universality of the entropy counting observed in
\cite{Carlip:2007ph,Carlip:2008rk}. Moreover, comparing the LQG derivation with the approach to black hole entropy
developed by Carlip \cite{Carlip:1998wz,Carlip:1999cy} raises the second question.
In that approach the entropy follows from the counting of states in some CFT appearing
as a symmetry of the near horizon geometry. But the directions of spacetime supporting this CFT
are orthogonal to the ones relevant for the LQG calculation. Whereas in the latter case
these are spacelike directions along the horizon, in the former case they form the orthogonal $(r,t)$-plane.
Thus, there is a fundamental difference between these two approaches,
which also explains why one is universal and the other is stuck to 4 dimensions.

\subsubsection{Diffeomorphisms}

Finally, let us mention also a few issues with the imposition of
diffeomorphism and Hamiltonian constraints:
\begin{itemize}
\item
Although at first sight it seems that the spatial diffeomorphisms
reduce the degrees of freedom related to the embedding of the spin network graph to just
information about its topology, this not quite true.
In fact, there still remain some continuous moduli
depending on the relative angles of edges meeting at a vertex of sufficiently high valence \cite{Grot:1996kj}.
Due to this the Hilbert space ${\cal H}_{GDiff}$ is not separable and if one does not want that
the physics of quantum gravity is affected by these moduli,
one is led to modify this picture. To remove this moduli dependence, one can extend $Diff(M)$ to a subgroup
of homeomorphisms of $M$ consisting of homeomorphisms which are smooth except at
a finite number of points \cite{Fairbairn:2004qe} (the so called ``generalized diffeomorphisms'').
If these points coincide with the vertices of the spin networks,
the supposed invariance under this huge group
will identify spin networks with different moduli and solve the problem.
However, this procedure has different drawbacks. First, the generalized diffeomorphisms are
not symmetries of classical general relativity. Moreover, they
transform covariantly the volume operator of Rovelli--Smolin but not the one of Ashtekar--Lewandowski which
is favored by the triad test \cite{Giesel:2005bk}. This analysis indicates that
these generalized diffeomorphisms should not be implemented as symmetries at quantum level and,
as a result, we remain with the unsolved problem of continuous moduli.
In $2+1$ dimensions this problem does not appear because in this case the
Hamiltonian constraint fixes the connection to be flat and on the flat solutions
the evaluation of a spin network does not depend on these continuous moduli.
This shows that a problem which cannot be cured at the kinematical level
might sometimes be resolved on the physical Hilbert space.

\item
In all canonical approach to quantum gravity, the manifold $M$ is fixed and the states of geometry on $M$ are given by
embedded spin networks. So the claim that ``spin networks are not embedded in space but are quantum states of space'' is not
completely true because  it forgets the topological degrees of freedom of the spin networks coming from
their embedding (knotting) in $M.$
The relevance for physics of these topological degrees of freedom is not very well understood
in the present formalism of LQG.

\item
In the Dirac formalism the constraints $H_i$ only generate diffeomorphisms which are connected
to the identity. Therefore, there is a priori no need for defining ${\cal H}_{GDiff}$
to be invariant under large diffeomorphisms. On the other hand, in LQG these transformations,
forming the mapping class group, are supposed to act trivially. This is justified
in \cite{Thiemann:2007zz} (section I.3.3.2) to be the most practical option
given that the mapping class group is huge and not very well understood.
However, in 2+1 quantum gravity in the Hamiltonian
picture on $\Sigma\times \mathbb{R}$ described in the Chern-Simons theory formalism \cite{Alekseev:1995rn},
one is led to first define a Hilbert space of states which are only invariant under the group of
diffeomorphisms connected to the origin and then to define a unitary projective
representation of the mapping class group of $\Sigma$ on this space of states.
Thus, the simplest option taken by LQG might be an oversimplification missing important features of
the right quantization.

\item
The construction of the Hamiltonian constraint operator has a lot of intrinsic arbitrariness.
It appears in the choice of representations associated to the holonomies
defining the operator \eqref{qHE} \cite{Gaul:2000ba,Perez:2005fn},
and in the choice of particular regularization procedure \cite{Alesci:2010gb}.
Moreover, a huge arbitrariness is hidden in the step
suggesting to replace the classical Poisson brackets, as for example \eqref{extr_vol},
by quantum commutators.  In general, this is true only up to corrections in $\hbar$ and
on general ground one could expect that the Hamiltonian constructed
by Thiemann may be modified by such corrections.
This is a bit disappointing situation for a would be fundamental quantum gravity theory.

In principle, all this arbitrariness should be fixed by the requirement that
the quantum constraints reproduce the closed Dirac constraint algebra.
However, the commutators of quantum constraint operators are not under control,
although a weak closure of the algebra in the form
\begin{equation}
\langle\Psi_{[\Gamma']},[\hat{H}[N_1],\hat{H}[N_2]]\Psi_{\Gamma}\rangle=0
\end{equation}
has been demonstrated in \cite{Thiemann:1996aw}.
This is the place where we expect some anomalies to appear, as is suggested
by our covariant analysis (see also section 6.2 of \cite{Nicolai:2005mc} on this issue).

\item
Although some solutions of the Hamiltonian constraint can be found in \cite{Thiemann:1997rv},
one does not know yet how to construct the physical scalar product on them.
And, of course, the most important problem is how to extract the classical limit and
whether it will have something to do with general relativity.
\end{itemize}

\subsection{Summary}

Let us recapitulate our main conclusions concerning the canonical loop approach to quantum gravity.

Trying to incorporate the full Lorentz gauge symmetry into the standard LQG framework based on the $SU(2)$ group,
we discovered that LQG is only one possible quantization of a two-parameter family of inequivalent
quantizations. All these quantizations differ by the choice of connection to be used in
the definition of holonomy operators --- the basic building blocks of the loop approach.
LQG is indeed distinguished by the fact that the corresponding connection is commutative.
Nevertheless, a more physically/geometrically motivated requirement selects another connection,
which gives rise to the quantization called CLQG.
Although the latter quantization has not been properly formulated yet,
it predicts the area spectrum which is continuous
and independent on the Immirzi parameter, whereas LQG gives a discrete spectrum dependent on $\im$.

We argued that these facts lead to suspect that LQG might be an {\it anomalous} quantization of general relativity:
in our opinion they indicate that it does not respect the 4d diffeomorphism algebra at quantum level.
If this conclusion turns out indeed to be true, LQG cannot be physically accepted.
At the same time, CLQG is potentially free from these problems. But due to serious complications,
it is far from being accomplished and therefore the status of the results obtained so far,
such as the area spectrum, is not clear.

We also pointed out that some of the main LQG results are incompatible either with other approaches
to the same problem or with attempts to generalize them to other dimensions. We consider these facts as
supporting the above conclusion that LQG is not, in its present state, a proper quantization of general relativity.

\newpage

\section{Path integral approach}
\label{sec_path}

\subsection{Spin foam models}

The second part of this critical review is devoted to the spin foam approach to quantum gravity.
It is very closely related to the loop approach discussed in the first part,
but they should not be mixed up. In a nutshell, LQG is supposed to give an Hamiltonian picture of quantum
gravity based on the use of specific variables (connections), whereas spin foam models are certain type
of discretized path integral approach to the quantization.
A priori these are different approaches using different methods and leading to different results.
Of course, in the best case their predictions should coincide
and they should be just equivalent quantizations. But at present such an agreement
has not been achieved yet.

We start by describing what a spin foam and a spin foam model are. Then we present
the basic strategies to derive SF models. In the following sections, we discuss the most important models
of 4-dimensional general relativity existing in the literature, their derivation, self-consistency,
and relation to the canonical quantization.
Some nice reviews on this subject can be found in \cite{Baez:1997zt,Oriti:2001qu,Perez:2003vx}.

\subsubsection{Basic concepts}
\label{subsubsec_concepts}

A spin foam is an oriented 2-dimensional complex colored with some group theoretic data like
representations and intertwiners. The representations are assigned to the faces of the complex,
whereas the intertwiners are associated to the edges. Note that the vertices are not colored. This is because
the coloring can be thought as a representation of a kinematical information, whereas
the vertices encode the dynamics (see below).

\begin{figure}
\psfrag{G1}{$\Psi_{\Gamma}$}
\psfrag{G2}{$\Psi_{\Gamma'}$}
\centering
\includegraphics[scale=0.4]{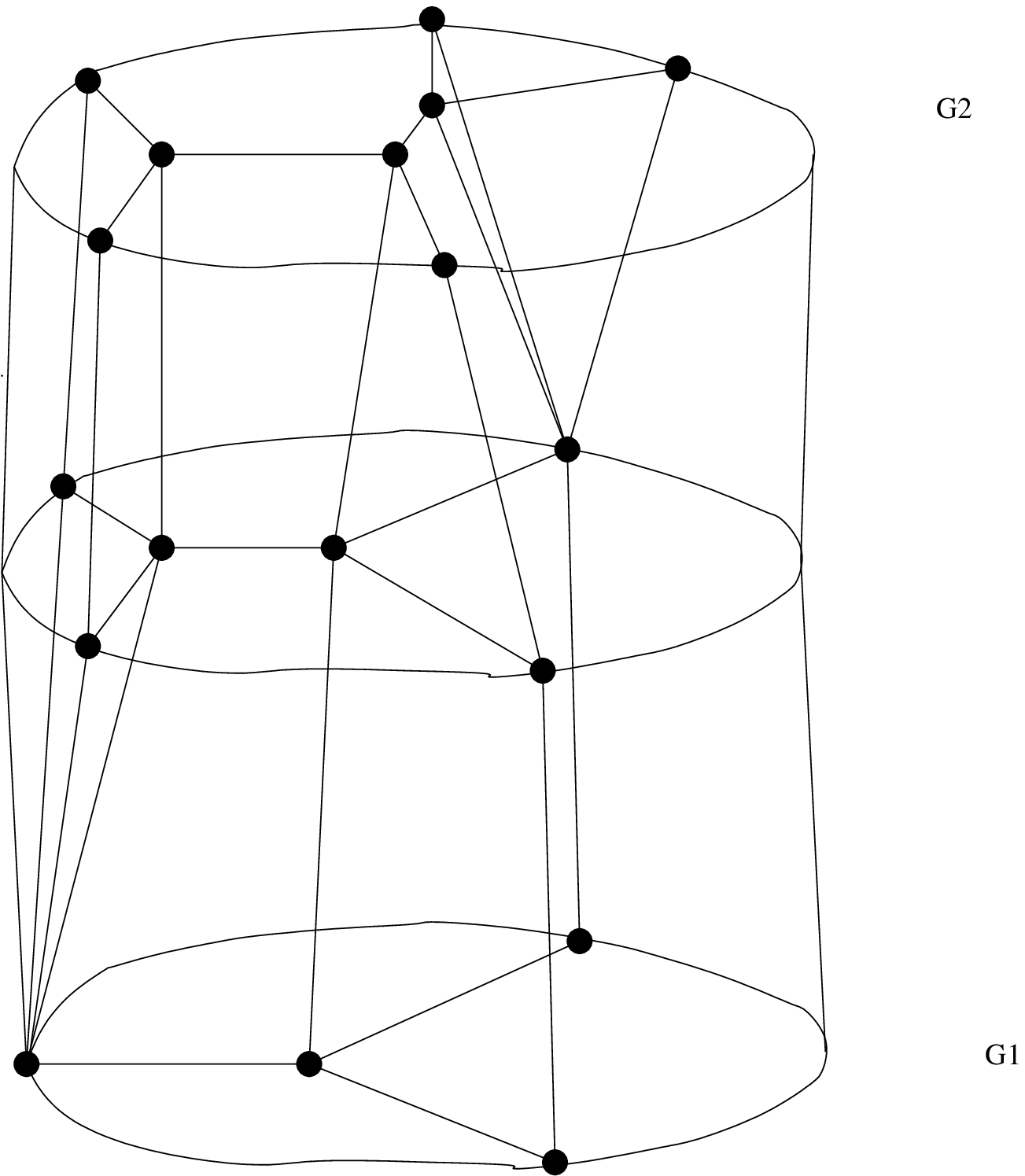}
\caption{A spin foam representing a transition amplitude between two spin networks.}
\label{figspinfoam}
\end{figure}

Spin foams can have boundaries. It is clear that on each connected component of the boundary
the spin foam induces a spin network such that its labeling is consistent with the labeling of the foam.
Reversing this picture allows to view spin foams as quantum histories of spin networks (Fig. \ref{figspinfoam}).
Since the latter are supposed to represent a quantum space, spin foams are thus considered as
a representation of quantum spacetime interpolating between given boundary data.
Moreover, depending on the number of boundaries, they can describe processes of either
splitting of space into different components, or disappearing of space into nothing, {\it etc.}

The appearance of spin networks at the boundaries points toward
a connection with the loop approach \cite{Reisenberger:1996pu}.
To get such a connection in a more precise way, one should start from the physical scalar
 product between
two kinematical states. The physical scalar product can be rewritten in terms of the
kinematical one
by means of insertion of a certain projection operator, which is closely related to the
evolution operator
and is written as exponential of the Hamiltonian \cite{Rovelli:1998dx}. Expanding the
exponential,
one obtains a series of terms given by expectation values of finite powers of $\hat H$ between
spin network states. It is easy to see that
each such term can be represented as a particular spin foam with the number of vertices
given by the power of the Hamiltonian operator. The vertices mark points in the evolution of spin networks
where an interaction takes place.
The full transition amplitude is represented in this way as a formal sum over all spin foams or,
given their interpretation, as a sum over all possible histories of spin networks.

This provides a qualitative relation between the two background independent approaches.
In fact, it is clear that given a loop quantization supplied with a Hamiltonian constraint
operator there should exist a spin foam model corresponding to it
and, following the above procedure, one may try to find such a model.
However, due to the complicated form of the Hamiltonian constraint,  one usually starts from
the opposite side. Namely, one tries to derive a spin foam model
by means of other  techniques and then compare the result with what is expected from the loop side.
This is why the SF approach should be distinguished from the loop quantization.
And as we will see, the results of the two approaches are indeed different so far.

But let us return to the definitions and define what a general SF model is.
Any spin foam model is supposed to associate a complex amplitude $Z_\Psi(\CM)$ to a spacetime region $\CM$
with boundaries provided the latter are characterized by fixed spin networks $\Psi$. For example, in the case of two
disconnected boundaries, such amplitudes are interpreted as transition amplitudes from one
spin network state to the other. Besides, they should satisfy a set of natural properties
such as
\be
Z_\Psi(\CM_1\cup\CM_2)=\sum_{\Psi'}Z_{\Psi_1}(\CM_1)Z_{\Psi_2}(\CM_2),
\ee
where $\Psi'$ is the restriction of $\Psi_1$ and $\Psi_2$ on $\p\CM_1\cap\p\CM_2$ and $\Psi$ is the union of the
remaining parts.

In all SF models the amplitudes can be represented in the following general form
\be
Z_\Psi(\CM)=\sum\limits_{C :\p C=\Gamma_\Psi} w(C) \sum\limits_{J,\, \Int} \prod_f A_f \prod_e A_e \prod_v A_v.
\label{defstat}
\ee
Here the sum goes over all $2$-complexes $C$ fitting the given graph of the spin networks at the boundaries and
over all colorings $(J,\Int)$ of each $C$ fitting the coloring of the spin networks,
possibly with some additional restrictions on allowed representations and intertwiners.
The weight $w(C)$ is usually some symmetry coefficient
and $A_f,\ A_e,\ A_v$ are {\it face}, {\it edge} and {\it vertex amplitudes}, respectively.
These three quantities are the main ingredients defining the model.
To give a SF model essentially means to provide explicit expressions of these amplitudes and the allowed
set of representations and intertwiners.

Usually, the amplitudes are supposed to be local, {\it i.e.}, they depend only on
the coloring of adjacent simplicial elements. Thus, $A_f$ is a function of the representations
located on the face $f$, $A_e$ is a function of the intertwiner assigned to $e$ and of the
representations on the faces containing $e$, whereas $A_v$ depends on the representations
on the faces and on the intertwiners on the edges containing the vertex $v$.

The fact which makes possible a geometric interpretation of a spin foam is that
the 2-complex  can be viewed as the dual 2-skeleton of a 4-dimensional triangulation of spacetime.
More precisely, to each triangulation of space time one can associate a 2-complex, the dual 2-skeleton,
by associating to each simplex a point in its interior (the vertices of the skeleton), by connecting
these points by edges when the corresponding $4$-simplices have a common tetrahedron, and by associating
to each face of the triangulation a face of the 2-complex corresponding to all the tetrahedra having this
face in common. In the other way around, one draws a 4-simplex around each vertex of the spin foam
such that edges and faces intersect tetrahedra and triangles of the triangulation, respectively.
In case where there are more than 5-valent vertices involved (or the boundary spin networks have more than
4-valent vertices), simplicial decomposition has to be replaced by a more general one.\footnote{See, for example,
\cite{Baratin:2008du} for a spin foam construction using cubulations.}

This ``duality'' opens two main ways to derive SF models. The first relies on a quantization of
 the geometry of 4-simplex. Note that the representation \eqref{defstat} implies that
the contributions of each 4-simplex
to the total amplitude are factorized. Therefore, it is sufficient to consider just one 4-simplex,
to find the corresponding amplitude and then to glue such several contributions together.

This approach makes also clear the meaning of all ingredients defining the model.
First, the allowed set of representations and intertwiners selects the allowed boundary states
and therefore one can say that it defines a kinematical Hilbert space.
For example, if one wants to have a model consistent with LQG, the corresponding boundary states must
be the usual $SU(2)$ spin networks parameterized by $SU(2)$ spins and $SU(2)$ invariant intertwiners.
Next, the vertex amplitude determines the dynamics and therefore it is the most important quantity
which one has to look for. According to \cite{Reisenberger:1996pu},
in the canonical picture it would correspond to an expectation value of the Hamiltonian
operator. Finally, the face and edge amplitudes are responsible for a consistent gluing of different
simplex contributions.

The approach based on the geometric quantization can be very useful and illuminating in determining
the state space of the model, but it is difficult to use it to find the right gluing factors and even
the vertex amplitude.
A more rigorous and powerful approach is the one based on the discretized path integral \cite{Freidel:1998pt}.
It arises very naturally in this context because from our discussion above, it follows that
the amplitude \eqref{defstat} can be seen as a sum over discretized spacetimes.
This is precisely what the discretized path integral for gravity is supposed to be.

By itself, however, the passage to a discretization does not solve any problems except in
$3$-dimensions where it is exact.
The discretized path integral is usually even more complicated than its continuous cousin.
This is why some additional hints are required to be able to extract a SF model.
Below we discuss a strategy which allows to do that and which is widely accepted
in most of the SF derivations.

\subsubsection{The strategy}
\label{subsubsec_strategy}

Our aim is to describe the SF approach to 4-dimensional general relativity.
But let us start from a much more simple 3-dimensional case where spin foams appeared for the first time.
In this case, gravity with cosmological constant $\Lambda$ is described by the simple action
\be
S_{\rm 3d}=\int \eps_{IJK} \(e^I\wedge F^{JK}(\omega)+\frac{\Lambda}{6}\, e^I\wedge e^J\wedge e^K\).
\label{3dimgr}
\ee
An important feature of this theory is that it is topological, {\it i.e.},
it does not have local propagating degrees of freedom (locally all solutions to the equations of motion are pure gauge).
This property implies that the theory can be discretized without losing any information
and moreover it does not depend on the chosen discretization if the latter is sufficiently refined
to take into account all global degrees of freedom.
Due to this, 3-dimensional gravity is amenable to the spin foam quantization and
transition amplitudes can be written in the form \eqref{defstat}, but {\it without} the sum over all possible
2-dimensional complexes. This is a crucial simplification allowing to find the Ponzano--Regge model
($\Lambda=0$) \cite{Ponzano:1968dq} and the Turaev--Viro model ($\Lambda>0$)
\cite{Turaev:1992hq}, both of them describing Riemannian gravity.
The discretization independence of these models follows from the pentagon relation satisfied by
the vertex amplitude given in terms of $6j$ coefficients of $U_q(su(2))$ with
$q=e^{i\sqrt{\Lambda}}$.
Note that when $\Lambda>0$ the sum over the coloring is a finite sum because of the infrared cutoff
$j\leq {\Lambda}^{-\frac{1}{2}}$ on the allowed representations.

Trying to lift these results to four dimensions, one finds that there is a model which takes a somewhat
intermediate place between 3 and 4-dimensional gravity. This is the so called BF theory,
which has been first studied in \cite{Horowitz:1989ng}.
It exists in any number of dimensions and is always given by the following action
\be
S_{\rm BF}=\int B_{IJ}\wedge F^{IJ}(\omega),
\label{actBF}
\ee
where $B^{IJ}$ is understood as a $d-2$-dimensional form taking values in the Lie algebra of the local symmetry
group $SO(\eta)$.
One can also include a cosmological constant term
which in four dimensions is represented by a term quadratic in the $B$-field.

It is clear that 3-dimensional gravity is a particular case of BF theory where $B_{IJ}=\eps_{IJK}e^K$.
Moreover, it turns out that its main property, namely, that this is a topological theory, extends to BF theory in
any dimension. Similarly, BF theory can be quantized by spin foam methods leading to a discretization
independent model \cite{Crane:1993if}.

Why is all this important for us? The reason for that is the form of the action of general relativity
in the first order formulation. If one denotes
\be
B^{IJ}=*(e^I\wedge e^J),
\label{Bee}
\ee
the Hilbert--Palatini action becomes the action of 4-dimensional BF theory \eqref{actBF}!
Of course, these are not the same theories because in general relativity the independent variable is the tetrad
one form $e^I$ and not the 2-form bi-vector $B^{IJ}$.
However, if one ensures that the latter is restricted to be given in terms of the tetrads as in \eqref{Bee},
the BF theory will reduce to general relativity.

This idea is accomplished in Plebanski formulation \cite{Plebanski:1977zz,Capovilla:1991qb,DePietri:1998mb}
which differs from the action \eqref{actBF} by
the presence of an additional term
\begin{eqnarray}\label{Plebanskiaction}
S_{\rm Pl} =    \int  \( B_{IJ}\wedge F^{IJ}(\omega)
+\hf\,\varphi_{IJKL}  B^{IJ}\wedge B^{KL}\) ,
\end{eqnarray}
where $\phi_{IJKL}$ is an independent field satisfying suitable symmetry constraints
which in particular include $\eps^{IJKL}\phi_{IJKL}=0$.
The additional term is introduced to impose the so called
simplicity constraints obtained by varying with respect to $\phi_{IJKL}$.
In the non-degenerate case, {\it i.e.}, when $\CV=\frac{1}{4!}\,\tr (B\wedge B)$ does not vanish,
they are equivalent to
\be
 B^{IJ}\wedge B^{KL}= \sigma{\cal V} \, \eps^{IJKL}
\ \ \Leftrightarrow\ \
\eps_{IJKL} B^{IJ}_{\mu\nu} B^{KL}_{\rho\sigma}=
\sigma{\cal V} \, \eps_{\mu\nu\rho\sigma} ,
\label{simplicityconditions1}
\ee
where $\sigma$ is as usual the sign distinguishing Riemannian and Lorentzian
cases. Of these two equivalent forms,
it is the second form that will be important for our purposes.

The role of the simplicity constraints is precisely to reduce  BF theory to general relativity.
This is done because, again for a non-vanishing $\CV$, they have two sets of solutions.
Both of them are written in terms of a tetrad. The first coincides with \eqref{Bee}
and the second is given by its Hodge dual $B^{IJ}=e^I\wedge e^J$. Plugging the first solution
into Plebanski action \eqref{Plebanskiaction}, one reproduces the usual Hilbert--Palatini formulation,
whereas the second solution gives the action coinciding with the second term of the Holst action \eqref{Sd1}.
The latter does not have local degrees of freedom and therefore the corresponding sector of Plebanski
theory is called topological.

Most of the constructions of SF models of 4-dimensional general relativity
heavily rely on the Plebanski formulation and  translate the classical relation between
BF theory and gravity directly to
the quantum level. In other words they all employ the following strategy:
\begin{enumerate}
\item discretize the classical theory putting it on a simplicial complex;
\item quantize the topological BF part of the discretized theory;
\item impose the simplicity constraints at the quantum level.
\end{enumerate}
Thus, instead of quantizing the complicated system obtained after imposing the constraints,
they {\it first quantize and then constrain}.

This strategy is behind all the progress achieved in the construction of 4-dimensional SF models.
However, at the same time, this is a very dangerous strategy and, as we believe, it is the reason
why most of these models cannot be satisfactory models of quantum gravity.
As we will show, it is inconsistent with the Dirac rules of quantization and is somewhat misleading.
But before explaining what is wrong with it, let us present the most important and elaborated
spin foam models.

\subsection{Barrett--Crane model}
\label{subsubsec_BC}

The Barrett--Crane model was the first spin foam model for 4-dimensional gravity and
remained the leading proposal during 10 years. It exists in both Riemannian \cite{Barrett:1997gw}
and Lorentzian versions \cite{Barrett:1999qw}, but the logic of the derivation
does not depend on the signature. Therefore, we can treat simultaneously both cases.

There are various ways to derive this model, all of which perfectly fit the strategy described above.
Therefore, let us start by discretizing the basic variables which are the spin-connection $\omega^{IJ}$
and the 2-form $B^{IJ}$. On a triangulation, the former naturally
gives rise to holonomies $g_t$ along edges of the 2-complex (spin foam)
dual to the triangulation which connect two 4-simplices and are in one-to-one correspondence
with tetrahedra. The $B$-field associates a bi-vector to each triangle
through
\be
B_f^{IJ}=\int_{\Delta_f}B^{IJ}.
\label{Bfdef}
\ee
Since triangles are dual to faces of the dual 2-complex, one can use the latter to label the bi-vectors.
When these bi-vectors come from the metric structure (the tetrad) they have a clear geometric meaning:
their norm gives the area of the corresponding triangle and the tensor structure encodes the direction of
its normal.

This provides the kinematical variables for the discretized BF theory and Plebanski formulation
of general relativity. But the latter contains in addition the simplicity constraints \eqref{simplicityconditions1}
which should also be discretized. For that purpose, one smears their second version over
two triangles of the discretization belonging to the same tetrahedron.
Depending on the relative position of the triangles one obtains three types of constraints:
\begin{itemize}
\item
diagonal simplicity: $\eps_{IJKL}B_f^{IJ}B_f^{KL}=0$ --- if two triangles are the same;
\item
cross simplicity: $\quad\ \, \eps_{IJKL}B_f^{IJ}B_{f'}^{KL}=0$ --- if two triangles share an edge;
\item
volume constraint: $\ \eps_{IJKL}B_f^{IJ}B_{f'}^{KL}=\pm \CV_v$ --- if two triangles meet only at the vertex.
\end{itemize}
It is worth to distinguish these three constraints because they have different implications on spin foams.

There is also an additional constraint which is imposed in the BC and many other models.
It appears from the geometric interpretation of the bi-vectors mentioned above.
If $B_f$ come from the tetrad,
they also should satisfy the {\it closure} constraint
\be
\sum_{f\subset t}B_f=0.
\label{closure}
\ee
In fact, these constraints are not independent from the previous ones.
It was shown \cite{Livine:2007ya} that the closure constraint
together with the diagonal and cross simplicity implies the volume constraint.

The next step is to quantize BF theory. This can be done, for example, via path integral.
Since the discretized action
\be
S_{BF}=\sum_f \tr\(B_f g_f\),
\label{SBFd}
\ee
where $g_f$ is the full holonomy around a dual face $f$,
is linear in $B_f$, the integral over the bi-vectors in the partition function is easily evaluated
yielding a product of delta-functions imposing the flatness condition on the curvature
\be
Z_{BF}=\int \prod_t \de g_t \prod_{f}\delta\(\prod_{t \supset f} g_t\).
\label{ZBFd}
\ee
This flatness condition is also one of the equations of motion of
the continuous BF theory and it is the main reason why this theory is topological.
Of course, in gravity this condition should be relaxed which is achieved by implementing
the simplicity constraints.

A spin foam representation of the partition function is obtained from \eqref{ZBFd} by using
the Plancherel decomposition of the $\delta$-function on the group. It gives
\be
Z_{BF}=\sum_{\lambda_f}\int \prod_t \de g_t \prod_{f}d_{\lambda_f}\tr_{\lambda_f}\(\prod_{t \supset f} g_t\),
\label{ZBFdecom}
\ee
where $\lambda$ denotes an arbitrary unitary irreducible representation
appearing in the Plancherel measure of the group $G$
and $d_\lambda$ is its dimension.
In the Lorentzian case where the gauge group is non-compact the sum over representations with the weight given by
$d_\lambda$ should be replaced by the integral with the Plancherel measure.
As a result of this decomposition, one associates an irreducible representation to each face.
Finally, the integral over $g_t$ can be evaluated assigning invariant intertwiners to the edges
of the dual 2-complex. The partition function is thus given by the product of spin networks
dual to the boundary of a 4-simplex and constructed from the representations $\lambda_f$ and invariant intertwiners,
which are evaluated on flat connections, with some additional factors determined by $d_{\lambda_f}$.
The resulting representation is of the type \eqref{defstat} where one omits the sum
over discretizations due to the topological nature of the theory.

If the original gauge group was $SU(2)$ then we could label the space of intertwiners
$\bigotimes_{k=1}^4 \CH^{j_k}\rightarrow {\mathbb C}$ by a spin $j$ and denote them by $\Int_j$.
In this case one gets a SF model with the following amplitudes:
\be
A_f^{SU(2)}(j)=d_j,
\qquad
A_e^{SU(2)}(j_1,...,j_4;\Int_j)=d_j^{-1}, \hspace{10cm}
\ee
\hspace{7cm}
\unitlength 0.3mm
\linethickness{0.2pt}
\ifx\plotpoint\undefined\newsavebox{\plotpoint}\fi 
\begin{picture}(0,129.5)(0,0)
\put(90,35){\circle*{6}}
\put(190,35){\circle*{6}}
\put(60,110){\circle*{6}}
\put(220,110){\circle*{6}}
\put(140,165){\circle*{6}}
\put(60,110){\line(1,0){160}}
\multiput(60,110)(.04904966278,.03372164316){1631}{\line(1,0){.04904966278}}
\multiput(140,165)(.04904966278,-.03372164316){1631}{\line(1,0){.04904966278}}
\put(220,110){\line(-2,-5){30}}
\put(190,35){\line(-1,0){100}}
\put(90,35){\line(-2,5){30}}
\multiput(60,110)(.058479532164,-.033738191633){2223}{\line(1,0){.058479532164}}
\multiput(190,35)(-.03373819163,.08771929825){1482}{\line(0,1){.08771929825}}
\multiput(140,165)(-.03373819163,-.08771929825){1482}{\line(0,-1){.08771929825}}
\multiput(90,35)(.058479532164,.033738191633){2223}{\line(1,0){.058479532164}}
\put(78.5,29){\makebox(0,0)[cc]{$\Int_{j_1}$}}
\put(47,111.25){\makebox(0,0)[cc]{$\Int_{j_2}$}}
\put(145.25,172.5){\makebox(0,0)[cc]{$\Int_{j_3}$}}
\put(235,111.25){\makebox(0,0)[cc]{$\Int_{j_4}$}}
\put(204.75,29){\makebox(0,0)[cc]{$\Int_{j_5}$}}
\put(64.25,69){\makebox(0,0)[cc]{$j_{12}$}}
\put(88.75,143.5){\makebox(0,0)[cc]{$j_{23}$}}
\put(191.5,143.5){\makebox(0,0)[cc]{$j_{34}$}}
\put(216.25,69){\makebox(0,0)[cc]{$j_{45}$}}
\put(140,26.5){\makebox(0,0)[cc]{$j_{51}$}}
\put(140,117.75){\makebox(0,0)[cc]{$j_{24}$}}
\put(127.25,82.5){\makebox(0,0)[cc]{$j_{25}$}}
\put(153.25,82.5){\makebox(0,0)[cc]{$j_{14}$}}
\put(104,98){\makebox(0,0)[cc]{$j_{13}$}}
\put(176.75,98){\makebox(0,0)[cc]{$j_{53}$}}
\put(-50,105){\makebox(0,0)[cc]{$ A_v^{SU(2)}(\{j_{ab}\};\{\Int_{j_{a}}\})\ =$}}
\end{picture}

\noindent
where $a,b=1,\dots,5$ label the tetrahedra of a 4-simplex, $j_{ab}$ is the spin associated to the triangle
shared by tetrahedra $a$ and $b$, and we used the graphical representation of the $15J$ symbol.
This representation shows the spin network which gives the $15J$ symbol  being evaluated on a flat connection.
Its graph corresponds to the structure of the 4-simplex and is dual to its boundary.

However, the BF theory relevant for our purposes has the gauge group $G=SO(\eta)$.
In the Euclidean case $G=SU(2)\times SU(2)$ so that one simply has to double the $SU(2)$ BF spin foam model.
Since the irreducible representations of $SO(4)$ are labeled by a couple $(j^+, j^-)$, the
intertwiner $\Int_{(j^+,j^-)}=\Int_{j^+}\otimes \Int_{j^-} $ and the different weights factorize as
$A_v^{SO(4)}(j_i^+,j_i^- )=A_v^{SU(2)}(j_i^+ )A_v^{SU(2)}(j_i^- ).$
In the Lorentzian case, the result is formally the same, provided one associates the couple
$2j^\pm+1=n \pm i\rho$ to a unitary principal representation $(n,\rho)$ and
uses the factorized formula from the Euclidean case.

Considering a discretization with a boundary, one immediately infers that
the state space of the model is spanned by $G$-spin networks.
The vertex amplitude is then obtained as evaluation of a boundary state of a 4-simplex
on a flat connection. In fact, we will see that the last point is the general feature of
all SF models derived using the strategy which starts from quantizing BF theory.

Finally, we now come to the crucial step of implementing the simplicity constraints at the level
of the SF model of BF theory. For this purpose one needs to find a quantum version of these constraints.
In the BC model this is achieved by identifying the bi-vectors $B_f^{IJ}$ with generators $\hT^{IJ}$
of the Lie algebra  in the representation $\lambda_f$ and by requiring that the operators
obtained from the simplicity conditions using this identification annihilate the boundary states
of any spin foam.

Then the diagonal simplicity constraint gives a restriction on
representations $\lambda_f$
\be
C^{(2)}_G(\lambda_f)=-\frac{1}{4}\,\eps^{IJKL}\,\hT^{(\lambda_f)}_{IJ}\hT^{(\lambda_f)}_{KL}=0,
\label{qdiag}
\ee
where $C_G^{(2)}$ is the second Casimir operator of the group $G=SO(\eta)$.
On representations $\lambda=(\jp,\jm )$ of $SO(4)$
and on unitary principal representations of type $\lambda=(n,\rho)$ of $SO(3,1)$, it is given by
\be
C^{(2)}_{SO(4)}(\jp,\jm)=2\jp(\jp +1)-2\jm(\jm+1),
\qquad
C^{(2)}_{SO(3,1)}(n,\rho)=2n\rho
\ee
leading to $\jp=\jm$ and $n=0$\footnote{The solutions of \eqref{qdiag} with $\rho=0$ have been disregarded
in the initial model \cite{Barrett:1999qw}, but incorporated later in \cite{Perez:2000ep}.}
for Riemannian  and Lorentzian cases, respectively. The irreducible representations satisfying
this condition are called {\it simple} representations.
Thus, only simple representations are associated to faces in the BC spin foam model.

The cross simplicity constraint is imposed already at the level of a tetrahedron and induces
a restriction on possible intertwiners. Taking into account that the intertwiners couple
simple representations, it is easy to conclude that the restriction means that,
given a decomposition of any two $\lambda_f$'s, the intertwiner has support only
on simple intermediate representations (see Fig. \ref{figBCint}).
In \cite{Barrett:1997gw} an intertwiner satisfying this condition has been constructed explicitly
and in \cite{Reisenberger:1998bn} it has been proven that it is actually unique.

\TEXfig{
\unitlength 0.33mm 
\linethickness{0.2pt}
\ifx\plotpoint\undefined\newsavebox{\plotpoint}\fi 
\begin{picture}(271,136.25)(0,0)
\put(-50,115){\line(1,-1){70}}
\put(20,115){\line(-1,-1){70}}
\put(100,45){\line(1,1){35}}
\put(281,136.25){\line(1,-1){35}}
\put(135,80){\line(1,0){50}}
\put(316,101.25){\line(0,-1){50}}
\put(-56,117){\makebox(0,0)[cc]{$\lambda_2$}}
\put(-56,42){\makebox(0,0)[cc]{$\lambda_1$}}
\put(29,117){\makebox(0,0)[cc]{$\lambda_3$}}
\put(29,42){\makebox(0,0)[cc]{$\lambda_4$}}
\put(94,117){\makebox(0,0)[cc]{$\lambda_2$}}
\put(94,42){\makebox(0,0)[cc]{$\lambda_1$}}
\put(229,117){\makebox(0,0)[cc]{$\lambda_3$}}
\put(229,42){\makebox(0,0)[cc]{$\lambda_4$}}
\put(280,142){\makebox(0,0)[cc]{$\lambda_2$}}
\put(280,12){\makebox(0,0)[cc]{$\lambda_1$}}
\put(358,142){\makebox(0,0)[cc]{$\lambda_3$}}
\put(358,12){\makebox(0,0)[cc]{$\lambda_4$}}
\put(75,75){\makebox(0,0)[cc]{$=\sum\limits_{\rm simple\ \lambda}d_\lambda$}}
\put(160,87){\makebox(0,0)[cc]{$\lambda$}}
\put(265,73){\makebox(0,0)[cc]{$=\sum\limits_{\rm simple\ \lambda}d_\lambda$}}
\put(326.25,77){\makebox(0,0)[cc]{$\lambda$}}
\put(185,80){\line(1,1){35}}
\put(316,51.25){\line(1,-1){35}}
\put(185,80){\line(1,-1){35}}
\put(316,51.25){\line(-1,-1){35}}
\put(135,80){\line(-1,1){35}}
\put(316,101.25){\line(1,1){35}}
\end{picture}\vspace{-0.4cm}
}{The Barrett-Crane intertwiner}{figBCint}

A very simple expression for this intertwiner, which is naturally generalized to any dimension,
has been given later in \cite{Freidel:1999rr}.
It is based on the fact that any simple representation has a vector invariant with respect to
the proper maximal compact subgroup $H$ (in the case of $G=SO(4)$, it is the diagonal $SU(2)$ subgroup).
Let the intertwiner couples representations $\lambda_k, \ k=1,\dots, L$ (in the case of 4d simplicial
decomposition one always has $L=4$), $p$ labels the basis elements in the representation space
and $p=0$ corresponds to the invariant vector mentioned above. Then the matrix elements of
the BC intertwiner can be represented as an integral over the factor space $X=G/H$
\be
\Int_{{\rm (BC)}p_1\dots p_L}=\int_X \de x \, \prod_{k=1}^L R^{(\lambda_k)}_{p_k 0}(g_x),
\label{matNBC}
\ee
where $g_x$ is a representative of $x\in X$ in $G$ and $R^{(\lambda)}_{pq}(g)$ is the matrix element
of $g$ in the representation $\lambda$.

Now one can implement these restrictions at the level of the partition function.
For this it is sufficient to take the spin foam representation of the BF partition function and restrict
the sum over $\lambda_f$ to only simple representations and to remove the sum over intertwiners
substituting for them $\Int_{\rm (BC)}$.

In principle, one still has to impose the volume simplicity constraint.
However, it is ignored in the BC model. This is justified by the fact that it is a consequence of
the previous constraints supplemented by the closure \eqref{closure}.
The latter is also required to hold on the boundary states.
But since we started from the state space of BF theory where the closure was already satisfied,
it does not produce any further conditions.
In fact, it is ensured by the integral over $x$ in the definition \eqref{matNBC} of the BC intertwiner.

Once we determined the state space, the vertex amplitude of the BC model is obtained
by using the above prescription as the boundary state of a 4-simplex evaluated
on a flat connection. This leads to the famous BC vertex (also called $10j$ symbol),
which can be expressed through the vertex amplitude of BF theory as follows
\be
A_v^{\rm BC}(\{j_{ab}\})=A_v^{SO(4)}(\{ (j_{ab}, j_{ab})\};\{\Int_{BC}\}).
\label{BCvertex}
\ee
Here we wrote the relation for the Euclidean case. In the Lorentzian case, it is sufficient
to replace the $SO(4)$ representations $(j_{ab},j_{ab})$ by the $SL(2,\Cmat)$ representations $(0,\rho_{ab})$.
However, this gives rise to an ill defined amplitude  since the integral over $X^{ 5}$ coming from the
definition of the BC intertwiner \eqref{matNBC} is divergent.
It can be easily regularized \cite{Barrett:1999qw} by integrating only over
$X^{ 4}$, which amounts to eliminate the infinite volume factor of $X$.
More generally, using the expression \eqref{matNBC}, one can show \cite{Barrett:1997gw}
that the evaluation on a flat connection of any spin network which edges and vertices
are colored by simple representations and the BC intertwiner, respectively,
can be expressed as a Feynman integral over $X^{n_V-1 }$
(where $n_V$ is the number of vertices) with a propagator $K(x,y )$ associated to each edge
given by $K(x,y )=R^{(\lambda_k)}_{0 0}(x y^{-1} ).$

The BC model was extensively studied during the years following its invention.
It has been also reformulated in terms of a group field theory \cite{Perez:2000ec},
which was later generalized to incorporate timelike bi-vectors \cite{Perez:2000ep}.
This overcomes the restriction of the initial Lorentzian BC model \cite{Barrett:1999qw}
that it only contains faces which are spatial, whereas this is clearly a non-generic configuration.
However, the properties of the model \cite{Perez:2000ep} have not been studied in detail and
it remains poorly understood. In particular, due to the fact that the propagator associated to a
representation $(k,0)$ develops a singularity for coincident points,
the corresponding vertex amplitude is infinite and no precise
regularization has been given up to date.

A great excitement about the BC model was caused by the finiteness results of \cite{Perez:2000bf,Crane:2001qk}.
Namely, it has been shown that the integration over the representations $\lambda_f$,
{\it i.e.}, the integration over the size of the dual triangles,
for a {\it fixed} triangulation of spacetime, gives a finite result.
This has been claimed to indicate a possible resolution of the non-renormalizability of
perturbative quantum gravity.
However, in fact, this result has nothing to do with the UV finiteness because it comes
from the absence of divergence when the area goes to infinity which is an IR regime.
Besides, it relies on three very important assumptions:
\begin{itemize}
\item the complex of the spin foam should be dual to a triangulation;
\item the faces of the triangulation are all of spatial type;
\item a special form of the face and edge amplitudes.
\end{itemize}
When one of these conditions is removed it is very unlikely that
the finiteness remains preserved. And on top of that, it is restricted to a fixed triangulation and
there is no clue on how to perform or to control the sum over
2-complexes, which is a central problem to be solved for building the physical scalar product.
As a result, it is not clear what is the actual meaning of these findings.

Unfortunately, despite big efforts, the face and edge amplitudes have not been uniquely determined.
Various attempts to fix them led to different results, whereas they
could affect enormously the physics of the resulting theory \cite{Baez:2002aw,Bojowald:2009im}.
It became more and more clear that the BC model suffers from different problems
and it is not able to capture the dynamics of general relativity.
One of these problems concerns the asymptotic expansion of the vertex amplitude when
the area of the faces gets large. In this limit the vertex amplitude is expected to
reproduce the semi-classical Regge
action through the formula
\be A_v^{\rm BC}\sim e^{\frac{\I}{\hbar}\,S_{\rm Regge}}.
\ee
However, it has been shown in different works \cite{Baez:2002rx,Freidel:2002mj,Barrett:2002ur} that the asymptotics is not
dominated by the semi-classical Regge action but by some degenerate configurations.
This result has questioned the ability of the Barrett-Crane model to be a viable model of quantum gravity.

\subsection{New vertices}
\label{subsec_newvert}

These doubts have become certain
after it was shown that the BC model is not able to reproduce
the structure of the graviton propagator \cite{Alesci:2007tx}. This last problem and some of the above
were related to the ultralocality of the BC model due to which the partition function factorizes
into a product of completely disentangled simplex contributions.
In turn this ultralocality is a consequence of the uniqueness of the BC intertwiner.
Such unique intertwiner is not sufficient to carry information from one simplex to another.

Since the uniqueness of the BC intertwiner is a consequence of the imposition of the
simplicity constraints, it became clear that one should modify the way these constraints are implemented.
This led to a revision of the BC model culminating in two new models,
\cite{Engle:2007wy} (EPRL) and \cite{Freidel:2007py} (FK), which we are going to discuss here.
The new models appeared as results of the two approaches to the derivation of spin foam models,
which we mentioned in section \ref{subsubsec_concepts}, based, respectively,
on the quantization of the simplex geometry and the discretized path integral.

Although the models of \cite{Engle:2007wy} and \cite{Freidel:2007py}
are in general different from each other and obtained using different ideas, they have several common
inputs. First, they both rely on the idea allowing to effectively linearize
the simplicity constraints \cite{Engle:2007uq,Alexandrov:2007pq}.
This linearization trivially follows from the geometric meaning of the diagonal and cross simplicity conditions
which state that four triangles belonging to the same tetrahedron $t$ and described by
bi-vectors $B_f$, $f\subset t$, lie in one hyperplane. In other words, there should exist a vector $x_t$,
normal to the tetrahedron, such that
\be
\(\star B_f\)^{IJ}(x_t)_J=0
\quad\Leftrightarrow\quad
I_{(Q)}(x_t) \cdot B_f=0,
\label{linearcross}
\ee
where the second form of the condition is written with help of one of the projectors \eqref{projxxx}.
The normal $x_t$ becomes an additional variable of our discretization.
It is easy to see that once \eqref{linearcross} is satisfied, the simplicity constraints, except their volume part,
trivially follow. But in fact \eqref{linearcross} is stronger because it excludes the topological sector
$B=\pm e\wedge e$ of Plebanski formulation. Thus, the linearization solves simultaneously the problem of
the BC model that it does not distinguish between the gravitational and the topological sectors.
The new constraint leads directly to the sector we are interested in. The corresponding
reformulation of Plebanski action with quadratic constraints replaced by the linear ones
was studied at the classical level in \cite{Gielen:2010cu}.

Second, both models suggest to quantize an extension of Plebanski formulation which
includes the Immirzi parameter.
This results in crucial deviations from the results of the BC model already
at the level of imposing the diagonal simplicity constraint.
Let us demonstrate this following the logic of the EPRL model.
For the FK model the result is identical.

The Immirzi parameter can be incorporated to Plebanski formulation \eqref{Plebanskiaction}
by modifying either the simplicity constraints \cite{Capovilla:2001zi}
or the BF part of the action. Here we take the second point of view.
Assuming that the bi-vector coupled to the curvature should be quantized as in the usual BF theory,
being represented by the generator of the symmetry algebra, one arrives at the quantization rule
\be
B_f+\frac{1}{\im}\star B_f\ \mapsto\ -\I \hbar\, \hT
\quad\mathop{\Leftrightarrow}\limits^{\im^2\ne \sigma}\quad
B_f\ \mapsto\ -\I \hbar\,\frac{\im^2}{\im^2-\sigma}\(\hT-\frac{1}{\im}\star \hT\).
\label{newquantB}
\ee
Plugging in this identification into the diagonal simplicity constraint, one
finds the following condition on representations
\be
\(1+\frac{\sigma}{\im^2}\)C^{(2)}_G(\lambda_f)-\frac{2\sigma}{\im}\,C^{(1)}_G(\lambda_f)=0.
\label{con_withoutsol}
\ee

At this point one should consider separately the Riemannian and Lorentzian cases.
Let us start with $\sigma=1$.
Then, as has been noted already in \cite{Livine:2001jt},
the constraint \eqref{con_withoutsol} does not have solutions except
some trivial ones.\footnote{In fact, the authors of \cite{Livine:2001jt} proposed to rotate the identification
of the $B$-field with the generators so that the change in the identification compensates the modification
of the simplicity constraints due to the Immirzi parameter. As a result, one arrives at the standard
constraint of the BC model \eqref{qdiag}. This strategy is however rejected in the new models which
insist on using the identification borrowed from the quantization of the BF part of the action.}
However, appealing to the ordering ambiguity, the authors of the model \cite{Engle:2007wy} adjusted the
operator in \eqref{con_withoutsol} so that the constraint does have solutions.
The resulting restriction on representations is
\be
\jp=\left|\frac{\im+1}{\im-1}\right|\jm.
\label{constr_repr}
\ee
In the second model \cite{Freidel:2007py} the fine tuning of the ordering is not needed since the diagonal
constraint is also linearized (see below).

The result \eqref{constr_repr} has striking consequences. Since $\jpm$ are (half-)integers,
the condition \eqref{constr_repr} has solutions only for a {\it rational} $\im$.
Thus, the approach described above leads to a ``quantization" of the Immirzi parameter!
This is a completely unexpected feature for which there are no signs in the canonical approach.

However, this is true only in the Euclidean theory. In the Lorentzian case $\sigma=-1,$
the spectra of both Casimir operators are continuous as the irreducible representations
are parameterized by one continuous and one discrete labels, $\rho$ and $n$, respectively.
Therefore, the condition \eqref{con_withoutsol} simply fixes the former in terms of the latter as \cite{Engle:2007wy}
\be
\rho=\im n, \qquad
{\rm or}
\qquad
\rho=-n/\im.
\label{constr_reprL}
\ee
But this constraint on the allowed representations  has important implications
because it turns the spectra of geometric operators given by combinations of
Casimir operators of the Lorentz group, which are normally continuous, into discrete spectra.
Below we will discuss this issue, which is one of the main conclusions of this model, to more extent.

In other aspects of the derivation the EPRL and FK models are different.
Nevertheless, it turns out that the final constructions coincide for the values of the Immirzi parameter $\im<1$.
Here we present the two constructions separately.

\subsubsection{EPRL model}
\label{subsubsec_EPRL}

The main suggestion of this model, which distinguishes it from the BC model and was first realized
in \cite{Engle:2007uq}, is that the simplicity constraints should be imposed only in
a {\it weak} sense that is instead of imposing the constraints on the allowed states by $\CC_n\vert\Psi>=0$ one only requires
$<\Psi'\vert  \CC_n\vert \Psi>=0$.
This is justified by  noting that after identification of the bi-vectors $B_f$ with generators of the gauge group
or a combination thereof \eqref{newquantB}, the simplicity constraints become non-commutative and
imposing them strongly leads to inconsistencies, as is well known for any second class constraints.
This does not concern the diagonal simplicity constraint which
lies in the center of the constraint algebra and therefore can still be imposed
strongly\footnote{The reader should not confuse the terms {\it weak} and {\it strong} with similar
terms in Dirac's theory. Here we use them in the sense described above and, in fact, imposing constraints
strongly corresponds to the weak imposition in Dirac's terminology.}
leading to the restriction \eqref{con_withoutsol} on the allowed representations.

The other idea already mentioned above is to use the linearized version \eqref{linearcross}
of the simplicity constraints.
The latter involves a four-dimensional vector $x_t$, the normal to tetrahedron $t$.
Let this tetrahedron be associated with the part of the boundary
state we are interested in. The approach employed in \cite{Engle:2007uq,Engle:2007qf,Engle:2007wy}
is to impose the cross simplicity for a fixed $x_t$ and then to average over all vectors.
This is achieved by inserting an integral over the gauge group and is viewed as implementation of
the closure constraint \eqref{closure}.
For the fixed $x_t$, \eqref{newquantB} leads to the condition
\be
\langl \Psi'|\, x_t \cdot\(\hT -\sigma\gamma\star \hT\)|\Psi\rangl=0
\label{constr_cross}
\ee
for any allowed boundary spin network states $\Psi,\Psi'.$
In fact, it should be stressed that \eqref{constr_cross} encodes both diagonal and cross simplicity.
However, in the EPRL approach the diagonal simplicity is treated on its own
leading to the condition \eqref{con_withoutsol} (see however footnote \ref{foot_weak}),
whereas the constraint \eqref{constr_cross} implies, as in the BC model, a restriction on intertwiners.
Namely, let us fix the maximal compact subgroup $H$ leaving the vector $x_t$
invariant. Then the space of all intertwiners between representations $\lambda_k$ can be parameterized
by representations $j_k$ appearing in the decomposition of $\lambda_k$ on $H$ and $H$-intertwiners
between these $j_k$. In this parametrization, the constraint \eqref{constr_cross} results in the following conditions
on $j_k$ \cite{Engle:2007wy}\footnote{From now on we restrict to positive $\im$, for which
the constraint \eqref{constr_repr} ensures that $\jp\ge\jm$.
Note that in \cite{Conrady:2010kc} a generalization of the Lorentzian EPRL model,
similar in the spirit to \cite{Perez:2000ep}, was suggested which shows how to include timelike bi-vectors
by considering $H=SU(1,1)$ as the isotropy subgroup of the normal $x_t$.}
\be
{\rm Euclidean:} \ \ j=\left\{
\begin{array}{c}
\jp +\jm \quad \im<1
\\
\jp -\jm \quad \im>1
\end{array}\right. ,
\qquad
{\rm Lorentzian:} \ \ j=n.
\label{constr_int}
\ee
Thus, one has to always choose either highest or lowest weight representations depending on the value of the Immirzi parameter
and the signature. But since they are all non-trivial except the case $\sigma=1,\im=\infty$, the solution
of the simplicity constraints is non-unique and there is more than one intertwiner per tetrahedron
due to the remaining freedom in the choice of $SU(2)$-intertwiners.

The two restrictions, \eqref{constr_repr} (or \eqref{constr_reprL}) and \eqref{constr_int},
define the state space of the EPRL model. The vertex amplitude then follows as usual from
evaluation of the resulting spin network on a flat connection and can be expressed through
$15J_G$ symbols. The corresponding SF model with certain face and edge amplitudes
has been proposed in \cite{Kaminski:2009cc}.

\subsubsection{FK model}

The second model \cite{Freidel:2007py} follows a different approach similar to the one
which was presented in section \ref{subsubsec_BC} for the BC model. One starts again from the partition
function for BF theory \eqref{ZBFdecom} where the simplicity constraints should be implemented
as restrictions on the representation labels. However, before doing that one makes a refinement
of the decomposition \eqref{ZBFdecom} using the coherent state techniques developed in \cite{Livine:2007vk}.

Here we concentrate on the Euclidean case. Although the Lorentzian case was also considered in \cite{Freidel:2007py},
the corresponding construction is much more complicated and even the Immirzi parameter has not been
incorporated in it so far. Thus, we have to deal with the coherent states for the group
$SO(4)=SO(3)\times SO(3)$. They can be constructed from the coherent states associated
with the two $SO(3)$ factors. The latter states are parameterized by a unit 3-dimensional vector $n\in S^2$ and
form an overcomplete basis in a representation space, which means that one can write the following
decomposition of the identity
\be
{\bf 1}_j=d_j \int_{S^2} \de n\, |j,n\rangle \langle j,n|.
\label{decunit}
\ee
An $SO(4)$ coherent state is just the tensor product of two $SO(3)$ coherent states and therefore is defined as
\be
\cohr{\lambda}{n}=|\jp,\np\rangle\otimes|\jm,\nm\rangle,
\ee
where ${\bf n}=(\np,\nm)$ is a pair of two normals, which can be thought also as $SU(2)$ elements defined up to
a phase since $SU(2)/U(1)=S^2$.

The idea of \cite{Livine:2007vk} is to split the trace in \eqref{ZBFdecom} into a product of
factors associated with different simplices without performing integration over $g_t$.
For that purpose one introduces ``half-holonomies" $g_{\sigma t}$ from the center of 4-simplex $\sigma$ to
the center of tetrahedron $t$ so that $g_t=g_{\sigma t}(g_{\sigma' t})^{-1}$.
Then one inserts the decomposition of the identity in terms of the $SO(4)$ coherent states analogous to \eqref{decunit}
between the two group elements for each tetrahedron and each dual face. As a result,
\eqref{ZBFdecom} becomes
\be
Z_{BF}=\sum_{\lambda_f}\prod_f d_{\lambda_f}
\int \prod_{(t,\sigma)} \de g_{\sigma t} \int \prod_{(t,f)} d_{\lambda_f} \de {\bf n}_{tf}
\prod_{(\sigma,f)}\cohl{\lambda_f}{n_{\it tf}} (g_{\sigma t})^{-1} g_{\sigma t'}\cohr{\lambda_f}{n_{\it t'f}}
\label{ZBFdecom-coh}
\ee
and the simplicity constraints should now be imposed as certain restrictions on representation labels
and the normals ${\bf n}_{tf}$.

To get such conditions, one associates to each coherent state $\cohr{\lambda_f}{n_{\it t f}}$
a bi-vector obtained by averaging the generators
\be
\Xbn{\lambda_f}{n_{\it t f}}^{IJ}=\cohl{\lambda_f}{n_{\it tf}} \,\hT^{IJ}\cohr{\lambda_f}{n_{\it tf}}.
\label{bivec-coh}
\ee
Since the coherent states are in a sense quasiclassical, their labels can be thought as
encoding an information about the geometry of a classical tetrahedron. In particular,
$\lambda_f$ encodes the area of the triangle dual to face $f$ and ${\bf n}_{tf}$ describes its normal viewed
from tetrahedron $t$. Then the bi-vector \eqref{bivec-coh} should be associated with the geometric
bi-vector \eqref{Bfdef}. However, in the presence of the Immirzi parameter, the correct relation
is done through \eqref{newquantB}. Thus, one arrives to the conclusion that
\be
\Xbn{\lambda_f}{n_{\it t f}}-\frac{1}{\im}\, \star \Xbn{\lambda_f}{n_{\it t f}}
\ee
must be a simple bi-vector. Above, in \eqref{linearcross}, we found a very convenient criterium for
this condition. In our case it can be equivalently rewritten as the requirement of existence of such
$X_f\in su(2)$ and $x_t\in SU(2)$ that the chiral decomposition reads as
\be
\Xbn{\lambda_f}{n_{\it t f}}=\(\(1+\frac{1}{\im}\)X_f, -\(1-\frac{1}{\im}\) x_t X_f x_t^{-1}\).
\label{simbfX}
\ee
Using $\langle j,n|\,\vec T|j,n\rangle =j\,\vec n$, one then finds the condition \eqref{constr_repr}.
But on top of that, the two normals $\npm_{tf}$ viewed as group elements get related as
\be
(\np_{tf},\nm_{tf})=(n_{tf}h_{\phi_{tf}}^{\left| 1-\frac{1}{\im}\right|}, x_t n_{tf}h_{\phi_{tf}}^{-\left( 1+\frac{1}{\im}\right)}\epsilon),
\label{cond_nn}
\ee
where the $U(1)$ elements $h_{\phi}=e^{i \phi \sigma_3}$ take care that $\npm$
are defined up to a phase\footnote{The powers of the phase elements cannot be fixed classically since
they can be changed by redefining $n_{tf}$. However, such redefinition would also affect the integration measure
in the partition function. We have chosen the powers in such way so that the partition function is non-trivial
for the standard choice of the measure. This issue is not discussed in \cite{Freidel:2007py} as
the solution \eqref{cond_nn} is written there only for $\im=\infty$.}
and
$\epsilon$ is either 1 or the matrix ${\scriptsize\begin{pmatrix} 0 &\!\! \!\!  1\\ -1 & \!\!\! 0\end{pmatrix}}$
such that $h\epsilon=\epsilon\bar h$ for $\forall h\in SU(2)$
depending on whether $\im$ is less or larger than 1.

The two restrictions, \eqref{constr_repr} and \eqref{cond_nn}, on the variables of the BF partition
function represent the simplicity constraints at the quantum level. They are to be inserted
into \eqref{ZBFdecom-coh} supplemented by integrations over $\phi_{tf},\ x_t$ and $n_{tf}$.
The first two are not important and can be easily performed. Assuming that the last integral
is performed with the standard measure on $S^2$, it produces for every pair
$(tf)$ the following factor \cite{Freidel:2007py}
\beq
\label{integr_n}
\im<1: & \quad &
\int \de n\, |\jp,n\rangle\otimes|\jm,n\rangle \langle \jp,n|\otimes \langle \jm,n|
=\overline{C^{\jp\jm\, \jp+\jm} } C^{\jp\jm\, \jp+\jm},
\\
\im<1: & \quad &
\int \de n\, |\jp,n\rangle\otimes\overline{|\jm,n\rangle} \langle \jp,n|\otimes \overline{\langle \jm,n|}
=\sum_{j=\jp-\jm}^{\jp+\jm} d_j\left|C_{\jp\jm \, \jp-\jm}^{\jp\jm k} \right|^2
\overline{C^{\jp\jm\, j} } C^{\jp\jm \, j},
\nonumber
\eeq
where the Clebsch--Gordan coefficients are considered as maps between $SU(2)$ representation spaces
\be
C^{\jp\jm\, j}:\ \CH^{\jp}\otimes\CH^{\jm}\to \CH^j,
\qquad
\overline{C^{\jp\jm\, j}}:\ \CH^j\to \CH^{\jp}\otimes\CH^{\jm}.
\ee
{}From this result one can read off the possible intertwiners.
Again, it is convenient to parametrize them by $SU(2)$ representations $j_{tf}$ appearing
in the decomposition of $\lambda_f$ on the diagonal subgroup and by $SU(2)$ intertwiners $\Int_t$
coupling the resulting $j_{tf}$. From \eqref{integr_n} one finds that for $\im<1$ only the highest
weight representation plays the role so that the state space is exactly the same as in the EPRL model
(see \eqref{constr_int}). On the other hand, for $\im>1$ all representations appearing in the decomposition
of $\lambda_f=(\jp_f,\jm_f)$ contribute with some particular weights given by the Clebsch–Gordan coefficients.

The final partition function is obtained by integrating over holonomies and leads to the vertex amplitude
as usual given by evaluation of a simplex boundary state on a flat connection.
For $\im >1$, comparing to the EPRL vertex, it has
15 additional labels $j_{tf}$ arising from the sum over representations in \eqref{integr_n}.

\subsubsection{The new models and LQG}
\label{subsubsec_SFandLQG}

As should be clear from above, for $\im<1$ the two models are essentially identical: they possess
the same state space and vertex amplitude, whereas on the edge and face amplitudes there is no agreement anyway
even in the framework of one model ({\it cf.} \cite{Kaminski:2009cc} and \cite{Bianchi:2010fj}).
This is a somewhat surprising result given that the models are obtained by completely different methods, although
a bridge between these two approaches has been established in \cite{Livine:2007ya}.
It is even more surprising taking into account that for $\im>1$ they do differ from each other.
In the FK model the space of intertwiners is much larger because the representation $j$ of the diagonal
$SU(2)$ in \eqref{integr_n} is not fixed contrary to \eqref{constr_int}.

This difference allows to ask which of the models is more preferable for $\im>1$.
To answer this question, it is useful to look at the limit $\im\to\infty$,
which is perfectly smooth in the classical theory.
Then it is easy to see that in this limit the EPRL model reduces to the BC model.
But the latter is expected to be an incorrect quantization of gravity
from many different perspectives. In fact, in \cite{Freidel:2007py} it was shown that
the coherent state technique leads to the BC model if one decouples the normals $n_{tf}$
seen from different simplices. This amounts to integrating in \eqref{integr_n} independently over
the normal $n_{tf}^\sigma$ appearing in the bra-vectors and another normal $n_{tf}^{\sigma'}$
in the ket-vectors. This gives rise to the unique BC intertwiner, but clearly involves
an unjustified step --- there is no physical reason to take $n_{tf}^\sigma\ne n_{tf}^{\sigma'}$.
Thus, this consideration points in favor of the FK model.

In the opposite limit $\im\to 0$, both models are expected to describe the topological
sector of Plebanski formulation.\footnote{In fact, the initial version of the EPRL model
\cite{Engle:2007uq,Engle:2007qf} corresponds precisely to this case. The flipped symplectic structure
used there is nothing else but the symplectic structure of the topological theory given by the second
term in the Holst action \eqref{Sd1}.} However, it is not clear whether the limit should be actually smooth
as it is not the case in the continuum theory.

Let us now look at the state space of the EPRL model (or FK for $\im<1$) for generic rational Immirzi parameter.
The relations \eqref{constr_int} and \eqref{constr_repr} (or \eqref{constr_reprL}) can be inverted and
the representation labels $\lambda$ of the gauge group can be expressed through the $SU(2)$ labels $j$.
As a result, the states are labeled only by $SU(2)$ representations associated with dual faces (becoming links
of a graph on the boundary) and the usual $SU(2)$ intertwiners assigned to tetrahedra (vertices of the boundary graph).
Due to this fact it was claimed that the boundary states
of the new models are the ordinary $SU(2)$ spin networks \cite{Engle:2007uq}
and it is now widely believed that there is a perfect agreement between the new SF models
and LQG at the kinematical level \cite{Rovelli:2010wq}.

However, it is easy to see that this is just not true.
First of all, the states induced on the boundary of a spin foam are not the ordinary spin networks, but {\it projected} ones
considered in section \ref{subsubsec_loopq} --- the $SU(2)$ representations $j_{tf}$ parameterizing
the intertwiners are precisely the representations  $j_{ve}$ which one projects to in \eqref{projspnet}.
In fact, this is a particular case of a quite general statement. In \cite{Alexandrov:2008da}
it was proven that the boundary states of any spin foam model obtained by imposing the simplicity constraints
are always given by projected spin networks. The state spaces of different models are distinguished by
constraints on the labels. For example, it has been known for long time \cite{Livine:2002ak,Alexandrov:2002br}
that the states of the BC model are obtained from generic projected spin networks
by taking simple representations $\lambda_f$, all $j_{tf}=0$, and
integrating over the normals $x_t$ appearing as additional argument in the canonical quantization.
The EPRL and FK models simply provide a generalization of this construction.

However, this does not mean yet that there is no agreement between the new models and canonical
quantization. As we know, in certain cases the projected spin networks can reduce to the usual ones.
An example of such situation has been considered in section \ref{subsubsec_LQGcov}
where one gets precisely the kinematical Hilbert space of LQG.
However, the key point of that construction was the choice of connection in the definition
of holonomies. In particular, to get the Hilbert space of LQG this connection should be chosen as in \eqref{conSU2}
and then the constraints \eqref{su2con} are responsible for the reduction.
On the other hand, in the spin foam models all holonomies are defined with respect to the usual
spin-connection and therefore the projected spin networks describing their boundary states cannot
be reduced to the kinematical states of LQG.

In fact, a precise relation between the two types of states has been elucidated in \cite{Alexandrov:2010pg}.
It turns out that if one adjusts appropriately restrictions on representations
\eqref{constr_repr}, \eqref{constr_reprL} and \eqref{constr_int} of the EPRL model, which amounts to choosing
a different ordering for Casimir operators thereby fixing this ambiguity, the projection of
the spin connection on the representation of the subgroup fixed by the constraints
gives rise to the Ashtekar--Barbero connection
\be
\pr{j}\(\omega_i^{IJ} T_{IJ}^{(\lambda)}\) \pr{j}=\Ab{i}{a} L_a^{(j)},
\label{projcon}
\ee
where $L_a^{(j)}$ is a generator of boosts in representation $j$.
This happens, for example, if one takes in the Lorentzian theory
\be
\rho=\im(n+1), \qquad j=n.
\ee
Note also that this choice is favored by the fact that, contrary to \eqref{constr_reprL},
it provides an {\it exact} solution to the constraints \eqref{constr_cross},
which does not require any additional large-spin approximation or ordering fitting \cite{Ding:2010ye}.
If the relation \eqref{projcon} holds, the projected spin networks can be reduced to the usual ones by inserting
$\pr{j}$ at all points along edges of the graph. This is equivalent to
inserting infinitely many bi-valent vertices and amounts to considering ``fully projected holonomies"
of the type
\be
\Umat^{(\lambda,j)}_{\alpha}= \lim\limits_{N \rightarrow \infty}
{\cal P}\left\{ \prod\limits_{n=1}^{N}
\pr{j} R_G^{(\lambda)}\(U_{\alpha_n}[\omega] \)\pr{j} \right\},  \qquad\quad \alpha=\mathop{\cup}\limits_{n=1}^N \alpha_n,
\label{WLpfull}
\ee
first introduced in \cite{Alexandrov:2002xc}. Then the property \eqref{projcon} ensures that
\be
\Umat^{(\lambda,j)}_{\alpha}=
R_{SU(2)}^{(j)}\left( U_{\alpha}[\Ab{}{}]\right),
\label{WL-W2}
\ee
similarly to \eqref{wl}, and the resulting functionals reduce to the $SU(2)$ spin networks defined with
the Ashtekar--Barbero connection.\footnote{Strictly speaking, the above equations are written in the time gauge.
They are easily generalized to the general case where the role of $\Ab{}{}$, not surprisingly, is played by
the Lorentz connection $\SSA$ \eqref{conSU2}. However, it is clear that this construction works only if one does
not integrate over the normals $x_t$ at the vertices,
which corresponds to the relaxed closure condition as explained below.}
However, on one hand, there is no any fundamental reason to perform such a projection.
And on the other hand, this relation shows that the kinematical states of LQG and the boundary states of the EPRL model
are indeed physically different and the agreement between their labels is purely formal.\footnote{In fact,
in the Euclidean case when the Immirzi parameter must be a rational number,
there is no even one-to-one correspondence between the state space of the SF model and the $SU(2)$ spin networks.
Indeed, for example, for $\im<1$ let $\frac{1+\im}{1-\im}=\frac{p}{q}$ where $p,q$ are two coprime integers.
Then $j$ can take only values $\hf\,m(p+q)$, $m\in \Nint$. As a result, spin networks with representations
which do not belong to this set are missing.}

The claimed agreement is often justified by comparison of the spectra of geometric operators,
area \cite{Engle:2007wy} and volume \cite{Ding:2009jq}. By appropriately adjusting the ordering, the
spectra in spin foams and LQG can be made coinciding. However, the operators, which are actually evaluated in these
papers, are not the standard ones, but shifted by constraints. For example, the area of a face
is taken in the form
\be
\CS^2_f= \hf\, (\star B_f)\cdot (\star B_f)-\(x_t \cdot (\star B_f) \)^2=\hf\, B_f \cdot I_{(P)}(x_t)\cdot B_f.
\label{areagauge}
\ee
It explicitly depends on the normal $x_t$. But in the EPRL approach the boundary states are supposed to be integrated
over these normals so that the operator corresponding to \eqref{areagauge} is simply not defined!
On top of that, even if one drops the integration over $x_t$, as we argue below, and gets a well defined operator
on a modified state space, we see that the quantization of the geometric operators is not unique.
To get the coincidence with LQG requires ad hoc choice of the ordering and of the classical expression to be quantized.

In the next section it will become clear that this ambiguity is a reflection of serious
problems of the presented approach. They arise because the way the simplicity constraints
are incorporated into quantum theory does not respect quantization rules of constrained systems.
As a result, the new models cannot be considered as
proper quantizations of general relativity.

\subsection{Imposition of constraints revisited}
\label{subsec_imposeconstr}

All models presented in the previous section have been derived following the strategy
of section \ref{subsubsec_strategy}: first quantize and then constrain.
Now we want to reconsider the resulting constructions
taking lessons from the canonical approach.

As we showed in the previous section, the spin foam quantization
originates in Plebanski formulation of general relativity.
The canonical analysis of this formulation has been carried out in
\cite{Buffenoir:2004vx,Alexandrov:2006wt,Alexandrov:2008fs} and turns out to be
essentially equivalent to the Lorentz covariant canonical formulation of the
Hilbert--Palatini action \cite{Alexandrov:2000jw} once $\eps^{ijk}B_{jk}$ is identified with $\tP^i$.
The Immirzi parameter is also easily included and appears in the same way.
Thus, the canonical structure to be quantized can be borrowed from section \ref{subsubsec_covcan}.
In particular, the role of the simplicity constraints is played by the constraints \eqref{phi}.

What is however important is that there are also secondary constraints \eqref{psi}, which
together with the usual simplicity form a system of second class constraints.
Due to this, as we extensively discussed in section \ref{sec_canan}, the correct symplectic structure
is provided by Dirac brackets. In particular, the commutation relation we are mostly interested in
can be found in \eqref{comomP} where $\tP$ and $\tQ$ should be replaced by the $B$-field and its Hodge dual.

Similarly to the Lorentz covariant formulation, one can define a shifted connection $\SA_i^{IJ}$
coinciding with the spin-connection on the surface of the Gauss constraint \cite{Alexandrov:2006wt}
(see section \ref{subsubsec_CLQG}).
This connection satisfies a simpler commutation relation
\be
\left\{ \SA_i^{IJ},\eps^{jkl} B_{kl}^{KL}\right\}_D=\delta_i^j\, I_{(P)}^{IJ,KL}.
\label{commPlebB}
\ee
Its ``rotational" part is non-dynamical in agreement with the secondary second class constraints
which now take the form
\be
I_{(Q)KL}^{IJ}\SA_i^{KL}=\Gamma_i^{IJ}(B).
\label{contA_B}
\ee

Our aim is to verify whether the new SF models provide a quantization of Plebanski formulation consistent
with this canonical analysis.
For that purpose we take a closer look at the imposition of constraints in the EPRL and FK models.
But first, to illustrate the situation, we suggest a very simple example \cite{Alexandrov:2010pg} which however contains
the essential features of kinematics relevant to our problem.

\subsubsection{A simple example}
\label{subsubsec_example}

Let us consider a system described by the following action:
\be
S=\int \de t \[ p_1 \dot q_1 +p_2 \dot q_2 -\hft p_1^2-\alpha\cos q_2+\lambda (p_2-\gamma p_1) \].
\label{example}
\ee
Here the coordinates $q_1$ and $q_2$ are supposed to be compact, so that we consider them as
living in the interval $[0,2\pi)$, and $\alpha,\gamma$ are numerical parameters.

The canonical analysis of this system is elementary.
The momenta conjugate to $q_1$ and $q_2$ are $p_1$ and $p_2$, respectively, so that
the only non-vanishing Poisson brackets are
\be
\{ q_1,p_1\}=1,
\qquad
\{ q_2,p_2\}=1.
\label{Poisson}
\ee
The variable $\lambda$ is the Lagrange multiplier for the primary constraint
\be
\label{primconst}
\phi=p_2-\gamma p_1\approx 0.
\ee
Commuting this constraint with the Hamiltonian
\be
H=\hft p_1^2+\alpha\cos q_2-\lambda\phi,
\ee
one finds a secondary constraint
\be
\psi=\sin q_2\approx 0,
\label{secconst}
\ee
which has two possible solutions
$
q_2=0 \ {\rm or} \ q_2=\pi.
\label{twosol}
$

Since the two constraints, $\phi$ and $\psi$, do not commute,
they are of second class. A way to take this into account is to
construct the Dirac bracket. It is easy to find that the only
non-vanishing Dirac brackets between the original canonical variables are
\be
\{ q_1,p_1\}_D=1,
\qquad
\{ q_1,p_2\}_D=\gamma.
\label{Dirac}
\ee
The second bracket here is actually a consequence of the first one provided
one uses $p_2=\gamma p_1$. The Hamiltonian is given (up to a constant) by
\be
H=\hft p_1^2.
\ee
As a result, classically the system reduces to the very simple system
describing one free particle on a circle.

As for the quantum theory, first, we would like to quantize the action \eqref{example} following
the strategy adopted by the spin foam approach. This is not a problem since
the analogy of this model with the kinematics of Riemannian general relativity is rather direct:
$q_1$ and $q_2$ are analogous to the right and left parts of the $SO(4)$ spin-connection under chiral decomposition,
$p_1$ and $p_2$ correspond to the chiral parts of the $B$-field, $\phi$ is similar to the diagonal simplicity
constraint, $\psi$ is its secondary partner, and $\gamma$ plays the role of the Immirzi parameter
(or rather of its combination $\frac{\gamma+1}{\gamma-1}$).

At the first step, we ignore the second class constraints and quantize
the unconstrained symplectic structure represented by the Poisson commutation relations \eqref{Poisson}.
Thus, in the coordinate representation the canonical variables are realized by the following operators
\be
\hat q_1=q_1,
\qquad
\hat q_2=q_2,
\qquad
\hat p_1=-\I \hbar \p_{q_1},
\qquad
\hat p_2=-\I \hbar \p_{q_2}.
\label{repr_P}
\ee
Since the configuration variables are compact,
the state space is spanned by linear combinations of
\be
\Psi_{j_1,j_2}(q_1,q_2)=e^{\I j_1q_1+\I j_2q_2},
\label{enstates}
\ee
where $j_1,j_2$ are two integers. These states form an orthonormal basis with respect to the scalar product
\be
\langle\Psi|\Psi' \rangle  =\frac{1}{(2\pi)^2}\int \de q_1 \de q_2\, \overline{\Psi}(q_1,q_2)\Psi'(q_1,q_2).
\label{ex_scprP}
\ee

At the second step, one should impose the primary constraint. As in the spin foam models,
we do this by requiring that the states should satisfy
\be
\hat\phi \,\Psi=(\hat p_2-\gamma\hat p_1)\Psi=0.
\label{impose_phi}
\ee
Being applied to the basis states \eqref{enstates}, this condition leads
to the following restriction on the labels
\be
j_2=\gamma j_1,
\label{jjgam}
\ee
so that the physical states are spanned by
\be
\Psi_{j}(q_1,q_2)=e^{\I j(q_1+\gamma q_2)}.
\label{resst_P}
\ee
They are similar to the $SU(2)$ spin networks with
$q_1+\gamma q_2$ being an analogue of the Ashtekar--Barbero connection.
In particular, this provides an illustration for the reduction mechanism, presented in \eqref{projcon},
of a generic spin connection to the Ashtekar--Barbero connection.
Moreover, since the representation labels in \eqref{jjgam} are integer,
one gets the quantization condition on $\gamma$
that this parameter should be a rational number. This is precisely the same result which
one has for the Immirzi parameter in the EPRL and FK spin foam models.

Let us compare this construction with the standard Dirac quantization.
Although it can be done in a straightforward way by passing immediately
to the phase space reduced by the second class constraints, we present it in
a longer version, which is closer to the procedure one follows
in the covariant loop approach.
Namely, we start with an ``enlarged" Hilbert space
consisting of square-integrable functions of $q_1$ and $q_2$, quantize
the Dirac commutation relations \eqref{Dirac}, and impose the constraint \eqref{secconst}
to remove a degeneracy of the original Hilbert space.

Our enlarged Hilbert space is spanned by the linear combinations of \eqref{enstates}
and thus coincides with the starting point of the previous approach.
On the other hand, the canonical variables are represented now by the following operators:
\be
\hat q_1=q_1,
\qquad
\hat q_2=q_2,
\qquad
\hat p_1=-\I\hbar \p_{q_1},
\qquad
\hat p_2=-\I\hbar \gamma\p_{q_1}.
\label{repr_D}
\ee
This representation differs from \eqref{repr_P}
since it is obtained by quantizing the Dirac algebra \eqref{Dirac}.
Together with the quantum operators, one should also adjust the scalar product, which
must now include one of the second class constraints.
Namely, the physical scalar product is given by\footnote{Here we
neglected the contribution from the second
solution of \eqref{secconst} which has essentially the same form.
On the other hand, in the path integral \eqref{corelfun} considered below
only one of the two solutions of $\psi$ can contribute
because of the continuity of the classical trajectories one sums over. Which of the two solutions contributes
depends on the boundary conditions at the initial and final moments.
For simplicity we assume that this is $q_2=0$.}
\be
\begin{split}
\langle\Psi|\Psi' \rangle_{\rm ph} & =\frac{1}{(2\pi)^2}\int \de q_1 \de q_2\, \delta(\psi)\overline{\Psi}(q_1,q_2)\Psi'(q_1,q_2)
=\frac{1}{(2\pi)^2}\int \de q_1 \, \overline{\Psi}(q_1,0)\Psi'(q_1,0).
\label{ex_scpr}
\end{split}
\ee
The resulting Hilbert space is degenerate along $q_2$ as it is fixed by
the secondary constraint $\psi$.
This degeneracy can be removed by restricting to the states with $j_2=0$ because
\be
\Psi_{j_1,j_2=0}(q_1,q_2)=\left.\Psi_{j_1,j_2}(q_1,q_2)\right|_{\psi}.
\ee
As a result, we end up with the Hilbert space spanned by the states
\be
\Psi_{j_1}(q_1)=e^{\I j_1q_1},
\label{resst_D}
\ee
and all operators can be built from $\hat q_1$, $\hat p_1$ defined in \eqref{repr_D}.
In particular, the Hamiltonian is represented as
\be
\hat H=-{\textstyle\frac{\hbar^2}{2}}\,\p_{q_1}^2.
\ee
This result reproduces the quantization of a free particle on a circle
and thus agrees with the reduced phase space quantization.
It is easy to see that this agrees also with the path integral method, which starts from
the phase space path integral:
\beq
\langle\CO\rangle&=&\int \de q_1 \de q_2 \de p_1 \de p_2 \,|\det\{\phi,\psi\}|\, \delta(\phi)\delta(\psi)\,
e^{\frac{\I}{\hbar} \int \de t\( p_1 \dot q_1 +p_2 \dot q_2 -\hf p_1^2-\alpha\cos q_2 \)} \CO(q_1,q_2,p_1,p_2)
\nonumber \\
&=& \int \de q_1\de p_1 \,
e^{\frac{\I}{\hbar} \int \de t\( p_1 \dot q_1 -\hf p_1^2 \)} \CO(q_1,0,p_1,\gamma p_1).
\label{corelfun}
\eeq

Comparing the results of the two approaches, one observes a drastic discrepancy:
$\gamma$ is either quantized or not. In the former approach for a non-rational
$\gamma$ the quantization simply does not exist, whereas there are no any
obstructions in the latter.
Besides, the resulting states, \eqref{resst_P} and \eqref{resst_D}, are also different.
In principle, they can be made identical by identification $q_1+\gamma q_2 \to q_1$,
which however may affect correlation functions.
In contrast, the problem with the quantization condition on $\gamma$ cannot be avoided by any tricks.
It shows that the two quantizations are indeed inequivalent. Taking into account
that the second approach represents actually
a result of several possible methods, which all follow the standard quantization rules,
it is clear that it is the second quantization that is more favorable.
The quantization of $\gamma$ does not seem to have any physical reason behind itself.

In fact, it is easy to trace out where a mistake has been done in the first approach:
it takes too seriously the symplectic structure given by the Poisson brackets,
whereas it is the Dirac bracket that describes the symplectic structure
which has a physical relevance. In particular, in the presented example, the Poisson
structure tells us that $p_2$ is the momentum conjugate to $q_2$, whereas
in fact it is conjugate to $q_1$.

It is easy to see that this leads to inconsistency of the first quantization.
For example, the Hamiltonian, which reads in this case as
\be
\hat H_{\rm sf}=-{\textstyle\frac{\hbar^2}{2}}\,\p_{q_1}^2+\alpha\cos q_2,
\label{sfHam}
\ee
is simply not defined on the subspace spanned by linear combinations of \eqref{resst_P}.
This problem is caused by the second term involving $q_2$.
It is impossible to ignore this term by requiring that
$\hat \psi$ vanishes on the physical states since it would be in contradiction with
the commutation relations \eqref{repr_P}.
Moreover, even if one manages to define a Hamiltonian operator on the physical subspace,
it would not have eigenstates there.
Indeed, assuming that $\Psi$ is a simultaneous eigenstate of $\hat\phi$ and $\hat H$, one
finds $[\hat H, \hat\phi]   \Psi=\I\hbar \alpha \hat \psi\Psi=0$ so that $\Psi$ is annihilated also by $\hat\psi$.
But as we already mentioned, this is not consistent with \eqref{repr_P}.

However, one can take a ``minimalistic" point of view and do not require the existence of
a well defined Hamiltonian on the constrained state space.\footnote{We thank Carlo Rovelli for discussion of this possibility.}
After all, spin foam models are designed to compute transition amplitudes.
Therefore, we are really interested not in the Hamiltonian itself, but in its matrix elements
and the latter can be defined by using the Hamiltonian and the scalar product on the original
unconstrained space. Moreover, one has the following encouraging property
\be
^{\rm sf}\!\langle j|\hat H_{\rm sf}|j'\rangle^{\rm sf}=\langle j|\hat H|j'\rangle_{\rm ph},
\label{propertyH}
\ee
where the states appearing on the l.h.s. $|j\rangle^{\rm sf}\equiv|j,\gamma j\rangle$
are those which arise in the first, spin foam like quantization and
are represented by \eqref{resst_P}, whereas the states on the r.h.s. result from the standard Dirac's approach
and are defined in \eqref{resst_D}.
This gives a hope that the transition amplitudes computed by SF models may nevertheless be compatible
with a self-consistent canonical picture. In particular, the vertex amplitude is believed to be
given by a matrix element of the Hamiltonian operator \cite{Reisenberger:1996pu} and, if this is the case,
a property like \eqref{propertyH} would ensure its agreement with the results of a more rigorous Dirac's approach.

However, this expectation turns out be wrong. As is clear from the derivations in \cite{Freidel:2007py,Engle:2007qf}
and has been explicitly demonstrated in a simple cosmological model \cite{Ashtekar:2009dn},
the vertex amplitude actually appears as a matrix element of the {\it evolution operator}.
This requires to consider expectation values of higher powers of the Hamiltonian for which
the property \eqref{propertyH} does not hold anymore. This leads to deviations of results obtained by
the spin foam strategy from those which are based on the well grounded canonical quantization.

Let us demonstrate this disagreement on the example considered above. To this end, we put it on a lattice,
construct an analogue of the vertex amplitude, and investigate its dependence on the coupling constant $\alpha$.
It is clear that any results compatible with the canonical quantization should be $\alpha$-independent
as this coupling multiplies the term fixed by the second class constraints to a constant.

\begin{figure}
\centerline{\hfill \includegraphics[height=2.7cm]{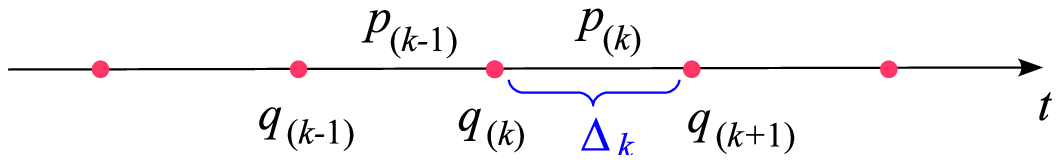}\hfill}
\vspace{-1.0cm}
\caption{The assignment of coordinates and momenta to the discretized time line. \label{figtime}}
\end{figure}

At the first step, one introduces a discretization of our ``spacetime", which amounts to
replacing the continuous time-line by a set of points. The discretized coordinates $q_{(k)}=(q_{1,(k)},q_{2,(k)})$
are assigned to these points, whereas their conjugate momenta $p_{(k)}$ are associated to the intervals
$\Delta_k$ between two points (see Fig. \ref{figtime}).
A possible discretization of the action \eqref{example} then reads
\be
S_{\rm disc}=\sum_k S_{\Delta_k}(q_{(k)},q_{(k+1)},p_{(k)})
\ee
where
\be
S_{\Delta}(q,q',p)=p_{1}\sin(q'_{1}-q_{1})+p_{2}\sin(q'_{2}-q_{2})-\hf\, p_{1}^2
-\frac{\alpha}{2}\(\cos q'_{2}+\cos q_{2}\),
\ee
and the kinetic term is written in terms of the $\sin$-function to take into account the compactness
of the configuration variables.
The vertex amplitude should be associated with the intervals $\Delta_k$ since they are simplices of the
highest dimension. Due to this, the coordinates $q$, living at the end points of the intervals,
are naturally considered as boundary variables
and the vertex in ``$q$-representation" can be defined by a ``path integral" over internal
momenta\footnote{Note that the $\delta$-function factor is reminiscent the flatness condition of BF-theory.
In this respect, our model is not a complete analogue of general relativity in Plebanski formulation
because it contains non-linear terms in $p_1$ which prevent from getting a similar ``flatness condition" for $q_1$.}
\be
A(q,q')=\int \de p_{1}\de p_{2}\, e^{\I S_{\Delta}}=\delta(q'_{2}-q'_{2})\,
e^{\frac{\I}{2}\,\sin^2(q'_{1}-q_{1})-2\alpha\cos q_{2} }.
\label{verqq}
\ee
To get the usual representation in a ``spin-network basis", it is enough to evaluate the matrix element of
\eqref{verqq} between two states \eqref{resst_P}. This gives the following result
\be
A(j,j')=\, ^{\rm sf}\!\langle j'|A(q,q')|j\rangle^{\rm sf}=A_c(j,j')A_a(j-j'),
\ee
where
\be
\begin{split}
A_c(j,j')& =\int \de q_1\de q'_1 \, e^{\frac{\I}{2}\,\sin^2(q'_{1}-q_{1})+\I jq_1-\I j'q'_1},
\\
A_a(j) &= \int \de q_2 \, e^{-\I\alpha \cos q_2+\I\gamma j q_2}\sim J_{\gamma j}(\alpha).
\end{split}
\ee
The first factor here, $A_c(j,j')$, is the amplitude which we would obtain
by following the standard canonical quantization equivalent to
working with the reduced phase space only.
The second factor is an anomalous contribution which marks the difference between the two approaches.
It depends explicitly on the coupling $\alpha$ and thus shows incompatibility of the spin foam strategy
with the canonical quantization.

Had we worked with a non-compact version of the model, we could
get a trivial $\alpha$-dependence (as an overall factor) due to the factor $\delta_{jj'}$ coming from $A_c$.
But such trivialization happens only in free models leading to the trivial dynamics ($j'=j$)
and cannot be expected to hold for such a non-linear theory as general relativity.
As soon as the spin dynamics is non-trivial, the dependence on the coupling constant is non-trivial as well.

Let us summarize what we learnt studying the simple model \eqref{example}:
\begin{itemize}
\item
The strategy based on ``first quantize, then constrain" leads to a canonical quantization which
is internally inconsistent as the Hamiltonian operator is ill-defined on the constrained state space.

\item
The origin of the problem as well as the quantization of the parameter $\gamma$
can be traced back to the use of the Poisson symplectic structure
which does not take into account the presence of the second class constraints.

\item
Besides, this approach completely ignores the presence of the secondary second class constraint
which is crucial for suppressing the fluctuations of non-dynamical variables and
producing the right vertex amplitude in discretized theory.

\item
An attempt to interpret the results of such quantization only as an approach to compute
transition amplitudes using (unphysical) Hamiltonian \eqref{sfHam}
does not work as they turn out to be incompatible with the results of the standard
(path integral or canonical) quantization. As a result, the transition amplitudes computed in this way
do not have any consistent canonical representation.

\end{itemize}

In our opinion, all these problems are just manifestations of the fact that the rules of the Dirac quantization cannot
be avoided. This is the only correct way to proceed leading to a consistent quantum theory.

\subsubsection{A new look at simplicity and SF strategy}

The example presented above explicitly reveals the main problems of the new SF models
and their origin.
All these models start from the symplectic structure provided by the simple BF theory,
which ignores constraints of general relativity.
In particular, they all use the usual identification of the $B$-field
with the generators of the gauge group, or its $\im$-dependent version \eqref{newquantB},
when the constraints are translated into quantum level.
But this identification does not agree with the symplectic structure of general relativity.
Why this is so can be understood from the following consideration \cite{Alexandrov:2007pq}.

\lfig{The action of the smeared $B$-field on a holonomy.}{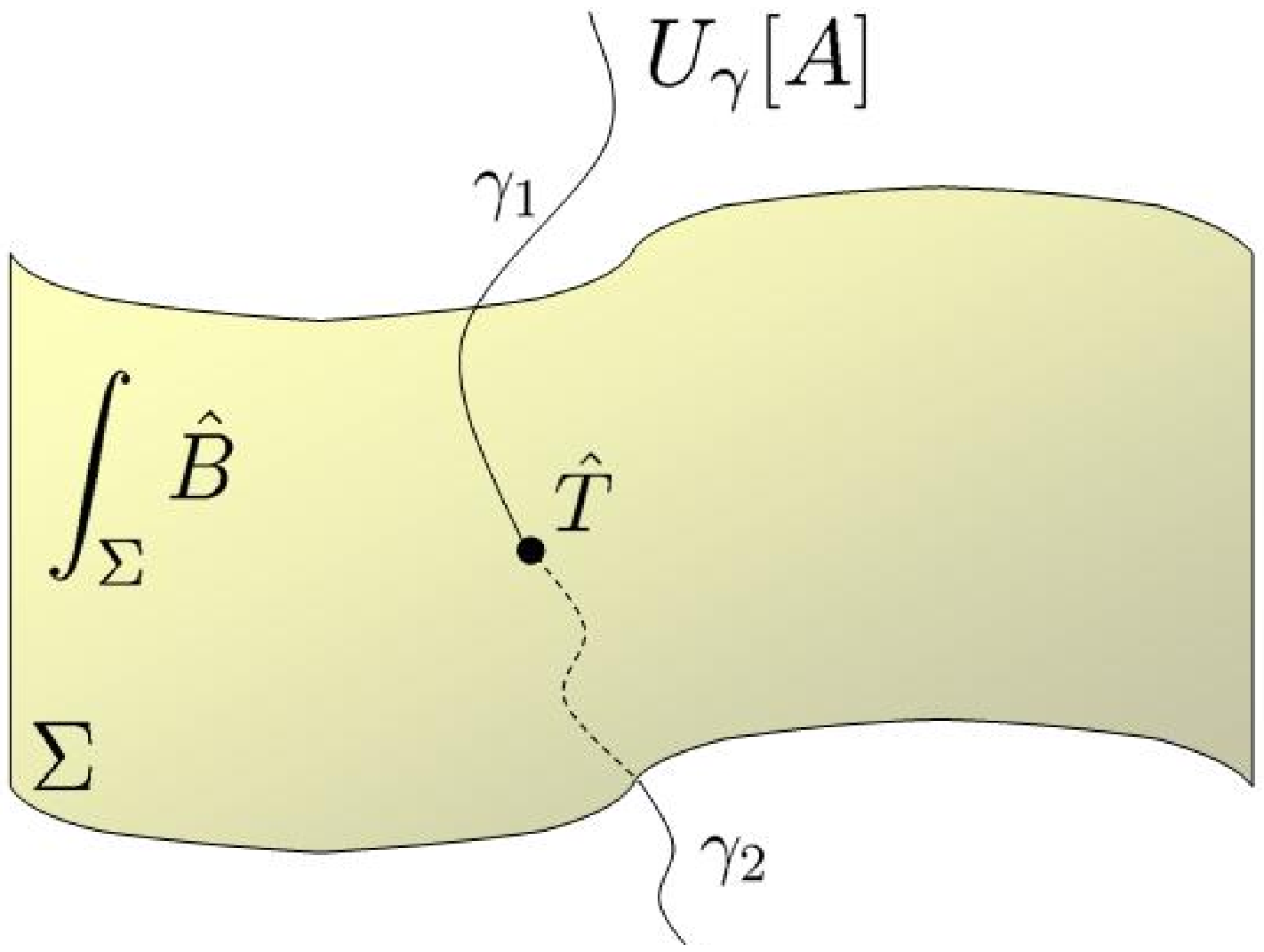}{5.5cm}{figsurf}

Let us consider the action of the operator
$\hat B$, smeared over a 2-dimensional surface, on a holonomy of a connection $A$. In BF theory the $B$-field is canonically
conjugate to $A=\omega$ and therefore the action simply brings down the generator $T$ of the gauge group
\be
\int_\Sigma  \hat B\, \cdot\, U_{\gamma}[A]=-\I \hbar\,U_{\gamma_1}[A]\cdot \hT\cdot U_{\gamma_2}[A],
\ee
where we assumed that a curve $\gamma$ is split into $\gamma_1 \cup\gamma_2$ by an intersection
with the surface $\Sigma$ (Fig. \ref{figsurf}).
It is this relation that allows to identify $B$ with the generator.
More generally, as soon as
\be
[\, \eps^{jkl}\hat B_{kl}^{IJ}(y),\hat A_i^{KL}(x)]
=-\I \hbar\,C^{IJ,KL}\delta_i^j\delta(x,y)
\label{commABC}
\ee
with some arbitrary function $C^{IJ ,KL}$, one obtains the same result with
$T^{IJ}$ replaced by $C^{IJ ,KL} T_{KL}$.
Thus, the above identification is true only if $C^{IJ ,KL}=\eta^{I[K}\eta^{L]J}$,
{\it i.e.}, $B$ and $A$ are canonically conjugate.

On the other hand, as we discussed in the beginning of this section,
the canonical analysis of Plebanski formulation \cite{Buffenoir:2004vx,Alexandrov:2006wt} shows
that it is not the case for general relativity because the symplectic structure
to be quantized is given by Dirac brackets having a non-trivial form.
In fact, the commutator of the $B$-field with the spin-connection \eqref{comomP} is even more
complicated than \eqref{commABC}. This form, needed to geometrize the $B$-field, can be achieved by
replacing the spin-connection by the shifted connection $\SA$.
But then the commutation relation \eqref{commPlebB} implies that
the identification of $B$ with the generators holds only for the boost part of the field,
whereas the rotation part effectively vanishes.

Anyway, it is clear that the quantization rule \eqref{newquantB} disagrees with the canonical structure
of Plebanski formulation.
As a result, one can apply the conclusions made in the previous subsection about the simple model \eqref{example}
directly to most of SF models in 4 dimensions.
Indeed, this example captures the $U(1)\times U(1)$ sector of the Euclidean theory constrained by the diagonal simplicity
which is imposed in a strong way as in \eqref{impose_phi}.
How the non-diagonal degrees of freedom are treated and how the cross simplicity is imposed
is not already relevant for the validity of these conclusions.
In particular, they are not affected by the progress in understanding
the cross simplicity in the new models and can be applied to both
EPRL and FK as well as to the original BC model.\footnote{One could think that the derivation
of the EPRL model presented in \cite{Ding:2010ye} is not captured by our analysis because
it does not distinguish the diagonal simplicity imposing all constraints weakly as in \eqref{constr_cross}.
However, it was overlooked that this approach leads to another model having a much larger boundary state space
since the condition \eqref{constr_reprL} is absent and $\lambda$ can take any values $\(j+k,\im \,\frac{j(j+1)}{j+k}\)$
with $k\in \Nint$. If one wants to get back the EPRL model, one still has to impose the diagonal simplicity
strongly. This issue is discussed also in \cite{Dupuis:2010iq} where it was concluded
that consistency requires the diagonal simplicity to be imposed only weakly. \label{foot_weak}}
Besides, we considered the compact case corresponding to the Riemannian models
just to elucidate the issue with the quantization condition on $\im$, but
all conclusions remain to be valid in the Lorentzian theory as well,
where the conditions on the representation labels
are obtained in the same way and look equally unnatural from a group-theoretic point
of view.

In fact, a special care which is paid to the diagonal simplicity, when it is imposed strongly whereas
the cross simplicity constraints are imposed only weakly, results from another common confusion.
As we explained in section \ref{subsubsec_EPRL}, this is done because the diagonal simplicity
is in the center of the non-commutative constraint algebra of all simplicity constraints
and thus interpreted as first class. But this classification would be correct only if there were no other constraints
to be considered. It completely ignores the presence of the secondary constraints.
The latter do not commute with all simplicity and in particular with the diagonal simplicity.
As a result, all these constraints are second class and should be quantized via
the Dirac bracket.

Given all this, we expect that the new SF models suffer from inconsistencies which we met in the previous subsection.
They can be summarized by saying that the statistical models defined by the SF amplitudes do not have
a consistent canonical quantization picture, where the vertex amplitude appears as a matrix element
of an evolution operator determined by a well defined Hamiltonian. In particular, there is no reason
to expect that the new models may be in agreement with LQG or any of its modifications.
Note that this incompatibly with the canonical quantization manifests itself in the issues involving the Hamiltonian.
This is why one does not see it in a semiclassical analysis or in any investigation
restricting to the kinematical level.

It should be stressed that this critics is not just about face or edge amplitudes, which
depend on details of the path integral measure but can be found in principle from
consistency on the gluing of simplices \cite{Bianchi:2010fj}.
In fact, the ignorance of the secondary second class constraints has much more profound implications
and, what is the most important, it affects the vertex amplitude (see the next subsection).
The standard prescription that the vertex is obtained by evaluating the boundary state of a 4-simplex
on a flat connection is a direct consequence of the employed strategy, which starts
by quantizing the topological BF theory, and should be modified to take into account
all constraints of general relativity.

Of course, the SF models are still well defined as statistical models.
But, in our opinion, this is not enough to consider them as candidates for quantum gravity.
A good candidate should allow a quantum mechanical representation in terms of wave functions, Hamiltonian, {\it etc.},
especially if one hopes to find a viable loop quantization of gravity.
The point we are making here is that the SF models derived using the strategy ``first quantize and then constrain"
do not satisfy this requirement.

There is however one special case when our reasoning, based on the use in the SF quantization
of a wrong symplectic structure, is not applied.
This is the FK model without the Immirzi parameter ($\im=\infty$) \cite{Freidel:2007py}.
Since in this model the simplicity constraints are imposed not directly on the generators,
but on their expectation values in a coherent state \eqref{bivec-coh},
for $\im=\infty$ the generators of the subgroup $SU(2)$, defined with respect to $x_t$, drop out
and only the boost generators survive.
In other words, effectively, the bi-vectors are quantized as
\be
B_f\longmapsto \hat B_{ft}= -\I \hbar \,I_{(P)}(x_t)\cdot \hT^{(\lambda_f)}.
\label{qmap}
\ee
Note that each bi-vector gets two realizations depending on the frame of one of the two tetrahedra
sharing the face.
The effective quantization rule \eqref{qmap} is consistent with the symplectic structure
of Plebanski formulation \eqref{commPlebB} written in terms of
the shifted connection $\SA$ where the second class constraints have been taken already into account
by means of the Dirac bracket.\footnote{For generic $\im$, \eqref{simbfX} implies that effectively
the bi-vectors are quantized as
$$
B_f\longmapsto \hat B_{ft}= -\I \hbar \,\frac{\im^2}{\im^2-1}\, I_{(P)}(x_t) \cdot \(\hT-\frac{1}{\im}\star \hT\).
$$
This is consistent with Dirac bracket \eqref{AP2} for $a=-\frac{\im}{1-\im^2}$ and $b=\frac{1}{1-\im^2}$.
However, the generic connection $\SAab$ appearing in this commutation relation
has nothing to do with the spin-connection defining the group elements
of the discretized theory. Therefore, there is no any reason why
the symplectic structure formulated in terms of this connection may be relevant.}
Thus, in contrast to all other models, the FK model without $\im$
may correctly incorporate the primary simplicity constraints.
Nevertheless, as we will see below, this model still suffers from ignoring the secondary constraints which
should affect it at the dynamical level.

\subsubsection{Secondary constraints and vertex amplitude}

The SF representation of quantum gravity can be seen as an outcome of a Lagrangian path integral
for discretized Plebanski formulation of general relativity.
However, the Lagrangian or a configuration space path integral is a derived concept. A more fundamental one
is the path integral over the phase space. Its measure can be rigorously derived and in particular
it contains $\delta$-functions of all second class constraints.
On the other hand, the Lagrangian path integral
can be obtained from the canonical one only at certain very special circumstances.
Usually, this amounts to integrating over some variables and to use some additional tricks.
We refer to \cite{Han:2009aw} for an extensive discussion of various obstacles appearing on this way.

Fortunately, here we deal with the first order formulations so that we do not need to integrate
over momenta which are a part of the configuration variables of the initial Lagrangian.
Thus, the main problem which remains in connecting the two path integrals is
to remove from the measure all secondary constraints. Indeed, in
the Lagrangian path integral the measure contains only $\delta$-functions resulting from
integrating over Lagrange multipliers, {\it i.e.}, of the primary constraints.
In the presence of second class constraints there is a nice trick suggested in \cite{Henneaux:1994jf}
which fulfils a transformation removing secondary constraints from the measure and leaving there just some local factor.
However, since it involves a canonical transformation of variables, it works only for the
partition function and fails for correlators \cite{Alexandrov:2008da}.\footnote{In \cite{Han:2009aw}
it was shown that the trick of \cite{Henneaux:1994jf} does work if one restricts the measured observables
to the set of reduced phase space coordinates. However, it is a hard task to relate them to the
initial phase space variables so that it is not clear that this can be accomplished in practice for such complicated systems
as general relativity. Even if one does that, one still finds a non-trivial (non-spacetime invariant)
contribution to the measure \cite{Buffenoir:2004vx,Engle:2009ba} which is not taken into account in SF models.
Therefore, most of our conclusions remain to be valid anyway.}
Therefore, if one wants to calculate transition functions as one does in SF models, one {\it must}
use the canonical path integral.

The main consequence of this conclusion is that, as we mentioned above, the secondary second class constraints
should appear explicitly in the integration measure. We believe that this is an important point
missed by the present-day SF models.
In \cite{Alexandrov:2008da} it was shown that it affects the expression for the vertex amplitude
so that it is crucial to take it into account to produce the right quantum dynamics.

The secondary constraints following from the canonical analysis are naturally written using orthogonal projectors
(see, for example, \eqref{contA_B}) and therefore will depend on the normals $x_t$ after discretization.
Thus, the measure involving the constraints should also depend on these normals
\be
\CD^{(x_t)} [\gl_{\sigma t}]\sim \delta(\psi_{\rm discr}\(\gl_{\sigma t},x_t)\)\de g_{\sigma t},
\label{measur}
\ee
where $\psi_{\rm discr}$ is a discrete version of the secondary constraints.
However, this dependence should be consistent with the following covariance property
\be
\CD^{(x_t)} \[\gl_{\sigma t}\, \gb\]=\CD^{(x_t^\gb)} [\gl_{\sigma t}], \qquad x_t^\gb=\gb\cdot x_t.
\label{trmeas}
\ee
This property is sufficient to prove that the corresponding vertex amplitude is given by \cite{Alexandrov:2008da}
\be
A(\lambda_f,j_{tf},\Int_t)=\int \prod_t \CD^{(x_t)} [\gl_t]\,
\Psi_\Delta^{(\lambda_f,j_{tf},\Int_t)}\[\gl_{d(f)}^{-1}\gl_{u(f)}^{\hphantom{1}};x_t\],
\label{amplit_simplex}
\ee
where $u(f),d(f)$ denote two tetrahedra sharing the triangle dual to face $f$, ``up" and ``down",
$\Int_t$ labels different $SU(2)$ intertwiners and $\Psi_\Delta^{(\lambda_f,j_{tf},\Int_t)}$ is
the projected spin network associated with the boundary of a 4-simplex.

This result represents a natural generalization of the standard prescription that the vertex is obtained
as the spin network evaluated on a flat connection. Indeed, if the measure is taken to be the usual
Haar measure on the group, $\CD^{(x_t)} [\gl_t]=\de \gl_t$, the integrals over the group elements
simply ensure the $G$-invariance of the spin network and can be neglected if the intertwiners are already invariant.
This leaves the unity in the argument producing the standard recipe,
which can be traced back to the flatness condition in \eqref{ZBFd}.
However, the result \eqref{amplit_simplex} shows that in the presence of second class constraints this prescription is incorrect.
The secondary constraints modify the measure and, in particular, restrict the integration
to a certain subspace of $G$ due to the presence of $\delta$-function in \eqref{measur}.

One might argue that since the secondary constraints appear as stability conditions for the primary ones
and the latter are imposed in the path integral at every moment of time, the secondary constraints
should follow automatically and need not to be imposed explicitly.
For example, in SF models based on Plebanski formulation
one could expect that all set of simplicity constraints ensures
the simplicity of bi-vectors at all times and thus it is enough.
However, this argument works only at the quasiclassical level where the equations of motion
are satisfied. Off-shell the quantum fluctuations of degrees of freedom fixed classically by the secondary constraints
are not suppressed if the constraints are not inserted in the path integral.
Moreover, in SF models the primary constraints are not imposed everywhere, but only on the boundary of 4-simplices,
because this is the place where the bi-vectors live.
At the same time, as is clear from the example of section \ref{subsubsec_example},
the vertex amplitude can be seen as an {\it unconstrained} path integral {\it inside} the simplex.
This is why we claim that it represents the dynamics of BF theory
rather than the true dynamics of general relativity.

Due to this reason it should be clear that
the modification of the measure which we advocate affects only the dynamics of the
corresponding SF model, but is crucial to get the right one \cite{Han:2009aw}.
It is also not seen at the quasiclassical level since the missing constraint
is obtained on mass shell anyway.
Therefore, it is not in contradiction with the fact that the semiclassical asymptotics of the EPRL and FK amplitudes
reproduce the Regge action \cite{Conrady:2008mk,Barrett:2009mw,Barrett:2009gg},
{\it i.e.}, the correct classical limit.
The problem is that the secondary constraints are not imposed {\it strongly}
at the quantum level and as a result one might expect the appearance of additional
quantum degrees of freedom in the new models.\footnote{In \cite{Alexandrov:2008fs} it was noticed that
in a class of generalized gravity theories based on a generalization of the {\it non-chiral}
Plebanski action \eqref{Plebanskiaction} the secondary second class constraints are missing and as a result one finds
6 more propagating degrees of freedom.
The resulting theory was interpreted as a bi-metric gravity in \cite{Speziale:2010cf}.
The difference with our situation is that in these generalized gravities the constraints
are missing already at the classical level.}

Unfortunately, we were not able to find the right measure $\CD^{(x_t)} [\gl_{\sigma t}]$ at the discrete level.
The main difficulty is that the constraints \eqref{contA_B} contain a non-trivial, $B$-dependent
right hand side. Thus, a credible candidate for a spin foam model of
quantum gravity is still lacking.

\subsubsection{Closure constraint}

The necessity to insert the secondary second class constraints into the path integral measure
over holonomies has important implications for another constraint widely used in the SF approach,
the closure \eqref{closure}. The usual identification of the bi-vectors $B_f$ with the generators of
the gauge algebra implies that its quantum version requires the invariance of intertwiners assigned to tetrahedra
\be
\vphantom{A\over B}
\smash{\sum_{f\subset\, t}}\, \hT^{(\lambda_f)}\, \Int_G^{(t)}=0.
\label{clcon}
\ee
In the spin foam models considered above such invariance is achieved by averaging over the normals $x_t$
appearing after implementation of the cross simplicity constraints.
For example, in the BC intertwiner \eqref{matNBC} this is equivalent to the insertion of an
integral over the factor space (boosts) $X$.
However, in \cite{Alexandrov:2008da} it was argued that integrating the boundary states over $x_t$ is inconsistent
with the gluing of different simplices which respects the second class constraints.

Indeed, the gluing of two simplex contributions is performed by integrating over holonomies
associated to triangles common to the two simplices. If the correct measure includes the contribution
of the secondary constraints, it depends on the normal to the shared tetrahedron,
as explained in the previous subsection. It is clear that this is the same normal which
appears in the corresponding boundary states of the glued simplices. On the other hand, the insertion
of integrals over this normal in the states would make $x_t$'s appearing in the states and in the measure
unrelated. This is clearly inconsistent with the fact that the geometry (the $B$-field) assigns unique data to the
elements of the simplicial decomposition.\footnote{This may be compared to the derivation of the BC model
in the FK approach. As discussed in section \ref{subsubsec_SFandLQG}, it arises after
{\it incorrect} decoupling of the normals $n_{tf}^\sigma$ for different simplices.
Similarly, we argue that the normals $x_t^\sigma$ should be taken equal for all $\sigma\supset t$.}

In other words, the gluing should be done according to the following schematic formula
\be
\Psi_{12}\[g_f;x_t\]
=\int \prod_{f_{12}}\CD^{(x_{t_{12}})} [\rho_{f_{12}}]\,
\Psi_1\[g_{f_1},g_{f_{12}}\rho_{f_{12}};x_{t_1},x_{t_{12}}\]
\Psi_2\[g_{f_2},\rho_{f_{12}}^{-1};x_{t_2},x_{t_{12}}\],
\label{glue_simplex}
\ee
where $f=(f_1,f_2,f_{12}),t=(t_1,t_2,t_{12})$, the labels 1,2 refer to non-shared faces and tetrahedra of
the corresponding glued simplices and the label 12 marks the shared faces and tetrahedron.
What we explained above is that the normal $x_{t_{12}}$ should be the same in the measure, $\Psi_1$ and $\Psi_2$.
Thus, it must be an additional argument of the boundary state functional as shown explicitly in \eqref{glue_simplex}.
Moreover, using \eqref{trmeas}, it is easy to see that the dependence on $x_{t_{12}}$ drops out
from this integral. Thus, even in the complete partition function all normals should not be integrated over and
must be kept fixed as otherwise one would get an infinite factor in the Lorentzian case.
This has a simple interpretation of a gauge fixing of boosts in the discretized path integral.

Once it is forbidden to insert integrals over $x_t$ in the definition of the boundary states,
their intertwiners are not invariant anymore. Indeed, the boundary states are precisely given
by projected spin networks \eqref{projspnet} whose intertwiners can be written as \cite{Alexandrov:2007pq}
\be
\Int^{(\{\lambda_{k}\},\{j_{k}\},\Int_t)}_{(\Psi)\,p_1 \cdots p_L}(x_t)=\mathop{\sum}\limits_{\ell_{j_1}\cdots\ell_{j_L}}
\Int^{(t)\,\{ j_{k}\} }_{SU\!(2)\, \ell_{j_{1}} \cdots \ell_{j_{L}}} \prod_{k=1}^L R^{(\lambda_k)}_{p_k \ell_{j_{k}}}(g_{x_t}),
\label{genBC}
\ee
where $k=1,\dots,L$ (for one simplex $L=4$),
the indices $\ell_{j_k}$ label the basis of the subspace $H_{SU(2)}^{j_k}$ appearing in the decomposition
of the representation $\lambda_k$ on the subgroup $SU(2)$, and $\Int_t\equiv\Int_{SU\!(2)}^{(t)}$ is an $SU(2)$ invariant intertwiner.
The intertwiners $\Int_{(\Psi)}$ provide a natural generalization
of the BC intertwiner \eqref{matNBC} where one drops the integral over $x$
and replaces the projection to the trivial representations $j_k=0$ by a generic one.
It is easy to see that they satisfy the following property
\be
\mathop{\sum}\limits_{q_{1}\cdots q_{L}}
\(\prod_{k=1}^L \Db^{(\lambda_k)}_{p_k q_k}(\gx)\) \Int^{(\{\lambda_{k}\},\{j_{k}\},\Int_t)}_{(\Psi)\,q_1 \cdots q_L}(x_t)
=\Int^{(\{\lambda_{k}\},\{j_{k}\},\Int_t)}_{(\Psi)\,p_1 \cdots p_L}(\gx\cdot x_t),
\label{invN}
\ee
or in the infinitesimal form
\be
\vphantom{A\over B_B}
\smash{\sum\limits_{{f\subset\, t}}}\, \hT^{(\lambda_f)}\, \Int_{(\Psi)}(x_t)=\hT\cdot \Int_{(\Psi)}\(x_t\),
\label{clconrel}
\ee
where $\hT$ acts on functions of $x_t$ as the usual generator of rotations
\be
\hT_{IJ}\cdot f(x)=\eta_{IK} x^K \p_J f-\eta_{JK} x^K \p_I f.
\label{generator}
\ee

Thus, according to our analysis, the new condition \eqref{clconrel} replaces the usual constraint
on intertwiners \eqref{clcon}. This relaxed version of the closure condition has been proposed in \cite{Alexandrov:2007pq}
on the basis of comparison with the canonical approach and appeared also in recent group field theory
models \cite{Oriti:2009nd}. It expresses the {\it covariance} of intertwiners
under the action of the gauge group, rather than their invariance as is usually required.

Note that the relaxed closure condition is not in contradiction with the classical closure constraint \eqref{closure}.
In fact, the latter follows from the gauge invariance only provided the flatness condition on the connection is satisfied,
which arises only as an equation of motion. Therefore, in quantum theory it is expected to appear only in
the quasiclassical limit. As a result, one should distinguish two forms of the closure \cite{Dittrich:private}:
the strong condition \eqref{closure} holding only on shell and the usual Gauss constraint expressing
the gauge invariance. The latter is more general and our relaxed condition \eqref{clconrel}
shows how it should be incorporated at the quantum level.

\subsection{Comparison with the canonical approach}
\label{subsec_compar}

Summarizing our analysis of the spin foam models, we see that already the proper
consideration of the diagonal simplicity constraint shows that all attempts to include
the Immirzi parameter following the usual strategy lead to inconsistent quantizations.
All problems of the EPRL and FK models can be traced back to that they quantize the symplectic structure
of the unconstrained BF theory and impose (a half of) the second class constraints
after quantization. So we conclude that the widely used strategy ``first quantize and then constrain"
does not work and should be abandoned.

This leaves us with the FK model for $\im=\infty$ which appears to be the best proposal up to now.
But as we showed, even in this case not all constraints are properly incorporated.
Namely, it misses the secondary second class constraints affecting the path integral measure and as a result
influencing the model at the dynamical level.
In particular, the vertex amplitude should not be given anymore by evaluation of a spin network on a flat
connection, which is again an artefact of the misleading strategy,
but must be represented by a more complicated integral \eqref{amplit_simplex}.

In fact, the analysis of the discretized path integral shows \cite{Alexandrov:2008da} that
once all constraints are properly incorporated, the boundary states of a spin foam are always given
by projected spin networks. Moreover, the arguments of these spin networks should be constrained
by the secondary second class constraints so that
there is a perfect agreement at the kinematical level
with the covariant canonical loop approach of section \ref{subsubsec_loopq}.

However, a concrete implementation of the secondary constraints both in the boundary states
and in the vertex amplitude \eqref{amplit_simplex} remains problematic.
This is precisely the same problem which supplies the canonical approach.
In particular, it is closely related with the non-commutativity of the connection.
Thus, although it is possible to obtain a picture similar to the loop quantization, the goal, which is to provide
a credible spin foam model and, in particular, a vertex amplitude correctly taking into account
all constraints of general relativity, is far from being accomplished.

\newpage

\section{Roads to Quantum Gravity}
\label{sec_where}

How to overcome the problems which have been exposed in the previous chapters?
What ideas can give new insights on these problems?
Where is the mysterious pathway to quantum gravity?
Here we would like to suggest tentative answers to some of these questions and to speculate on
possible research directions which seem to be promising to us.

\subsection{Canonical way}
\label{subsec_canway}

As we argued in chapter \ref{sec_canan}, the idea of the loop quantization potentially
can be realized in various inequivalent ways. LQG is only one of them and not the best.
In our opinion this is the approach of CLQG described in section \ref{subsubsec_CLQG}
that has more chances to successfully quantize general relativity. But on this way
one encounters two major problems:
\begin{enumerate}
\item to find a unitary irreducible representation of the
quantization of the classical algebra of Dirac brackets;
\item to impose the secondary second class constraints \eqref{contA} at the level of the kinematical Hilbert space.
\end{enumerate}
These two problems are closely related because the first of them arises due to the non-commutativity
of the connection $\SA^{IJ}$, which is the central object of this approach playing the same role as
the Ashtekar--Barbero connection plays in LQG.
This non-commutativity can be traced back to the constraints \eqref{contA} which show that
the rotation components of the connection are expressed in terms of coordinates canonically conjugate
to its boost components. Therefore, it is very likely that a solution to one the above problems
will solve also the second.

The most straightforward way to find a representation of the Dirac algebra would be
to understand the geometric meaning of the commutator of two connections.
The easiest solution would be if it is related to some non-commutative
algebra structure, not too complicated to study, as it happens for
example in Chern--Simons theory \cite{Fock:1998nu} where it appears to be related to quantum groups.
But although this commutator was known for some time \cite{Alexandrov:2002xc,Alexandrov:2005ar},
no such structure has been discovered so far.

A different possibility would be to give up the connection representation and
to develop a representation which starts from a configuration space given by classically commuting variables.
Such set of variables is provided, for example, by triads which hints that it might be reasonable
to look for a triad representation \cite{Alexandrov:2005ng}. But unfortunately the classical commutativity
disappears as soon as one passes to the triads smeared over two-dimensional surfaces \eqref{smtriad} \cite{Ashtekar:1998ak},
which is the necessary step in the background independent loop approach.
Remarkably, in the simple case of LQG kinematics, this is not really a problem and the triad or {\it flux representation}
can be constructed and is formulated in terms of non-commutative functions on Lie algebras \cite{Baratin:2010nn}.
It is an exciting problem whether this LQG flux representation can be generalized to take into account
the non-commutativity of the CQLG connection.

Finally, an interesting possibility to deal with the non-commutativity
and to address the problem of how to impose the secondary constraints
on arguments of projected spin networks is opened by the coherent state technique developed
in the context of SF models \cite{Livine:2007vk}. It might help to disentangle
degrees of freedom responsible for different geometric information in a similar way
as has been done for the usual $SU(2)$ spin networks \cite{Freidel:2010aq,Freidel:2010bw}.
Once the geometric meaning of different elements is understood, the geometric quantization methods
may give insights on the right quantization.

\subsection{Path integral way}
\label{subsec_pathint}

In our opinion, it is a very encouraging fact that, once all constraints are properly taken into account,
the spin foam path integral leads to a picture consistent with the canonical approach.
In particular, one finds that the state space is the same and provided by projected spin networks with
suitably constrained arguments.

But this convergence of results implies that the problems of the two approaches
are also similar. Indeed, the main obstacle which prevents us so far from formulating a consistent and credible
SF model is again related to the secondary constraints.
This time the problem is how to incorporate them in the measure \eqref{measur} to be used in the vertex
amplitude \eqref{amplit_simplex}.
In fact, it can be split into two parts:
\begin{enumerate}
\item how to discretize the secondary second class constraints;
\item which measure the discretized constraints induce for the vertex amplitude.
\end{enumerate}

Very promising results concerning the first point appeared recently in
\cite{Dittrich:2008ar,Dittrich:2010ey} and \cite{Gielen:2010cu}
where they were identified with some geometric constraints
on the phase space of simplicial geometries. This geometric interpretation in the simplicial context
might be very helpful in incorporating them directly in the discretized path integral,
but this has not been done in full detail yet.

On the other hand, the problem of the right measure in the vertex amplitude does not seem to allow
a simple solution. The main difficulty again comes from the fact that the secondary constraints \eqref{contA_B}
depend on the $B$-field which, as we recall, was the origin of the non-commutativity of the connection in CLQG.
Due to this it is very difficult to integrate the $B$-field out, as is usually done in derivations
of SF models, and it is impossible to insert the constraints directly in the vertex amplitude,
since the integral formula \eqref{amplit_simplex} assumes that the measure is $B$-independent.

A promising research direction in this respect is an attempt to construct a spin foam model where
the bi-vectors appear on equal footing with holonomies.
Such models has been introduced in the framework of group field theories (GFT)
\cite{Oriti:2007vf,Oriti:2009nd,Baratin:2010wi}
and will be discussed in the next section.

We would like to finish this short section with a remark which might turn out to be important in trying
to construct a path integral quantization of general relativity.
It is usually of prime importance to choose an appropriate gauge fixing condition
which may either facilitate or complicate the quantization.
The connection representation widely used in LQG might suggest that it could be
convenient to impose conditions on the connection variables. However, in \cite{Alexandrov:1998cu}
it was shown in the context of the BRST path integral quantization that
such gauges lead to higher ghost terms in the effective action appearing due to
the dependence of the structure constants of the constraint algebra on the dynamical fields.
As a result, the ghosts cannot be integrated out to produce the usual Faddeev--Popov determinant.
This indicates that gauge conditions on connections should be systematically avoided.

\subsection{GFT way}

A group field theory (GFT) formulation of spin foam models appears from the observation that
SF amplitudes can be presented as evaluation of Feynman graphs
on product of groups or homogeneous spaces.
Therefore, it is very natural to build a QFT whose perturbative expansion produces these Feynman integrals.
This aim is fulfilled by GFTs, combinatorial non-local QFTs on $G^{d}$,
where $G$ is the group appearing in the corresponding SF model and $d$ is the dimension of spacetime.
The expansion of the partition function of such theory in a coupling constant
$\lambda$ gives precisely the expansion of the spin foam type
\begin{equation}
Z_{\rm GFT}=\sum\limits_{C} \frac{\lambda^{n_V(C)}}{sym(C)} \sum\limits_{J,\, \Int} \prod_f A_f \prod_e A_e \prod_v A_v.
\label{defstatGFT}
\end{equation}
The group field theories were first invented as higher dimensional
extensions of matrix models quantizing two-dimensional gravity.
The first model in three dimensions was defined by Boulatov \cite{Boulatov:1992vp}
and in four dimensions by Ooguri \cite{Ooguri:1992eb}.
Later they were generalized in \cite{DePietri:1999bx} to get a GFT reformulation of the BC spin foam model.
For recent reviews on GFT see, for example, \cite{Freidel:2005qe,Oriti:2009wn}.

An advantage of the GFT formulation is that it generates not only spin foam amplitudes, but also
organizes them in a sum with certain weights.
Therefore, it is expected that the ``non perturbative'' study of these GFTs can
be a promising direction for giving insights on the construction of the physical
scalar product in the Hamiltonian approach and on the particularly obscure issue of the summation over 2-complexes.
Moreover, there is a viewpoint that GFTs are not just auxiliary QFTs, which role
is to eventually solve these problems, but they are fundamental formulations of quantum gravity.
One hopes that they can, for example, address fundamental issues
such as the topology change, which are poorly studied by means of other techniques.

It might be useful to recall the basic ideas of GFT. Let $G$ be a Lie group and denote $\de x$ the
right invariant Haar measure. In the applications to spin foam models associated
to quantum gravity, the gauge group is usually $G=SO(\eta)$ (in that case the Haar measure is right and left invariant)
and the group field $\phi$ is a map $\phi:G^d \rightarrow {\mathbb R}$ denoted
$\phi(x_i),\ i=1,...,d$.
The general form of the GFT action is given by
\begin{equation}
S[\phi]=\frac{1}{2}\int \prod_{i=1}^d \de x_i \de y_i\,
\phi(x_i){\cal K}(x_i y_i^{-1})\phi(y_i) +
\frac{\lambda}{d+1} \int \prod_{i\not= j}^{d+1} \de x_{ij}\,{\cal V}(x_{ij}x_{ji}^{-1})
\prod_{i=1}^{d+1} \phi(x_{ij}),
\end{equation}
where $\phi(x_{ij})=\phi(x_{i1},..., x_{id+1})$,
${\cal K}$ is a kinetic kernel, ${\cal V}$ is an interaction kernel and $\lambda$
is a coupling constant whose precise physical meaning is still a matter of debate.
This action has a global symmetry given by $\phi(x_i)\mapsto \phi(x_i g),\ g\in G.$
If one assumes that ${\cal K}$ and ${\cal V}$ are bi-invariant under $G$, {\it i.e.},
${\cal K}(gx_i g')={\cal K}(x_i)$ and ${\cal V}(g_i x_{ij} g_j^{-1})={\cal V}(x_{ij}),\ \forall g,g', g_i\in G,$
one obtains an additional gauge symmetry which can be gauge fixed by
requiring that the group field satisfies $\varphi(gx_i)=\varphi(x_i).$
Then the link between Feynman graphs following from this action and spin foam amplitudes is precisely obtained
through this invariance: one expands the field $\varphi$ in Fourier modes given by matrix elements
of irreducible unitary representations and imposes the $G$-invariance through Clebsch-Gordan maps.
The rest is described by tensor matrix models whose Feynman graphs
can be easily computed and appear to be spin foam amplitudes.

In practice, however, very often it is more convenient to work not in terms of kernels, but
using certain projectors, especially if the actual space where the field lives is not the group,
but a homogeneous space $G/H$ where $H$ is a closed subgroup of $G$.
This happens, for example, for the BC model and is expected to be the case for any credible GFT
model of general relativity in four dimensions.
For example, GFTs for BF theory and the BC spin foam model can be constructed as follows.
Let us define two commuting projectors acting on a field $\varphi$ on $G^d$ as
\be
(P^{r}_H\varphi)(x_i)=
\int_{H^d}\prod_i \de h_i\, \varphi(x_ih_i),
\qquad
(P^{l}_G\varphi)(x_i)= \int_{G} \de g\, \varphi(gx_i).
\ee
Then the GFT action, which defines a theory on the homogeneous space $G/H$, reads as \cite{DePietri:1999bx}
\begin{equation}
S[\varphi]=\frac{1}{2}\int \prod_{i=1}^d \de x_i\,
\( (P^{r}_H P^{l}_G \varphi)(x_i)\)^2 +
\frac{\lambda}{d+1} \int \prod_{i\not= j}^{d+1} \de x_{ij}
\prod_{i=1}^{d+1} (P^{r}_H P^{l}_G\varphi)(x_{ij}).
\end{equation}
Choosing various subgroups $H$, one recovers different spin foam models.
In particular, the SF model of BF theory is obtained by taking $H=\{e\}$, whereas
taking $G=SO(4)$ and $H$ being its diagonal $SO(3)$ subgroup leads to the BC model
with a certain choice of face and edge amplitudes.
The corresponding kernels $\CK$ and $\CV$ are then given
by combinations of $\delta$-functions.

Since GFTs automatically generate expansions which have interpretation in terms of spin foams,
but can be formulated in a more compact and geometric way, one may hope that they propose
a natural framework to seek for a correct implementation of the simplicity constraints.
Although this goal has not been achieved yet, we would like to point out
an interesting development already mentioned in the previous section.
This is a new class of GFTs proposed in
\cite{Oriti:2007vf,Oriti:2009nd,Baratin:2010wi}
which is formulated in terms of fields living on $(G\times {\mathfrak g})^d$
where $\mathfrak g$ is the Lie algebra of $G$.
Such models are supposed to provide a ``GFT in a first order formalism" because, whereas the group variables
are usually used to encode holonomies $g$, the algebra-valued variables
should describe the $B$-field of Plebanski formulation.
The presence of both variables gives a possibility to work with the constraints of the type \eqref{contA_B}
which depend on the connection and on the $B$-field as well.
However, it still remains an unsolved problem how to incorporate these constraints as restrictions on
the group field and which consequences for the SF amplitudes this implies.

What has been already realized in the framework of these models is
that the closure constraint should be imposed in the relaxed form proposed in \eqref{clconrel} \cite{Alexandrov:2008da}.
Indeed, it was found that the only consistent way to implement the gauge invariance
for such generalized field is to require that
$\varphi(gg_i,g x_i g^{-1})=\varphi( g_i,x_i)$.
As a result, the gauge transformation affects both, the holonomies and the variable encoding the $B$-field,
which is of course absolutely natural from the continuum point of view. But this implies that
the intertwiners entering SF boundary states are no more invariant under gauge transformations,
but rather covariant, precisely as in \eqref{clconrel}.
We view this convergence of results as a promising indication.

However, we should mention that there is an issue common to all GFT models, which on one hand makes their results
non-conclusive, but on the other hand has a potential to solve some of the fundamental problems such as the summation
over 2-complexes or the classical limit. This is the problem of interpretation of the expansion parameter $\lambda$.
Is it a fundamental constant or  should it be fixed to some particular value? Up to now there is no
definite opinion on this issue.
In \cite{Freidel:2005qe} it was proposed that it should be sent to zero what will select only a particular class
of two-complexes, those which do not have some unwanted bubbles. But this proposal has not been put on a solid ground.
A different proposal has been put forward in \cite{Ashtekar:2009dn} where in the context of
a loop quantum cosmology model it was argued that $\lambda$ should be related to the cosmological constant.
On the other hand, recalling that GFT is a generalization of matrix models for two-dimensional gravity,
one might think that there should be some double scaling limit involved \cite{Brezin:1990rb,Douglas:1989ve,Gross:1989vs}.
In the case of matrix models this limit was crucial to perform the sum over surfaces, to get the classical limit
and to make the models integrable.
Therefore, it would be an exciting result if such a limit is found to govern GFT.
Unfortunately, GFT models have somewhat different structure than their matrix analogues
and so far no signs have been found that a double scaling limit is relevant for these models.

On the other hand, recently some progress has been done on the problem of summing over two-complexes.
For example, this problem was addressed in the works \cite{Freidel:2002tg,Freidel:2009hd,Magnen:2009at,Geloun:2010nw}
and some results for the 3d Boulatov's model (which is topological and based on $SU(2)$) were
obtained concerning questions such as: What is the planar limit of this model?
Can one give bounds on general Feynman graphs?
Can one show Borel summability for some quantities?
Besides, the question of topological singularities in the simplicial
complexes dual to the Feynman graph expansion
of GFT models has been studied in \cite{Gurau:2009tw,Gurau:2010nd} and an interesting proposal
how to avoid such singularities has been suggested.
However, being very important, these issues are still far from their complete resolution,
especially in the case of GFT models for four-dimensional gravity.

\subsection{Gravity as an effective theory}
\label{subsec_effect}

All the above approaches to quantum gravity are conservative in the sense
that they try to quantize various reformulations of the standard Einstein action of general relativity.
Although this is not {\it a priori} impossible, this is also one of the main points of criticism of these approaches:
it is not clear why one should believe that an action describing a classical theory in infrared remains
valid also in ultraviolet. Experience with quantum field theories indicates that this expectation is
in general unreliable as a fundamental ultraviolet action can acquire completely different form in
the infrared limit, where even degrees of freedom may change.
Therefore, considering the Einstein action as a fundamental action to be quantized is a huge assumption
which should be taken with great care.

Besides, now there are serious evidences that general relativity is only an effective theory,
coming first of all from its thermodynamic properties
(see, for example, \cite{Barcelo:2001tb,Padmanabhan:2009vy,Verlinde:2010hp}).
These properties most explicitly show up in the black hole entropy which scales as the horizon area, but this scaling in fact
is only a particular manifestation of a more general holographic behavior of gravity which is found
in various approaches. Quite remarkably, such behavior is already seen at the level of the classical action
and almost uniquely determines its form \cite{Jacobson:1995ab,Kolekar:2010dm}.
This shows that the holography is an important feature of at least classical gravity and
should likely play an important role in its quantum description.
On the other hand, neither in LQG nor in SF models the holographic behavior has been observed so far
(see however \cite{Smolin:1998qp}). In our opinion, this is an indication that something important is missing
in these approaches.

These results point in the direction that gravity is some kind of thermodynamic limit of a more fundamental theory.
Such scenario is realized, for instance, in string theory where the holography appears as open/closed string duality
which allows to describe gravity encoded in degrees of freedom of closed strings in terms of gauge theories
described by open strings. Moreover, the latter often have a description in terms of matrix models
which in turn can be reduced to a system of free fermions. Thus, gravity is found to be hidden in
the dynamics of these elementary degrees of freedom and arises as their collective field theory
\cite{Das:1990kaa}.

In fact, in the approaches where gravity arises as an effective theory, spacetime itself appears often as
an emergent phenomenon. Nowadays this is a huge research area and we do not aim to review it here.
Instead we would like to remark that the loop and spin foam quantizations fit this emergent strategy
because they both suggest that the continuous spacetime appears only in the classical limit
whereas the quantum geometry is intrinsically discrete.
Moreover, there is a close analogy with the situation in the above mentioned matrix models and their
fermionic description. Indeed, in both cases the vacuum represents a ``no space" state such that
it cannot be associated with any spacetime. To get the latter one should consider states with a
large number of excitations, which in matrix models corresponds to the famous large $N$ limit and its more refined version,
the double scaling limit \cite{Brezin:1990rb,Douglas:1989ve,Gross:1989vs}.

This analogy is not a surprise given that the GFT models, considered in the previous section and supposed
to be a fundamental formulation of spin foams, have been constructed as generalization of matrix models.
However, it also indicates that in LQG and SF there is a tension between two ingredients:
whereas spacetime is emergent, the fundamental action is taken to be the one which describes this emergent spacetime.
The above discussion suggests that it would be more natural to have some fundamental quantum theory
of spin networks or spin foams which knows nothing about the Einstein action,
except that it appears in the infrared limit, and is defined instead
using some natural symmetry or other principles. It is not clear how such a theory can be found as
it certainly requires additional insights. Note however that the recent research on GFT models
goes essentially in this direction \cite{Oriti:2007vf,Oriti:2009nd} and it has been argued that
the whole GFT framework can be seen as a realization of this idea \cite{Oriti:2007qd}.
Hopefully, it will allow to end up with a reliable model free of the inconsistencies
which have been discussed in this review.

\newpage

\section*{Acknowledgements}

The authors are grateful to B. Dittrich, L. Freidel, C. Rovelli and S. Speziale for useful discussions
and especially to K. Krasnov and D. Oriti for the careful reading of the manuscript and many valuable comments.
S.A. would like to thank Perimeter Institute for Theoretical Physics for the kind hospitality and the financial support.
This research is supported by contract ANN-09-BLAIN-0041 and in part by Perimeter Institute.

\providecommand{\href}[2]{#2}\begingroup\raggedright\endgroup

\end{document}